\documentclass[superscriptaddress,
longbibliography,
amsmath,amssymb,aps, prf, preprint]{revtex4-1}
\usepackage{graphicx}
\usepackage{epstopdf, epsfig}
\usepackage{graphicx}
\usepackage{dcolumn}
\usepackage{bm}
\usepackage{epsfig} 
\usepackage{amsmath}
\usepackage{float}
\usepackage[utf8]{inputenc}
\usepackage[T1]{fontenc}
\usepackage{nomencl}
\usepackage{CJKutf8}
\usepackage{xcolor}
\usepackage{caption}
\usepackage{subcaption}
\usepackage{makecell}
\usepackage{soul}
\usepackage{etoolbox}

\usepackage{graphicx}
\usepackage{adjustbox}
\usepackage{amssymb}
\usepackage{mathrsfs}

\usepackage{nomencl}
\makenomenclature
\begin{document}

\title{An investigation of anisotropy in the bubbly turbulent flow via direct numerical simulations}

\author{Xuanwei Zhang (\begin{CJK*}{UTF8}{gbsn}张轩玮\end{CJK*})}
\affiliation{Cluster of Excellence SimTech (SimTech), University of Stuttgart,
Pfaffenwaldring 5a, 70569 Stuttgart, Germany}

\author{Yanchao Liu (\begin{CJK*}{UTF8}{gbsn}刘雁超\end{CJK*})}
\affiliation{Institute of Aerospace Thermodynamics, University of Stuttgart,
Pfaffenwaldring 31, 70569 Stuttgart, Germany}

\author{Wenkang Wang (\begin{CJK*}{UTF8}{gbsn}王文康\end{CJK*})}
\affiliation{Max Planck Institute for Intelligent Systems, Heisenbergstraße 3, 70569, Stuttgart, Germany}

\author{Guang Yang (\begin{CJK*}{UTF8}{gbsn}杨光\end{CJK*})}
\affiliation{Institute of Refrigeration and Cryogenics, Shanghai Jiao Tong University, 200240 Shanghai, China}

\author{Xu Chu (\begin{CJK*}{UTF8}{gbsn}初旭\end{CJK*})}
\email{xu.chu@simtech.uni-stuttgart.de}
\affiliation{Cluster of Excellence SimTech (SimTech), University of Stuttgart,
Pfaffenwaldring 5a, 70569 Stuttgart, Germany}
\affiliation{Department of Engineering, University of Exeter, UK}


\begin{abstract}

This study explores the dynamics of dispersed bubbly turbulent flow in a channel using interface-resolved direct numerical simulation (DNS) with an efficient Coupled Level-Set Volume-of-Fluid (CLSVOF) solver. 
The influence of number of bubbles (96 and 192), flow direction, and Eötvös number was examined across eight distinct cases. The results indicate that in upward flows, bubbles tend to accumulate near the wall, with smaller Eötvös numbers bringing them closer to the wall and enhancing energy dissipation through increased turbulence and vorticity. This proximity causes the liquid phase velocity to attenuate, and the bubbles, being more spherical, induce more isotropic turbulence. Conversely, in downward flows, bubbles cluster in the middle of the channel and induce additional pseudo-turbulence in the channel center, which induce additional turbulent kinetic energy in the channel center. The study further examines budget of Turbulent Kinetic Energy (TKE) and the exact balance equation for the Reynolds stresses, revealing that near-wall bubble motion generates substantial velocity gradients, particularly in the wall-normal direction, significantly impacting the turbulence structure.  

\end{abstract}
\maketitle

\section{Introduction}

Dispersed turbulent bubbly flow plays an important role in industrial applications, such as boiling water reactors in the nuclear industry, bubble columns in the chemical metallurgical industry, and heat exchangers in thermal power plants \citep{Lohse2018bubble, elghobashi2019direct}. The performance of these systems depends partly on the turbulence in the liquid phase, as it significantly influences the local distribution of the dispersed phase. Furthermore, bubble fragmentation and coalescence also dictate the distribution of bubble sizes. On the other hand, it relies on the distribution of the dispersed phase, which affects both the large-scale velocity fields and the small-scale turbulence \cite{Ma2020}, thereby facilitating interactions. Additionally, parameters such as bubble velocity, bubble diameter, and the capability for bubble deformation directly impact the transfer of momentum, heat, and mass \citep{Liao2015baseline}. In experiments, individual measurements of bubbles and liquid velocity fluctuations present a challenge. Optical measurement techniques such as Particle Image Velocimetry (PIV), Laser Doppler Anemometry (LDA), and Particle Tracking Velocimetry (PTV) are limited to bubbly flows with a gas fraction lower than 3\% due to the potential occlusion between bubbles. Overall, these methods can only provide effective measurements under limited conditions~\citep{Kim2016, Martinez2010}.

Numerical studies of bubbly flows can be conducted through various approaches, primarily classified into direct numerical simulations (DNS), Euler-Lagrange (EL), and Euler-Euler (EE) methods. Among these, DNS employs an interface resolving method to accurately treat the phase boundary and fully resolve the bubble geometry along with the interfacial effects. While DNS for multiphase flows requires substantial computational resources than those of single-phase \citep{Lozano2012,lozano2014effect, chu2018flow, chu2019direct, chu2020instability, pandey2018buoyancy, wang2021assessment}, many studies are therefore limited to the examination of individual or pairs of droplets/bubbles due to these computational demands \citep{hasslberger2018flow, pesci2018computational, yang2020droplet, yi2022numerical, Vaikuntanathan2022}. However, with the advancement of numerical techniques, the development of DNS for multi-phase flows has advanced rapidly \cite{Ervin1997, ROGHAIR2011, lu2013dynamics, BOLOTNOV2011}, and studies now encompass thousands of bubbles \citep{cifani2018highly, li2019mass, trautner2021conditional, zhang2021direct}, droplets \citep{dodd2016interaction, Cannon2021,liu2024simulation,yang2020droplet}, or fluid jets \citep{liu2021simulation,liu2023large}. Nevertheless, these techniques require significant computational resources and are not suitable for modeling large industrial systems. For large-scale simulations, the Euler-Euler (EE) approach coupled with Reynolds-averaged Navier-Stokes (RANS) modeling is the only viable framework \citep{IshiiHibiki2006}. The accuracy of the models largely depends on the closure models employed. Developing a suitable model for bubble-induced turbulence (BIT) is currently a central topic of investigation, providing crucial support for simulating multiphase flow engineering applications that involve bubbles.

Many studies are based on the rigorous derivation by \citet{Kataoka1989} of the basic equations of turbulence in gas-liquid two-phase flow. Through ensemble averaging, the local instantaneous conservation equation for averaged turbulent energy is obtained. This equation includes terms for diffusion, turbulent dissipation, turbulent production, and the interfacial transport term of turbulent energy, which incorporates the interfacial area concentration. These equations have also been utilized within the RANS modeling framework for bubbly flows, and various forms of the BIT model have been proposed, e.g., by \citet{Troshko2001}, \citet{Politano2003}, \citet{Bertodano1994}, \citet{COLOMBO2015}, and \citet{ma2017direct}. Based on DNS data, it is possible to accurately assess the budget of Turbulent Kinetic Energy (TKE), which provides a pathway for the modeling of BIT source terms and the evaluation of available models. However, the direct utilization of DNS data to assist BIT modeling is still rare \citep{ma2017direct}. Current BIT models frequently fail to deliver consistent accuracy across different conditions, making the use of DNS data to enhance BIT modeling a crucial area for further research.

However, EE k-$\epsilon$ type models for bubbly flows exhibit some issues. The numerical values required for modeling may fluctuate with changes in the void fraction, leading to deficient predictions. Simultaneously, the eddy viscosity models fail to account for the anisotropic velocity fluctuations caused by the buoyancy-generated rise of bubbles in the liquid. Measuring the anisotropy induced by bubbles based on DNS data is a key focus of this study.

 An alternative to eddy-viscosity models are differential second-moment closures (SMC). Unlike eddy-viscosity models that use the k-equation, SMC employs balance equations for the Reynolds stresses of the liquid. \citet{Kataoka1992} also rigorously derived the exact Reynolds-stress equations for two-phase flows, where additional terms can similarly serve as data for DNS analysis. So far, only \citet{Ma2020} has combined DNS data with the SMC framework to develop BIT models, focusing on the pressure–strain term in the application of bubbly flows.

The organization of the paper is as follows: Section II provides an overview of the DNS-related simulation details, including the governing equations, numerical methods, and the computational domain used in the study. Section III discusses key results and significant findings from the simulations. Finally, Section IV concludes the paper with a summary of findings.

\section{Simulation details}

\subsection{Governing equations and computational method}
\label{subsec:2.1}

In the analysis of incompressible flows of two immiscible fluids, the fundamental dynamics are described by the nondimensionalized Navier-Stokes equations, specified as follows:

\begin{equation}
\nabla \cdot\mathbf{u}=0,
\label{eq:1}
\end{equation}
\begin{equation}
\frac{\partial \mathbf{u}}{\partial t}+\nabla \cdot(\mathbf{u u}) = -\frac{1}{\rho}\nabla p + \frac{1}{\rho Re_{\tau}}\nabla \cdot\pmb{\tau} + \frac{1}{Fr}\mathbf{g} + \frac{1}{\rho We}\mathbf{F_{\sigma}}, 
\label{eq:2}
\end{equation}


Here $\pmb{\tau} = \mu(\nabla\mathbf{u} + \nabla\mathbf{u}^{T})$ symbolizes the dimensionless viscous stress tensor, $\mathbf{g}$ represents body force and $\mathbf{F_{\sigma}} = \kappa \delta (f) \mathbf{n}$ is surface tension force. Here, $f$ is identified as the Level-Set (LS) function, characterized as a signed distance from the interface. Regions with $f$ greater than 0 are designated as liquid, whereas regions with $f$ less than 0 are classified as gas. The collection of points where $f$ equals 0 implicitly outlines the free surface. $\mathbf{n} = \frac{\nabla f}{|\nabla f|}
$ denotes the unit normal vector pointing from the liquid phase to the gas phase.  $\kappa = - \nabla\cdot\mathbf{n}$ signifies the curvature of the interface. The purpose of the Dirac function $\delta(f)$ is to localize the surface tension forces $\mathbf{F_{\sigma}}$, ensuring that they act exclusively at the interface location. The symbols $t$, $\rho$, $\mu$, $\mathbf{u}$, and $p$  represent the dimensionless time, density, dynamic viscosity, velocity vector and hydrodynamic pressure, respectively. The variables are normalized by the characteristic length $h^\ast$ and the friction velocity $u_{\tau}^\ast$, which is defined as $\sqrt{\tau_w/\rho}$, $\tau_w$ the average wall shear,in the following manner:

\begin{equation}
t = \frac{t^\ast}{h^\ast/u_{\tau}^\ast}, \quad x_i = \frac{x_i^\ast}{h^\ast}, \quad \rho = \frac{\rho^\ast}{\rho_c^\ast}, \quad \mu = \frac{\mu^\ast}{\mu_c^\ast}, \quad u_i = \frac{u_i^\ast}{u_{\tau}^\ast}, \quad p = \frac{p^\ast}{\rho_c^\ast u_{\tau}^{\ast2}}.
\label{eq:3}
\end{equation}
Here the superscript $^*$ indicates dimensional quantities, while the subscripts $_c$ and $_d$ distinguish between the continuous and dispersed phases, respectively. Three non-dimensional numbers, Reynolds number $Re_{\tau}$, Froude number $Fr$ and Weber number $We$, are specified as follows:
\begin{equation}
Re_{\tau} = \frac{\rho^\ast_c u_{\tau}^\ast h^\ast}{\mu_c^\ast}, \quad Fr = \frac{u_{\tau}^{2\ast}}{g^\ast h^\ast}, \quad We = \frac{\rho_c^\ast u_{\tau}^{\ast2} h^\ast}{\sigma^\ast}, 
\label{eq:4}
\end{equation}
where $\sigma^\ast$ is the surface tension coefficient. Using the Heaviside function
\begin{equation}
H(f) = \begin{cases}
	    1 & f>0\\
	    0 & \text{otherwise}.
	  \end{cases}
\label{eq:5}
\end{equation}

the fluid density and the viscosity in Eqn. \eqref{eq:3} can be written as:

\begin{equation}
\rho = \rho_c H(f) + \rho_d (1-H(f)),
\label{eq:6}
\end{equation}
\begin{equation}
\mu = \mu_c H(f) + \mu_d (1-H(f)),
\label{eq:7}
\end{equation}

Following~\citet{Chang1996level}, to regularize the viscosity, the Heaviside function $H$ is substituted by the subsequent smoothed version as:
\begin{equation}
H(f) = \begin{cases}
	    0 & f<-\chi\\
	    \frac{1}{2}\left(1+\frac{f}{\chi}+\frac{1}{\pi}sin\left(\frac{\pi f}{\chi}\right)\right) & |f|\leq\chi\\
	    1 & f>\chi,
	  \end{cases}
\label{eq:8}
\end{equation}

where $\chi$ is a function of grid size (typically $\chi=1.5\Delta$). To accurately model the moving interface of the dispersed and continuum phases, both the Lagrangian and Eulerian methods are feasible options. In the Lagrangian context, as detailed by Tryggvason et al. \citep{tryggvason2001front}, the interface between the phases is tracked by the marker points that are defined at the interface. This method involves initially advancing the front, followed by the formulation of a grid density field tailored to align with the front's new location, effectively capturing the dynamic evolution of interfaces within the computational domain. Front-capturing methods are represented by the level-set (LS) method and the volume-of-fluid (VOF) method \citep{lafaurie1994modelling, SCHLOTTKE2008}. \citet{sussman2000coupled} combined these two methods into the Coupled Level-Set Volume-of-Fluid (CLSVOF) method, which exploits the strengths of both methods. A level-set function is employed to accurately compute the position of the interface, and a VOF function is responsible for volume and mass conservation. The equations governing the interface motion are

\begin{equation}
\frac{\partial f} {\partial t} + \mathbf{u}\cdot\nabla f = 0.
\label{eq:9}
\end{equation}
However, According to \citet{Rider1995}, the level-set method exhibits a significant mass loss problem. Therefore, the introduction of the Volume-of-Fluid (VOF) function, $\psi$, is motivated by the necessity to conserve mass while still preserving a sharp delineation of the interface. 
This term denotes the discrete volume fraction within a computational cell, characterized by the integral of the Level-Set (LS) function over the cell.
\begin{equation}
\psi(t) = \frac{1}{\Omega}\int_{\Omega}H(f(\mathbf{x},t))d\Omega,
\label{eq:10}
\end{equation}

where $\Omega$ denotes the volume of a computational cell. \(\psi = 0\) corresponds to the continuous phase, whereas \(\psi = 1\) corresponds to the dispersed phase. If the cell is transected by the interface, then \(0 < \psi < 1\). The advection of the volume fraction is performed by the following mass conservation
\begin{equation}
\frac{\partial \psi} {\partial t} + \nabla\cdot(\mathbf{u}\psi) = 0.
\label{eq:11}
\end{equation}

The level set function is updated with the exact signed normal distance to the reconstructed interface, thereby facilitating the coupling with the volume fraction described by the VOF function. The specific iterative details and processes are detailed in \citet{sussman2000coupled}.


The governing equations are solved using a second-order central difference scheme for spatial discretization on a staggered and equidistant Cartesian grid.
Typically used fast Poisson solvers, which employ a combination of fast Fourier transforms (FFT) and Gauss elimination, are expressed as follows:

\begin{equation}
\nabla\cdot\left(\frac{1}{\rho^{n+1/2}}\nabla p^{n+1/2}\right) = \frac{1}{\Delta t}\nabla\cdot\hat{\bf{u}}
\label{eq:12}
\end{equation}

Due to the significant variations in the density \(\rho\) at the interface between two phases in two-phase flows, the coefficient in the Poisson equation is no longer constant, causing the coefficient \(\frac{1}{\rho}\) to vary in space and time. Consequently, iterative methods are employed to solve it, and a fast pressure-correction method developed by \citet{dodd2014fast} is applied, as shown in the following equation:

\begin{equation}
\nabla^2\cdot p^{n+1/2} = \nabla\cdot\left(\left(1-\frac{\rho^{(0)}}{\rho^{n+1/2}}\right)\nabla \hat{p}\right) + \frac{\rho^{(0)}}{\Delta t}\nabla\cdot\hat{\bf{u}},
\label{eq:13}
\end{equation}
where $\rho^{(0)}=min(\rho_c,\rho_d)$ for numerical stability \cite{Dong2012}, $\hat{p} = 2p^{n-1/2}-p^{n-3/2}$ is a linear approximation of the pressure from the previous time levels. The predicted velocity field \(\hat{\mathbf{u}}\) can be computed using \(\mathbf{u}^n\) and \(\mathbf{u}^{n-1}\),

\begin{equation}
\hat{\bf{u}} = \mathbf{u}^n +\Delta t \left( -\frac{3}{2}\mathcal{A}(\mathbf{u}^n)+\frac{1}{2}\mathcal{A}(\mathbf{u}^{n-1}) \right)+\frac{\Delta t}{\rho^{n+1/2}} \left( \frac{3}{2}\mathcal{D}(\mathbf{u}^n)-\frac{1}{2}\mathcal{D}(\mathbf{u}^{n-1}) \right)+ \frac{\Delta t}{Fr}\mathbf{g} + \frac{\Delta t}{\rho We}\mathbf{F_{\sigma}}^{n+1/2}, 
\label{eq:14}
\end{equation}
where $\mathcal{A}$ and $\mathcal{D}$ are discretized forms of $\nabla \cdot(\mathbf{u u})$ and $\frac{1}{\rho Re_{\tau}}\nabla \cdot\pmb{\tau}$.

Under periodic boundary conditions, the right-hand side of equation \eqref{eq:13} is transformed into the frequency domain via FFT. In this Fourier space, the Poisson equation becomes an algebraic equation. This equation can be efficiently solved using the Gauss elimination method, enabling the computation of \(p^{n+\frac{1}{2}}\) in Fourier space. The numerical solver's validity and previous applications are documented in the works of \citet{talebanfard2019heat}, \citet{chu2020turbulence},  \citet{nemati2021direct}, and \citet{Liu2023}.


\subsection{Computational domain and boundary conditions}

\begin{figure*}        
        \includegraphics[height=7cm,keepaspectratio]{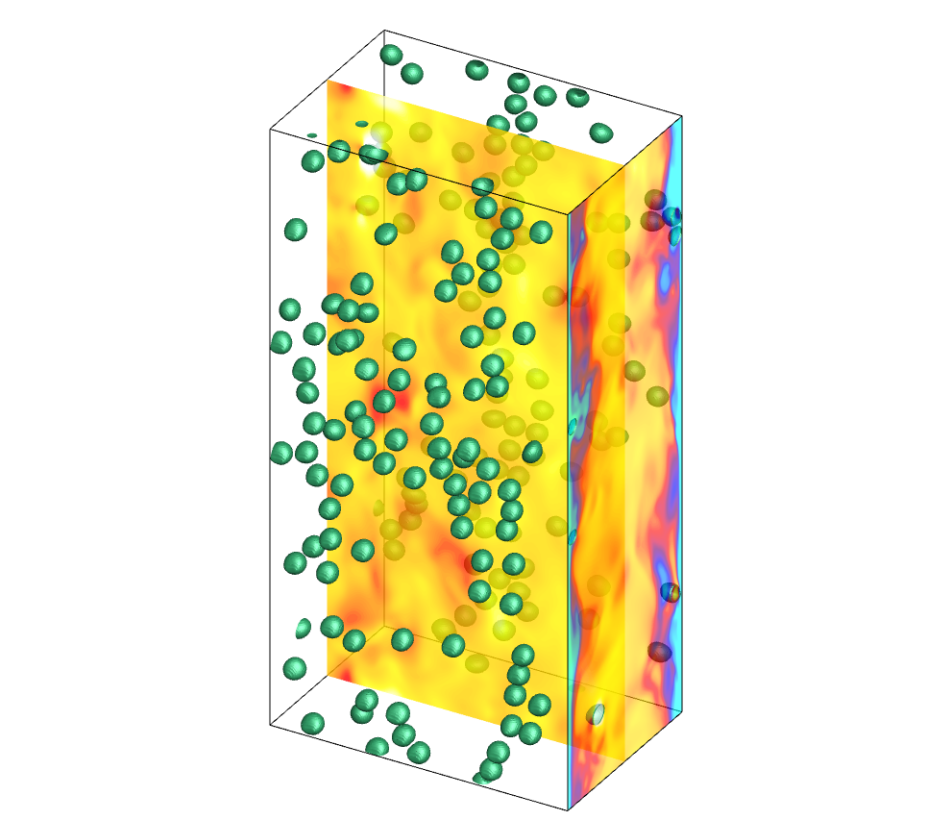}
        \includegraphics[height=7cm,keepaspectratio]{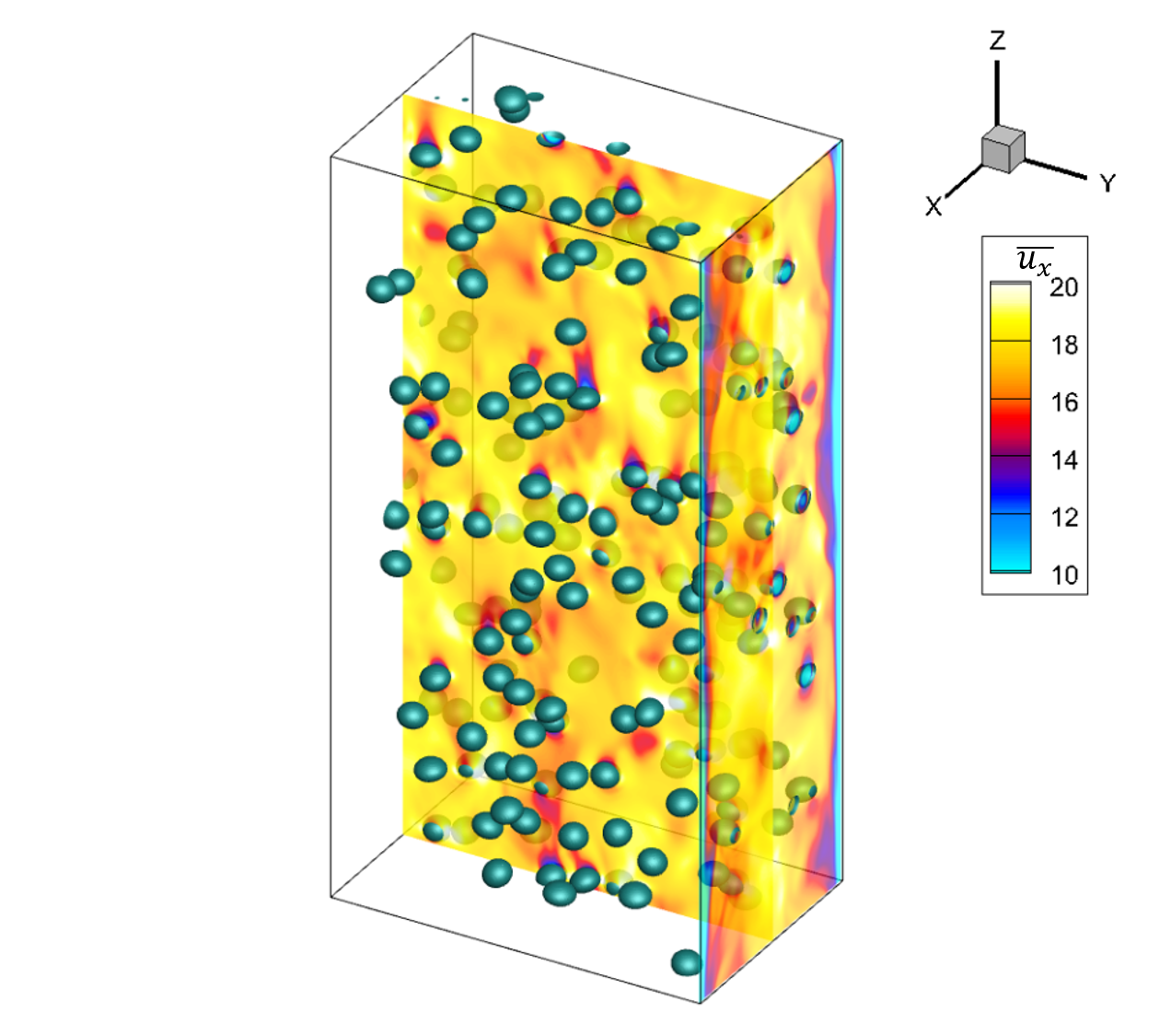}
    \caption{An illustration of the computational domain: (a) shows the Instantaneous DNS data for the case 192Eö0.5U,  (b) displays the case 192Eö0.5D. The vertical plane with the contour plot represents the instantaneous streamwise liquid velocity.}
    \label{fig:computational_domain}   
\end{figure*}

The DNS simulations were performed for both upward and downward flows in a rectangular channel bounded by two flat vertical walls, featuring periodic boundary conditions in the streamwise (x) and spanwise (z) directions, and enforcing a no-slip condition for the liquid phase on both walls. The size of the domain is \(L_x \times L_y \times L_z = 6h^\ast \times 2h^\ast \times 3h^\ast\) with the domain discretized for all DNS cases by a cubic mesh of \(576 \times 192 \times 288\) points in the streamwise, wall-normal, and spanwise directions with the same step size \(\Delta= \frac{2}{192}\). And $h^\ast$ is the half channel width. The domain length in the streamwise and spanwise directions is approximately twice that of \citet{lu2013dynamics}, specifically $\pi h^* \times 2h^* \times \frac{\pi h^*}{2}$, and considerably larger than that reported by \citet{ma2017direct} which was $4.41h^* \times h^* \times 2.21h^*$. Additionally, it is identical to the domain used in our previous publications \citep{Liu2023}. Hence, it is assumed that the domain size is sufficiently large to enable a fully-developed turbulent field. It includes two vertical no-slip walls perpendicular to the \(y\)-direction and utilizes periodic boundary conditions in the other directions.
Figure.~\ref{fig:computational_domain}(b) illustrates the domain and presents an instantaneous snapshot of the bubbly flow for one of the DNS upward cases, whereas FIG.~\ref{fig:computational_domain}(a) does the same for one of the DNS downward cases. The gravitational force is oriented in the negative z-direction. The chosen dimensionless parameters for the flow equations are as follows:

\begin{equation}
Re_{\tau} = \frac{\rho_c^\ast u_{\tau}^\ast h^\ast}{\mu_c^\ast}=180, \quad Fr = \frac{u_{\tau}^{\ast2}}{g^\ast h^\ast}=2.6422\times10^{-2}, \quad We = \frac{\rho_c^\ast u_{\tau}^{\ast2} h^\ast}{\sigma^\ast} 
\label{eq:15}
\end{equation}


In the simulations, a density ratio of \(\rho_d/\rho_c=0.03\) and a viscosity ratio of \(\mu_d/\mu_c=0.018\) are applied. The density of the continuous phase is specified as 1000~\(\mathrm{kg/m^3}\). To describe the deformability of bubbles, two dimensionless parameters, the Eötvös number and the Morton number, are defined as follows:

\begin{equation}
    Eo = \frac{(\rho_c^\ast - \rho_d^\ast)g^*D_0^{\ast2}}{\sigma^\ast}, \quad 
    Mo = \frac{(\rho_c^\ast - \rho_d^\ast)g^\ast\mu_c^{\ast4}}{\rho_c^\ast\sigma^{\ast3}}.
\end{equation}

The specifics of the DNS cases are presented in Table~\ref{table1}. The model is equipped with separate marker functions for each bubble within the domain, which prevents numerical coalescence in the solution; consequently, the number of bubbles remains constant over time. It is important to highlight that this work does not account for the effects of bubble breakup. A total of eight cases are primarily classified by the number of bubbles, E\"{o}tv\"{o}s number, and flow direction. The initial bubble radius is kept constant, equivalent to ten grid lengths \((10\Delta)\), and the non-dimensional initial bubble radius is defined as \(R_0 = R_0^{\ast}/h^{\ast} = 0.1042\), consistent across all eight cases. The number of bubbles is set to either 96, corresponding to a void fraction of \(\alpha_G=1.263\%\), or 192, corresponding to \(\alpha_G=2.525\%\), uniformly initialized within the channel. Only two E\"{o}tv\"{o}s numbers are considered: 0.5 and 2. The flow direction is categorized into upward and downward flows. For instance, '96Eo0.5D' indicates a case with 96 bubbles and an E\"{o}tv\"{o}s number of 0.5, with 'D' signifying downward flow, while 'U' denotes upward flow. 20 cell points are placed on each bubble diameter \(D_0^{\ast}\). As a reference, \citet{lu2013dynamics,lu2006numerical} allocated 16 grid points per bubble diameter; \citet{cifani2018highly} placed 20 grid points along the bubble diameter. Additionally, \citet{Liu2023} placed 31 grid points along the bubble diameter.

The driving force in the turbulent channel is represented by a constant nondimensional pressure gradient:

\begin{equation}
\frac{dp_m}{dx} = \frac{dp}{dx} + \frac{1}{Fr}\rho_{av} g,
\label{eq:16}
\end{equation}

where $\rho_{av}$ is the volume averaged density, and the applied driving force is $\frac{dp_m}{dx}=1$ for the downward flow. The simulations is initially conducted for single-phase flows and has been run for a sufficient duration to ensure the full development of turbulence. Subsequently, bubbles corresponding to different cases are uniformly dispersed into the flow field. The fluid flow is then sustained, and simulations continue for over fifteen flow-through times to achieve statistical convergence.



\begin{table}
\begin{tabular}{c|c|c|c|c|c|c}
\hline
Case 	& {\makecell[c]{bubble volume \\fraction $\alpha_G$ }}	& {\makecell[c]{ number of \\bubbles $N$}}		&{\makecell[c]{  Initial bubble \\radius ($R_0
$)}}	& {\makecell[c]{ Weber number\\$We$}} &{\makecell[c]{ Eötvös number\\ $Eo$}} & {\makecell[c]{ Morton number\\ $Mo$}}\\
\hline\hline
96Eö0.5D   &   1.263\%   	&  96		& 0.1042	    &	0.30746 	& 0.5 & 1.6209$\times10^{-8}$\\
96Eö0.5U 	&   1.263\%    	&  96		& 0.1042	    &	0.30746 	& 0.5 & 1.6209$\times10^{-8}$\\
96Eö2D  	&   1.263\%    	&  96 		& 0.1042	    &	1.2298	   & 2 & 1.0374$\times10^{-6}$\\
96Eö2U  	&   1.263\%    	&  96 		& 0.1042	    &	1.2298	   & 2 & 1.0374$\times10^{-6}$\\
192Eö0.5D  	&   2.525\%     	&  192 		& 0.1042	    &	0.30746 	& 0.5 & 1.6209$\times10^{-8}$\\
192Eö0.5U  	&   2.525\%     	&  192 		& 0.1042	    &	0.30746 	& 0.5 & 1.6209$\times10^{-8}$\\
192Eö2D  	&   2.525\%     	&  192 		& 0.1042	    &	1.2298	   & 2 & 1.0374$\times10^{-6}$\\
192Eö2U  	&  2.525\%     	&  192 		& 0.1042	    &	1.2298	   & 2 &1.0374$\times10^{-6}$\\
\hline
\end{tabular}
\caption{DNS case parameters for the current study are summarized as follows: The numerical labels '96' and '192' denote cases with 96 and 192 bubbles, respectively. 'Eö0.5' and 'Eö2' refer to Eötvös numbers of 0.5 and 2, respectively. 'U' signifies upward flow, while 'D' indicates downward flow.}
\label{table1}
\end{table}

\section{Turbulence statistics}

\begin{figure*}        
    \centering
    \begin{subfigure}[b]{0.5\textwidth} 
        \centering
        \includegraphics[height=6.4cm,keepaspectratio]{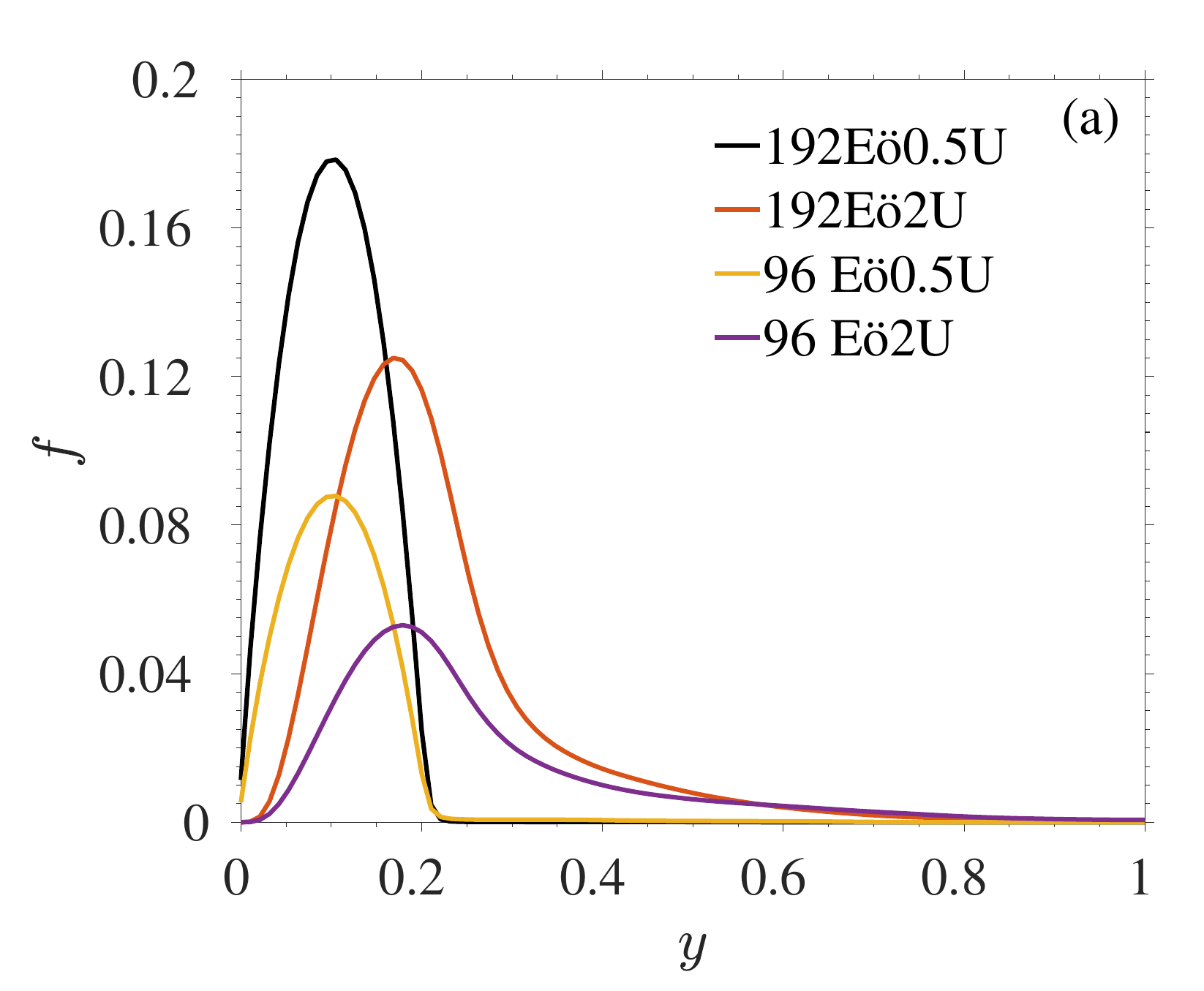}
        \label{fig:volume_fraction_downward}
    \end{subfigure}%
    \begin{subfigure}[b]{0.5\textwidth} 
        \centering
        \includegraphics[height=6.4cm,keepaspectratio]{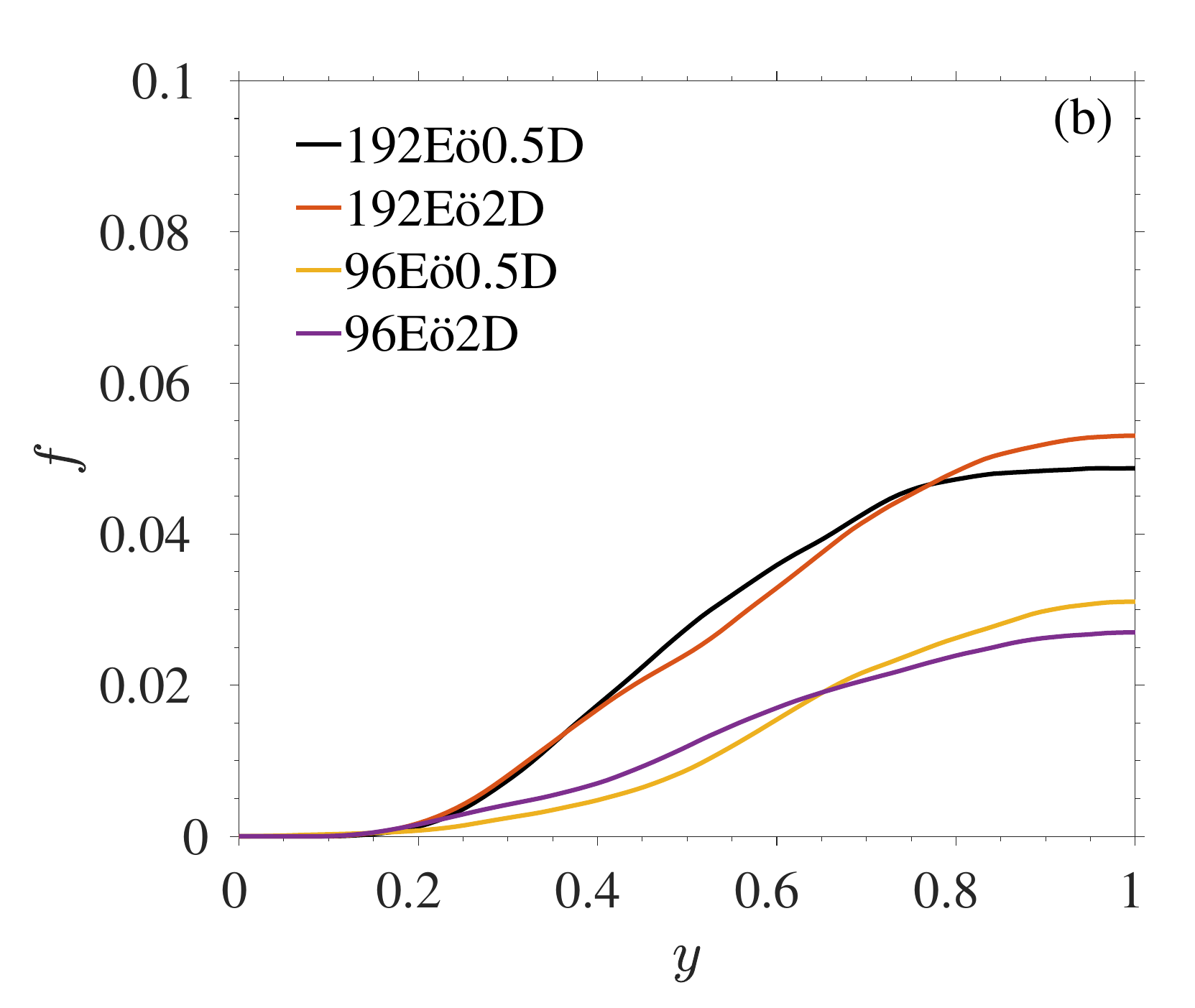}
        \label{fig:volume_fraction_upward}
    \end{subfigure}
    \caption{Bubble volume fraction $\alpha_G$ as a function of wall distance}
    \label{fig:volume_fraction}
\end{figure*}

Figure~\ref{fig:volume_fraction} demonstrates the relationship between the bubble volume fraction $\alpha_G$ and the distance to the wall $y$, with subfigures (a) and (b) depicting the upward and downward cases, respectively. The volume fraction $\alpha_G$ is determined by averaging over both time and spatial dimensions, including streamwise and spanwise directions. In a vertical shear flow, the lift force acting on a clean spherical bubble is directed towards the faster-moving fluid side relative to the bubble, in a frame of reference moving with the bubble. This dynamic affects the average mixture density, which depends on the number density of bubbles. Thus, lateral motion increases the density of the mixture where bubbles are lost and decreases it where bubbles accumulate. In upflow, the imposed pressure gradient that drives the flow exceeds the force of gravity, and the excess pressure gradient is balanced by the shear forces in the mixtures. As bubbles migrate from the center of the channel towards the walls, the mixture density increases until the weight of the mixture is balanced by the pressure gradient, leading to zero shear and stopping the lateral migration of the bubbles. 

\begin{figure*}    
    \centering
    \begin{subfigure}[b]{0.5\textwidth} 
        \centering
        \includegraphics[height=6.5cm,keepaspectratio]{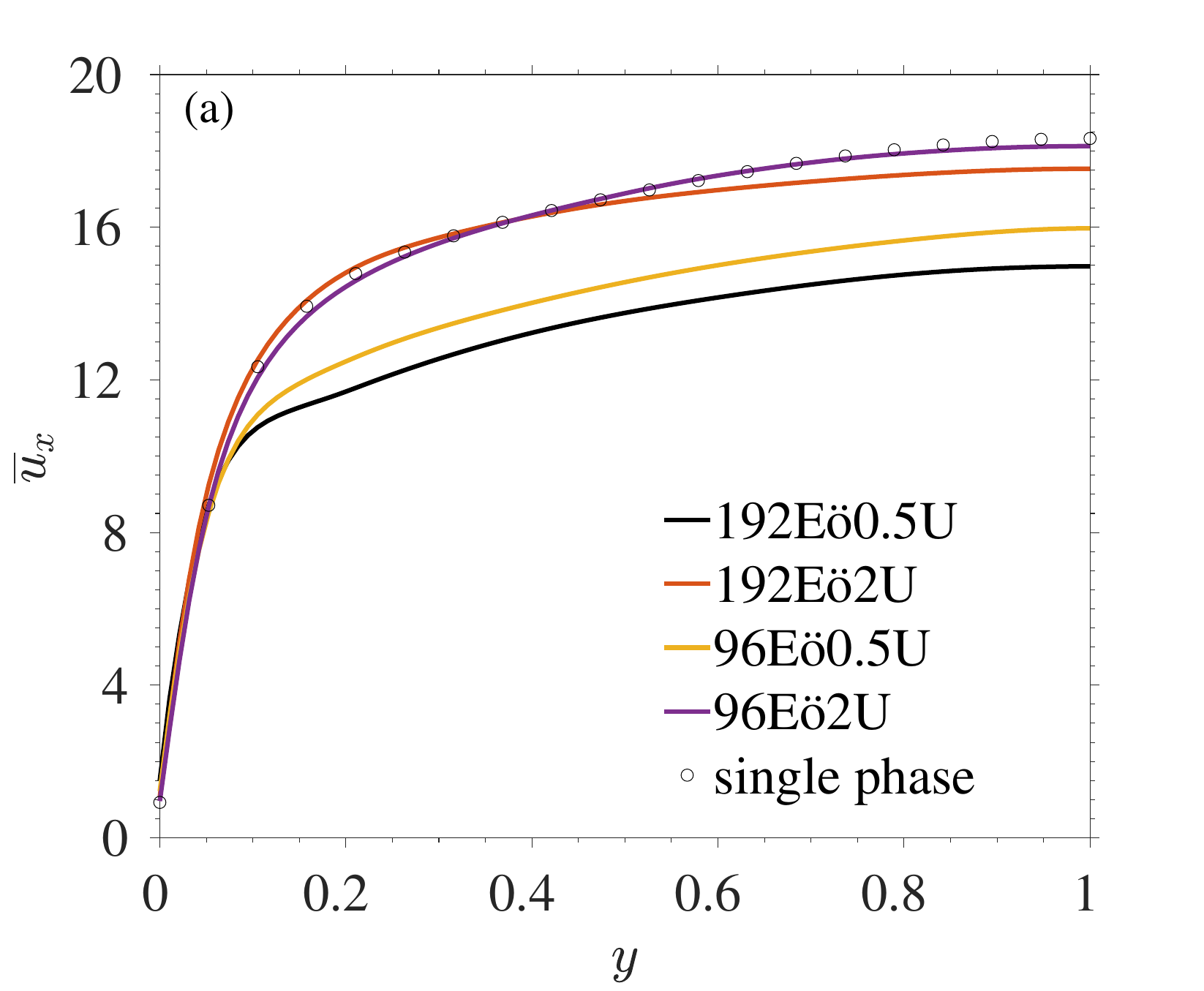}
        \label{fig:downward}
    \end{subfigure}%
    \hspace{-5mm} 
    \begin{subfigure}[b]{0.5\textwidth} 
        \centering
        \includegraphics[height=6.5cm,keepaspectratio]{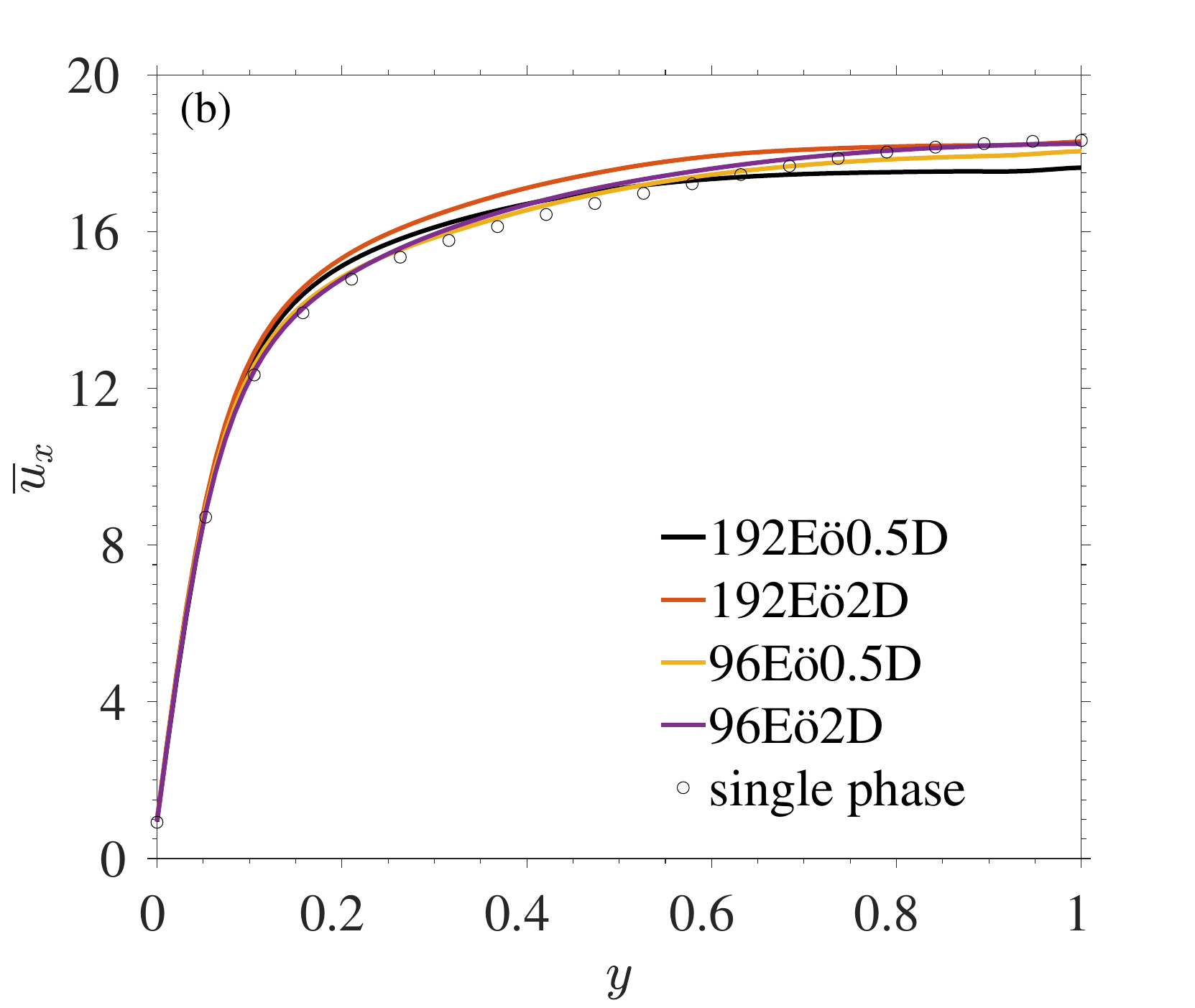}
        \label{fig:upward}
    \end{subfigure}
    \caption{Liquid phase-averaged mean velocity profile}
    \label{fig:Mean_velocity_liquid}
\end{figure*}

\begin{figure*}		
    \centering
    \begin{subfigure}[b]{0.5\textwidth} 
        \centering
        \includegraphics[height=6.5cm,keepaspectratio]{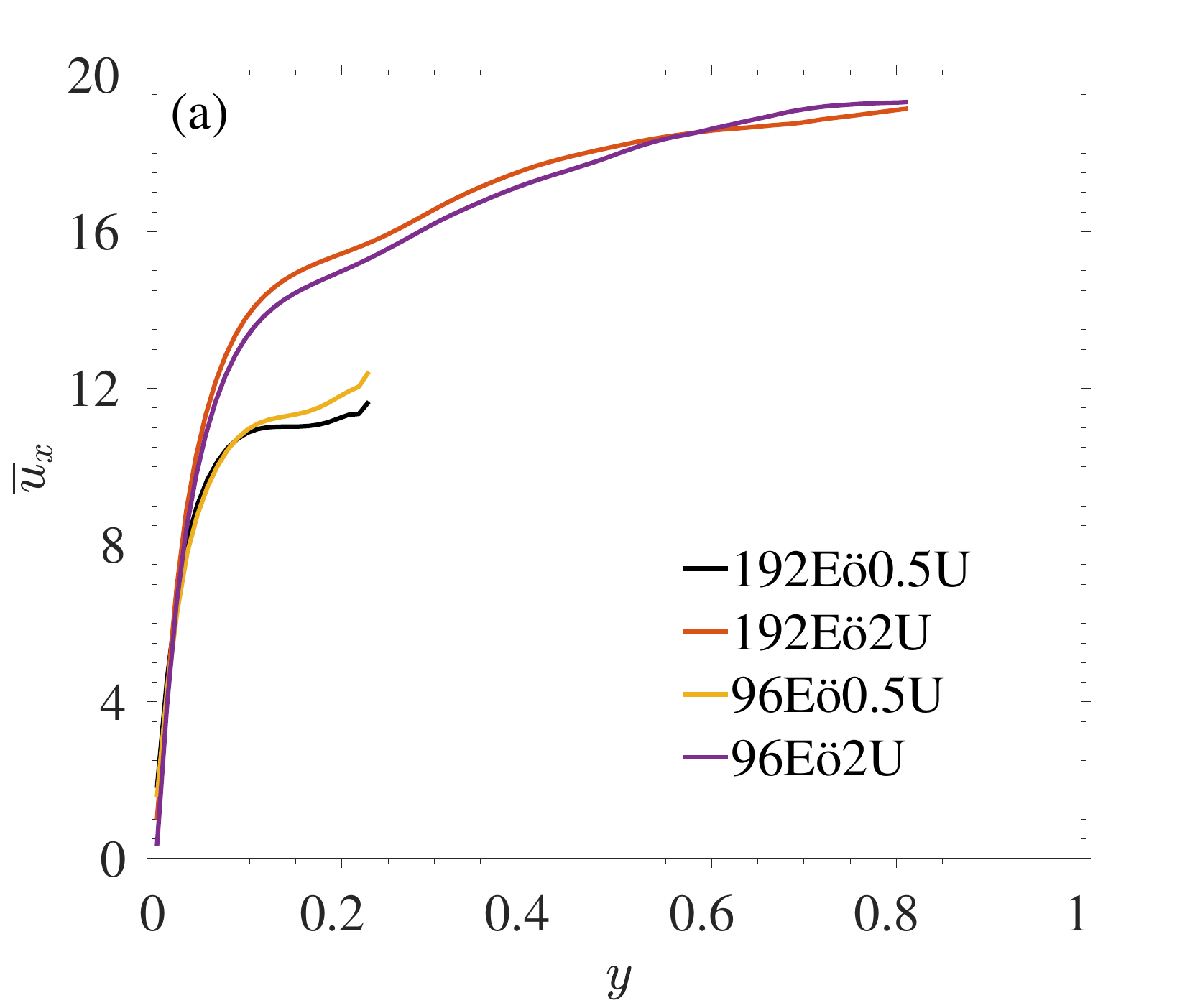}
        \label{fig:downward}
    \end{subfigure}%
    \hspace{-5mm} 
    \begin{subfigure}[b]{0.5\textwidth} 
        \centering
        \includegraphics[height=6.5cm,keepaspectratio]{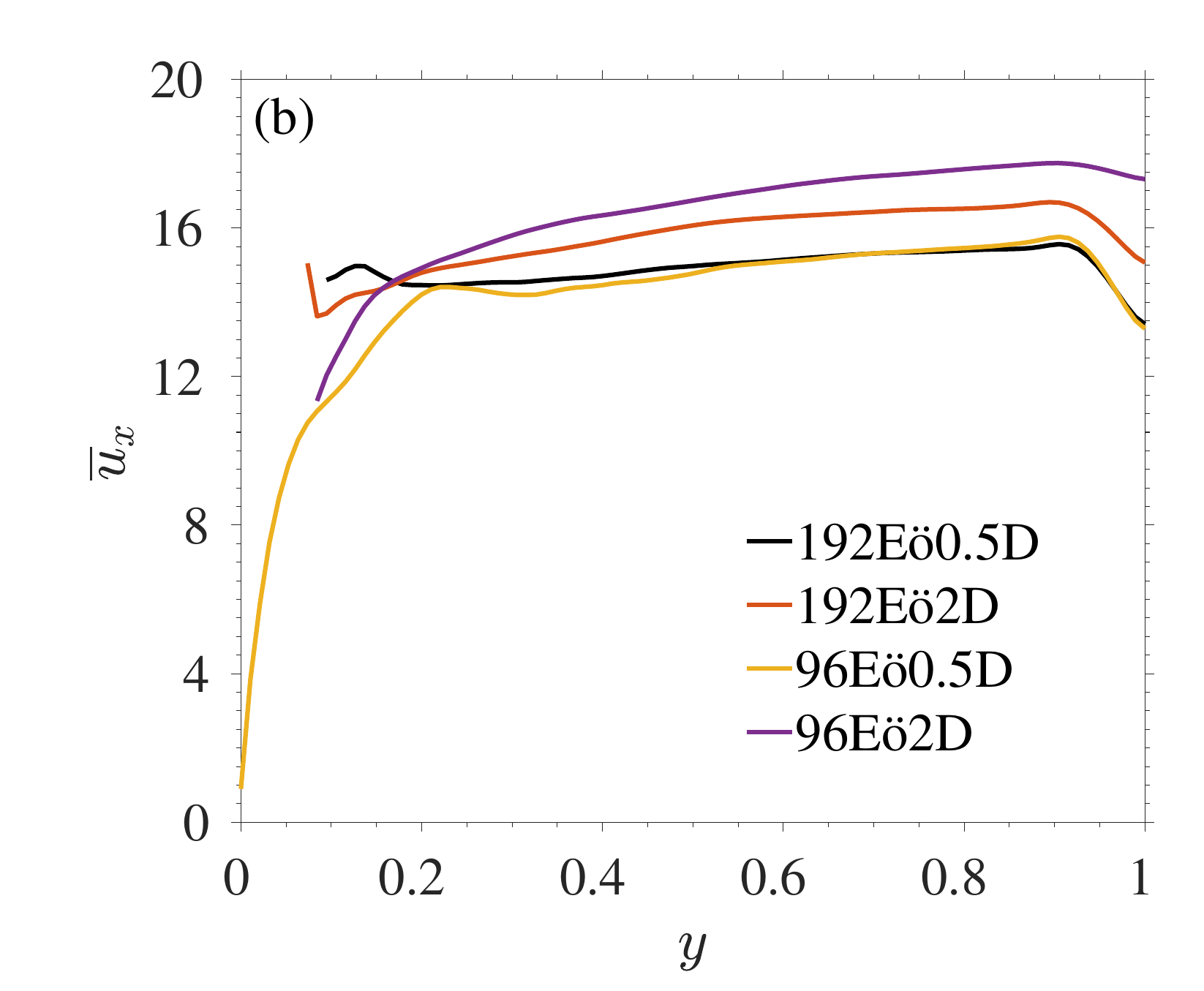}
        \label{fig:upward}
    \end{subfigure}
    \caption{Gas phase-averaged mean velocity profile}
    \label{fig:Mean_velocity_gas}
\end{figure*}

In the case of downward flow, the bubbles are observed to accumulate in the middle of the channel. Considering the eight cases exhibit low volume fractions, the phenomena observed align with the results presented in the studies conducted by \citep{chu2021turbulence} and \citep{lu2006numerical}. Closer to the wall (in the region \(0 < \textit{y} < 0.2\)), the gas volume fraction \(f\) is negligible for all cases. As \(\textit{y}\) continues to increase, particularly in the range of \(0.2 < \textit{y} < 0.8\), the volume fraction profiles for all cases show a more significant rise, eventually stabilizing as they approach the central region of the channel. Since in downflow, gravity and the downward pressure gradient are in the same direction, a large pressure gradient may not be required to maintain flow. As bubbles migrate towards the center and the density decreases, the shear forces in the central area diminishes. This phenomenon is illustrated by FIG.~\ref{fig:Mean_velocity_liquid}(b) and FIG.~\ref{fig:Mean_velocity_gas}(b), where both the liquid and gas velocities tend to flatten at the center of the channel, indicating that the shear force is related to the velocity gradient, i.e., the difference in speed between the bubbles and the surrounding fluid.

\begin{figure*}        
    \centering
    \begin{subfigure}[b]{0.5\textwidth} 
        \centering
        \includegraphics[height=6.4cm,keepaspectratio]{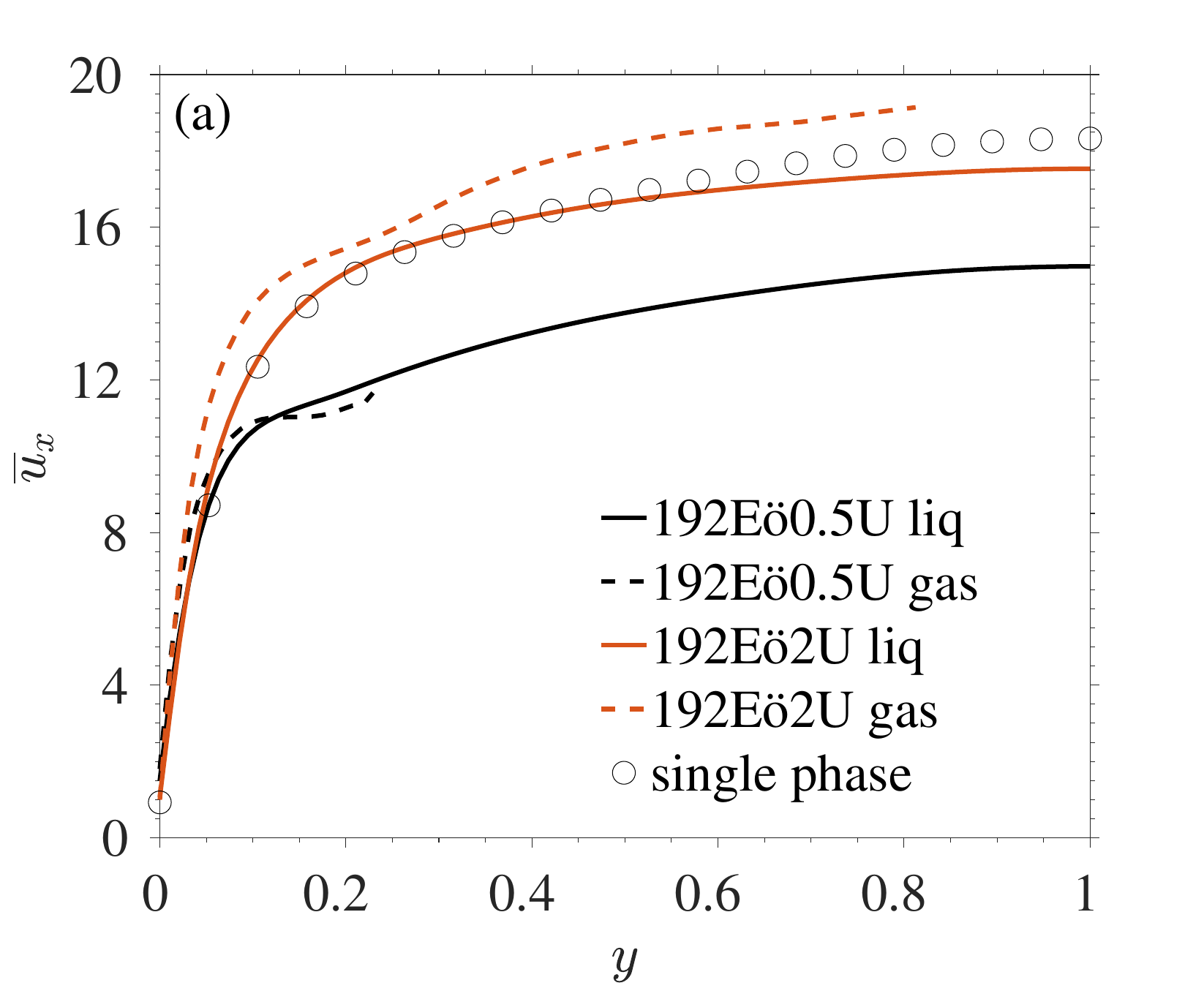}
    \end{subfigure}%
    \begin{subfigure}[b]{0.5\textwidth} 
        \centering
        \includegraphics[height=6.4cm,keepaspectratio]{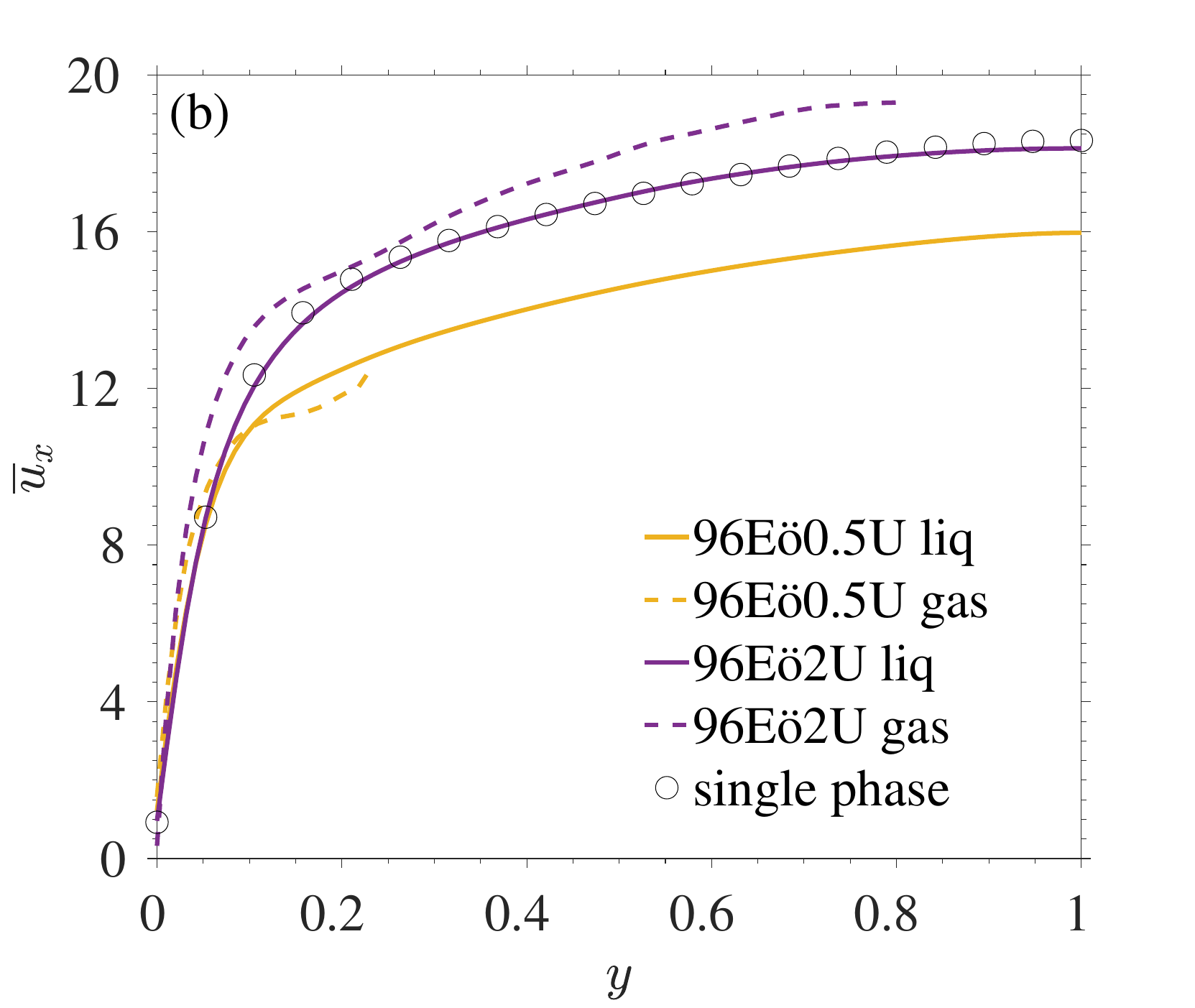}
    \end{subfigure}
    \caption{Comparison of velocity differences between liquid and gas phases in upward flow}
    \label{fig:difference_velocity}
\end{figure*}

In addition, cases with an Eötvös number of 0.5 have their peak void fraction closer to the wall relative to those with an Eötvös number of 2. The concentration of bubbles becomes more centralized within the channel as the Eötvös number rises. This phenomenon is consistent with the relationship between the Eötvös number and the void fraction described by \citet{Dabiri2013}. Figure~\ref{fig:difference_velocity} displays the differences between the liquid phase-averaged mean velocity and the gas phase-averaged mean velocity in upward cases. The figure shows that, in the case of 192Eo0.5U, the velocity of the gas is higher than that of the liquid near the wall, but at $y = 0.1$, the velocities of gas and liquid are the same, with this point of minimum difference coinciding with the peak of the bubbles. Similarly, in the case of 192Eo2U, the velocity of the gas is consistently higher than that of the liquid throughout, but the location where the difference between gas and liquid velocities is minimal also coincides with the peak of the bubbles.

The comparison of phase-averaged streamwise velocity profiles, $\overline{u}_{x}$, for the liquid phase in upward (a) and downward (b) flows, as shown in FIG.~\ref{fig:Mean_velocity_liquid}, suggests a trend towards uniform liquid velocity in the channel's midpoint~\cite{lu2006numerical}. double bar denotes the phase-weighted averaging, which is defined by

\begin{equation}
\overline{\overline{A}}_m = \frac{\varphi_m A_m}{\overline{\varphi}_m} \quad m = 1, 2),
\label{eq:18}
\end{equation}


The function $\varphi_k$ represents the characteristic function of phase $m$, which is defined as follows:

\begin{align}
\varphi_L (x, y, z, t) &= H(f(x, y, z, t)),
\label{eq:19}
\\
\varphi_G(x, y, z, t) &= 1 - H(f(x, y, z, t))
\label{eq:20}
\end{align}

In the upward cases, as shown in FIG.~\ref{fig:Mean_velocity_liquid}(a), the mean velocity incrementally increases in the range of 0.2 < \textit{y} < 0.8 and exhibits a plateau within the region of 0.8 < \textit{y} < 1. Remarkably, there is a significant reduction in mean velocity for the cases with an Eötvös number of 0.5 when compared to those with an Eötvös number of 2. As highlighted in \citet{lu2013dynamics}, upward flows with low Eötvös numbers not only exhibit a pronounced peak in the void fraction adjacent to the wall but also a substantial reduction in velocity, which reflects the trends observed. In the downward cases, as illustrated in FIG.~\ref{fig:Mean_velocity_liquid}(b), the relatively high density of bubbles generates significant buoyancy in the channel's bulk. This results in a smoothed-out velocity profile. The liquid phase's mean velocity approaches a uniform value across the region where \(0.4 < \textit{y} < 1\).

Figure~\ref{fig:Mean_velocity_gas} illustrates the mean velocity profiles along the streamwise direction, $\overline{u}_{x}$, for the gas phase in both upward (a) and downward (b) flows. In the downward case, the scenario with 192 bubbles, results in a flattened velocity profile in the bulk. This phenomenon occurs because the accumulation of bubbles in the bulk of the channel significantly increases buoyancy. According to the discussion in \citet{chu2020turbulence}, a high volume fraction results in a noticeably more uniform mean velocity profile. The velocity profiles are noticeably reduced compared to those from the liquid phase, predominantly in scenarios with 192 bubbles. This reduction is attributed to the buoyancy effect, which acts in opposition to the direction of the main flow. The average streamwise velocity in the gas phase remains uniform with the exception of the region where 0.1 < \textit{y} < 0.2. Here, the velocity gradient in the liquid phase significantly accelerates the bubbles. In the upward cases, as shown in Fig.~\ref{fig:Mean_velocity_gas}(a) the gas velocity with an E\"otv\"os number of 0.5 is significantly lower than that with an E\"otv\"os number of 2.


\begin{figure*}		
    \centering
    \begin{subfigure}[b]{0.5\textwidth}
        \centering
        \includegraphics[height=5cm,keepaspectratio]{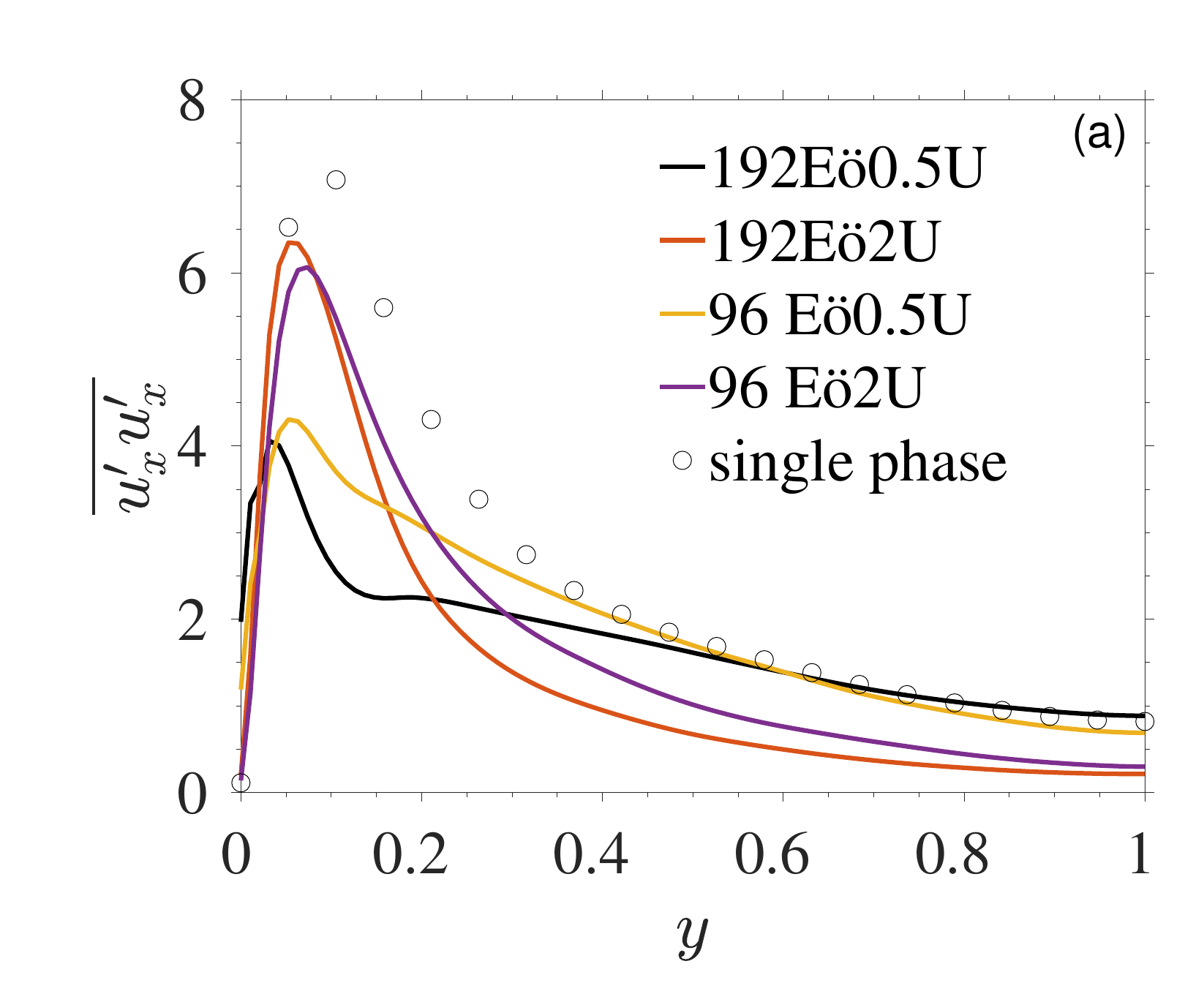}
    \end{subfigure}%
    \hspace{-5mm}
    \begin{subfigure}[b]{0.5\textwidth}
        \centering
        \includegraphics[height=5cm,keepaspectratio]{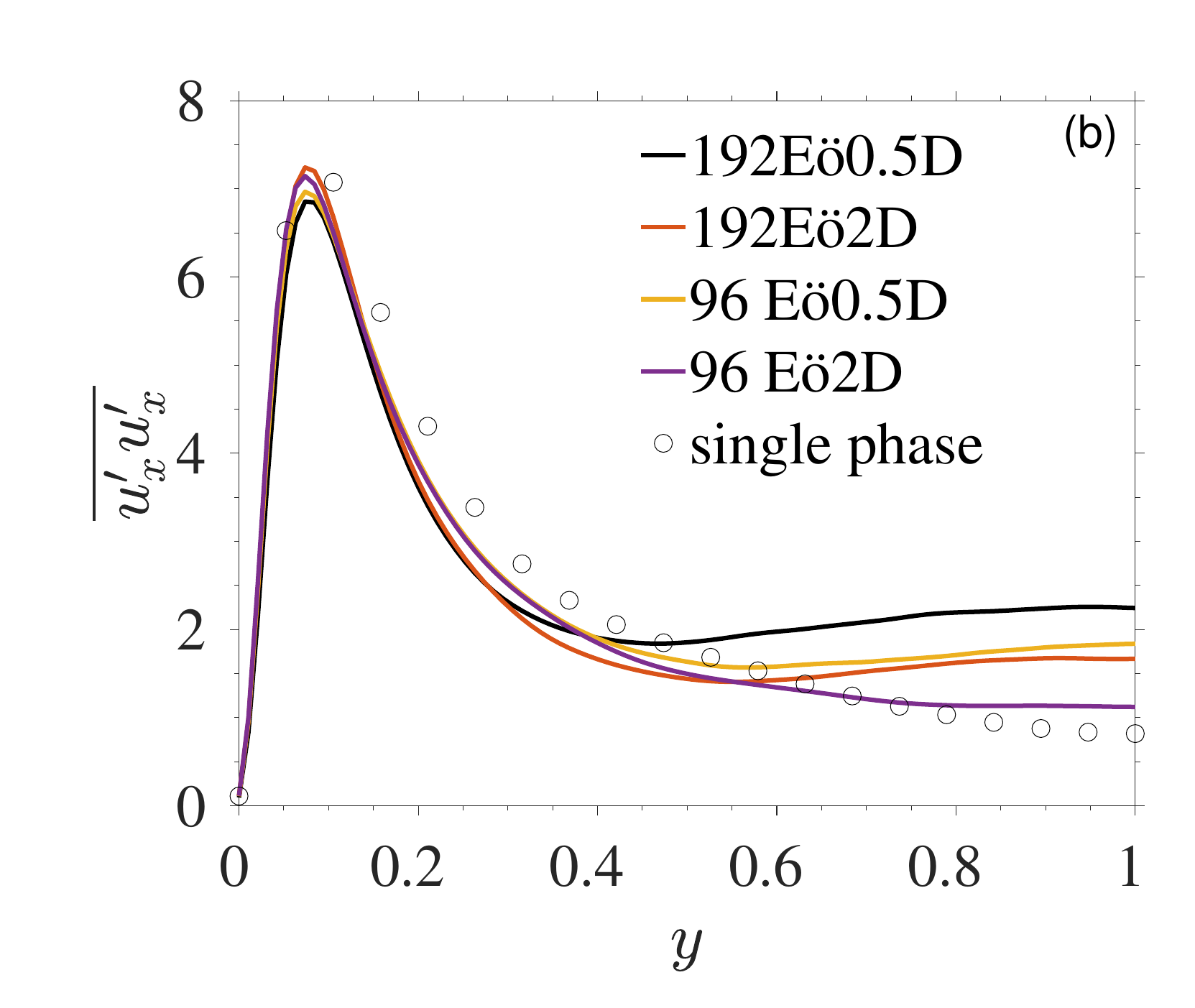}
    \end{subfigure}
    
    \begin{subfigure}[b]{0.5\textwidth}
        \centering
        \includegraphics[height=5cm,keepaspectratio]{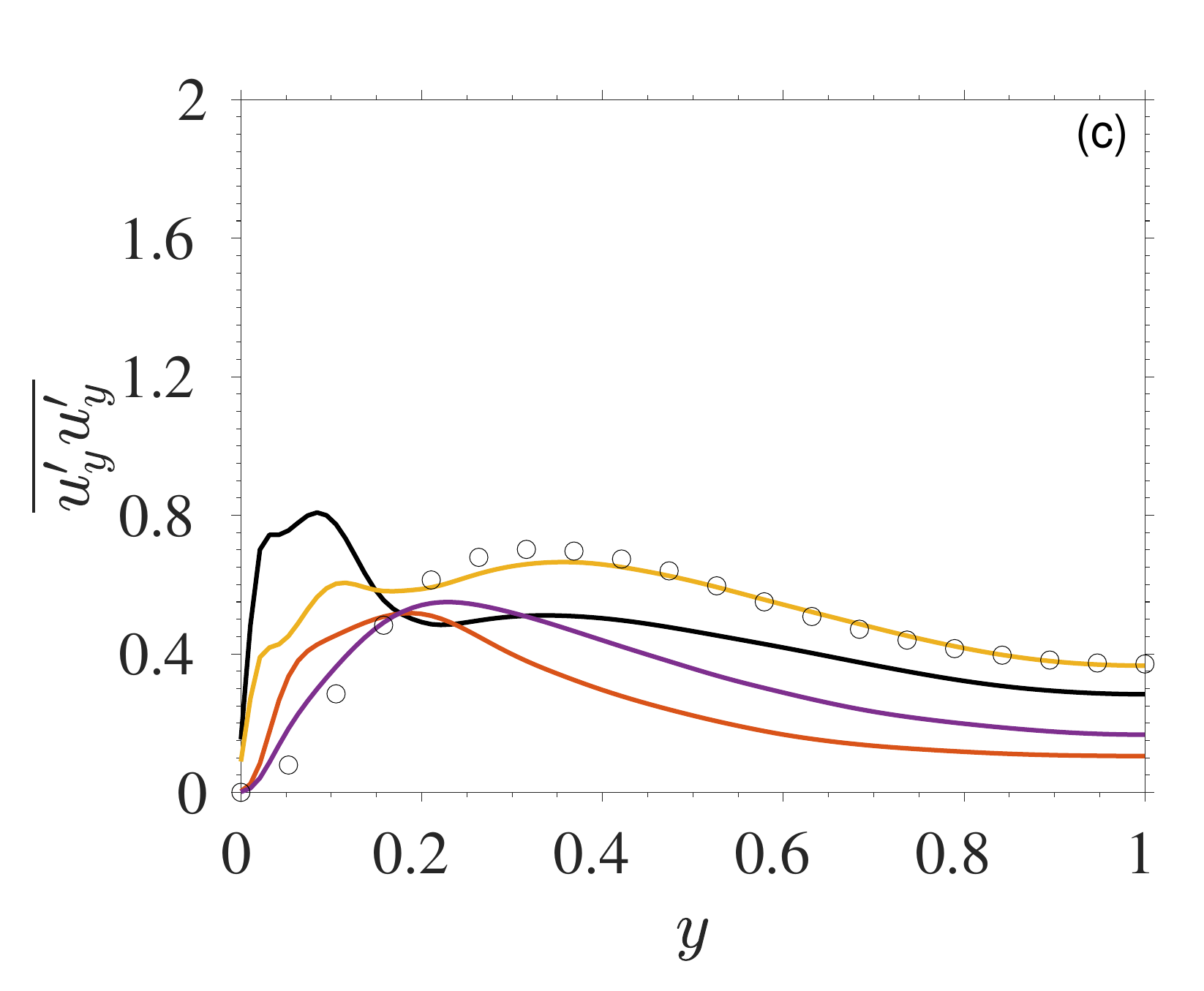}
    \end{subfigure}%
    \hspace{-5mm}
    \begin{subfigure}[b]{0.5\textwidth}
        \centering
        \includegraphics[height=5cm,keepaspectratio]{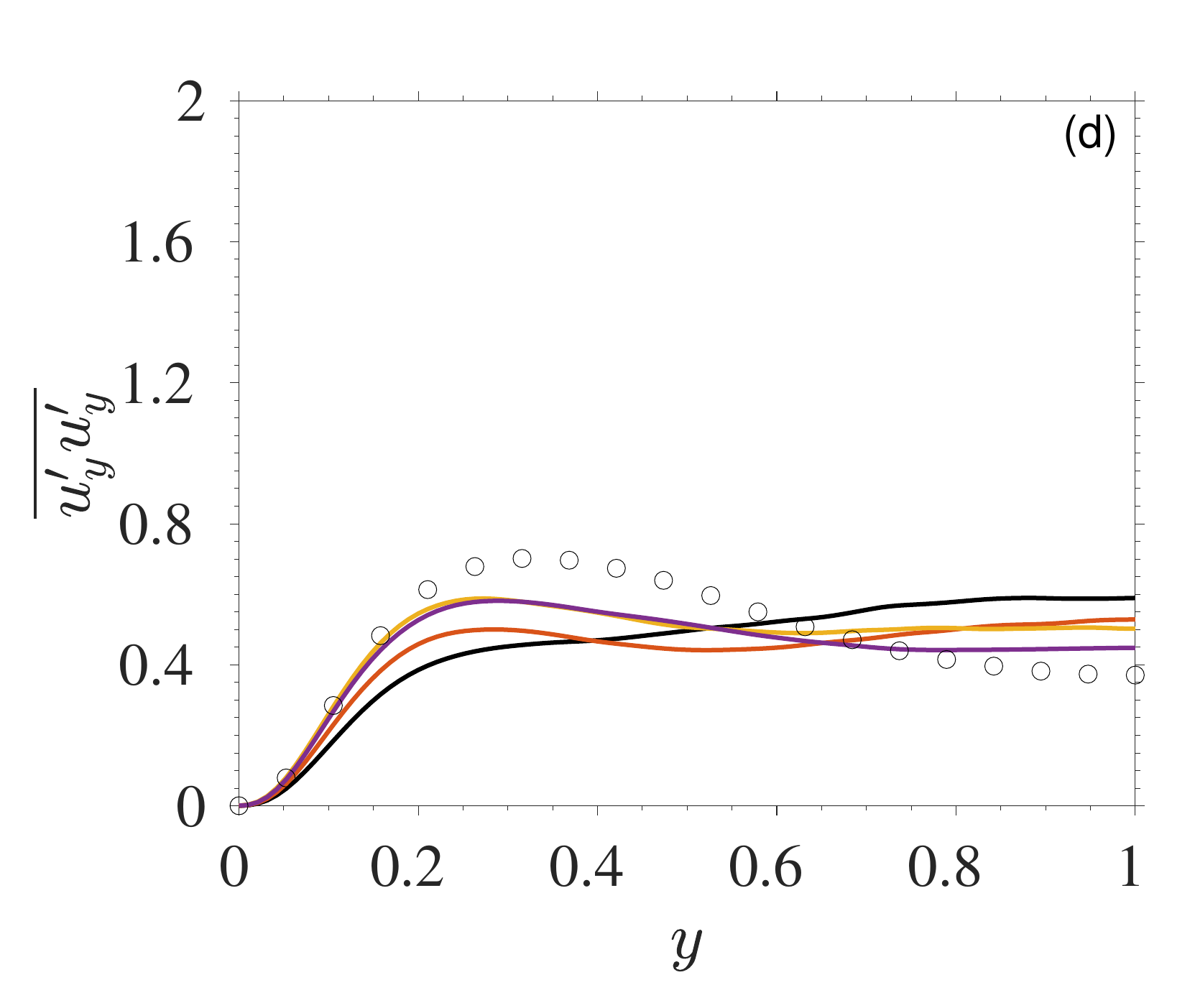}
    \end{subfigure}
    
    \begin{subfigure}[b]{0.5\textwidth}
        \centering
        \includegraphics[height=5cm,keepaspectratio]{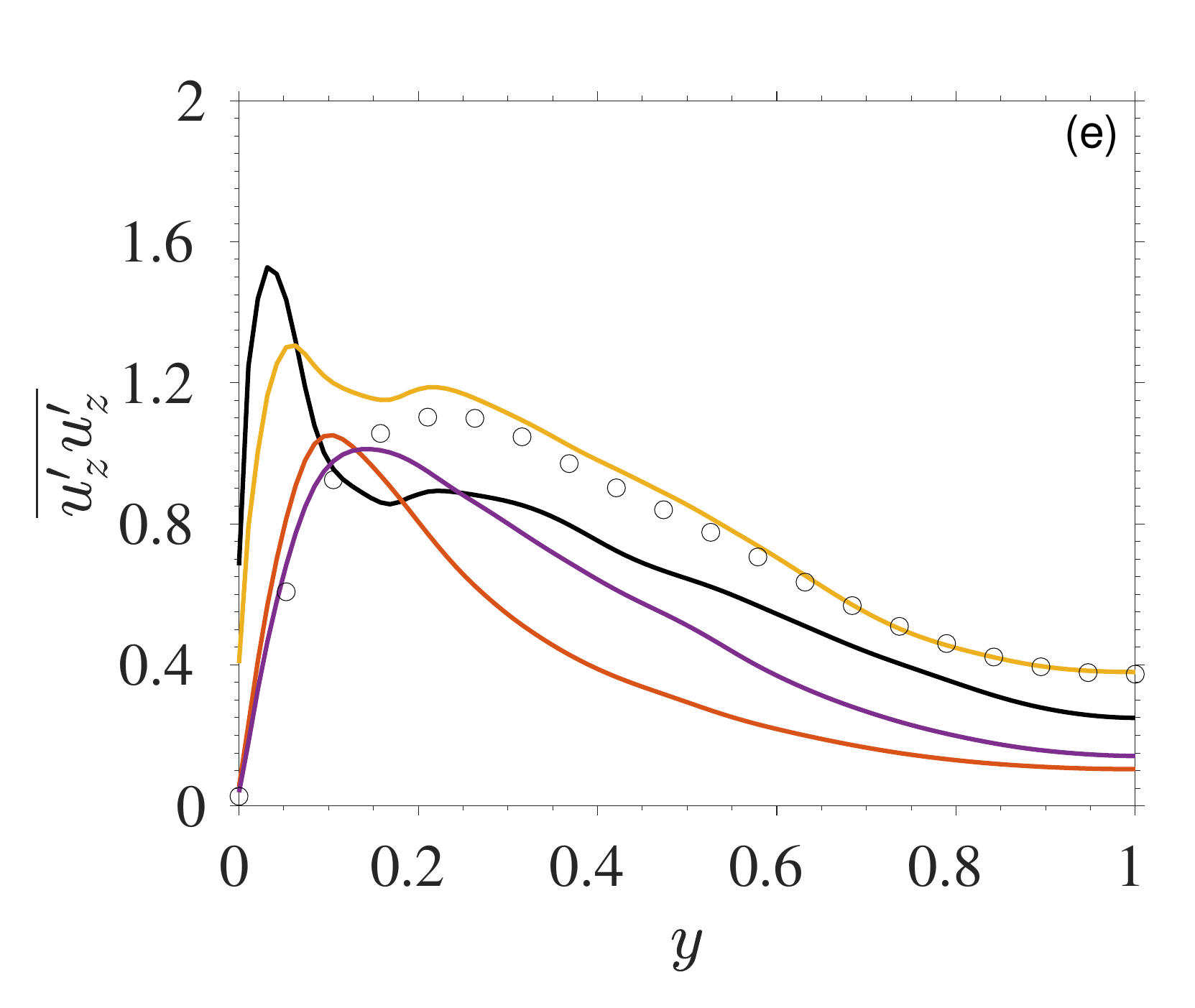}
    \end{subfigure}%
    \hspace{-5mm}
    \begin{subfigure}[b]{0.5\textwidth}
        \centering
        \includegraphics[height=5cm,keepaspectratio]{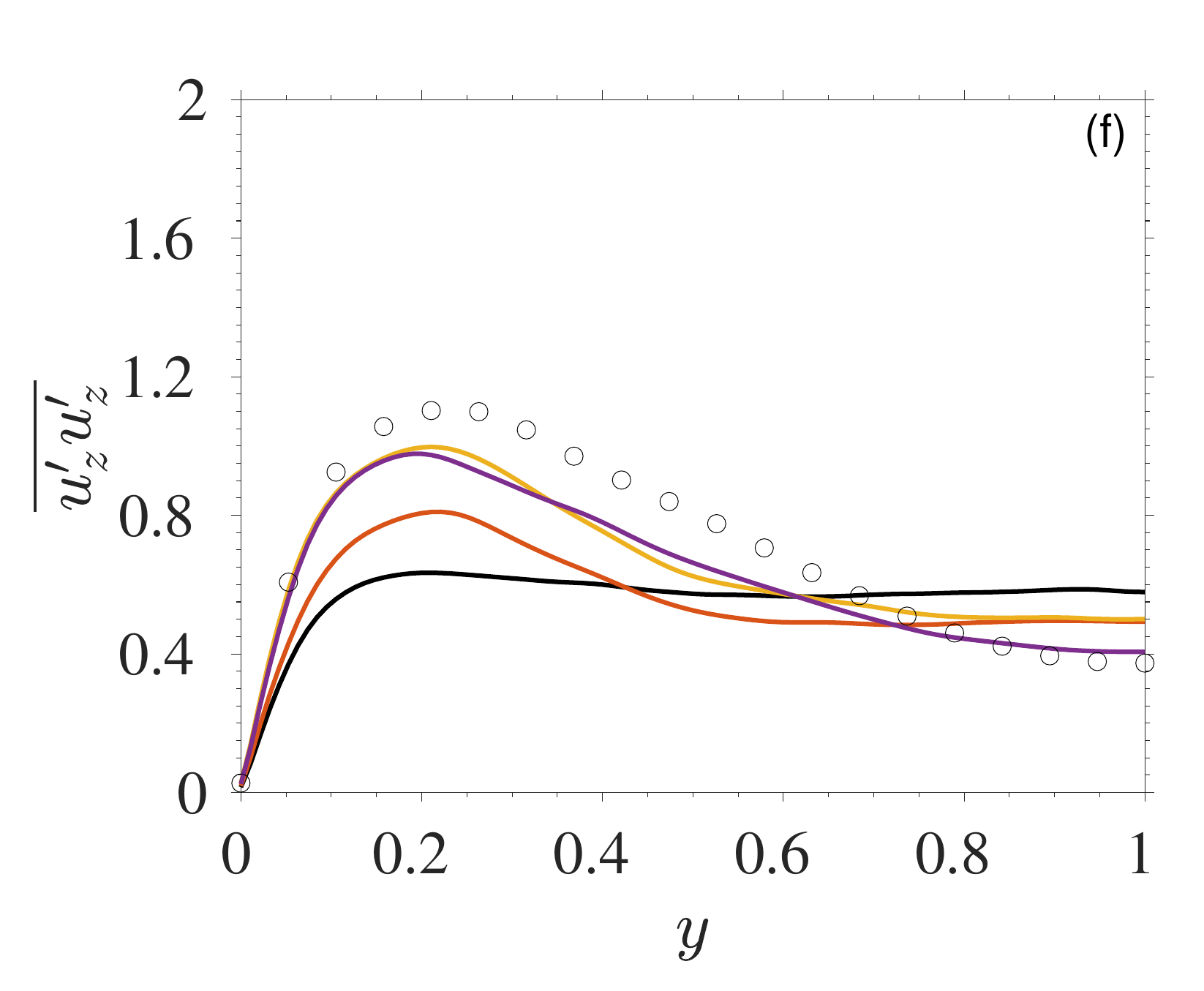}
    \end{subfigure}
    
    \begin{subfigure}[b]{0.5\textwidth}
        \centering
        \includegraphics[height=5cm,keepaspectratio]{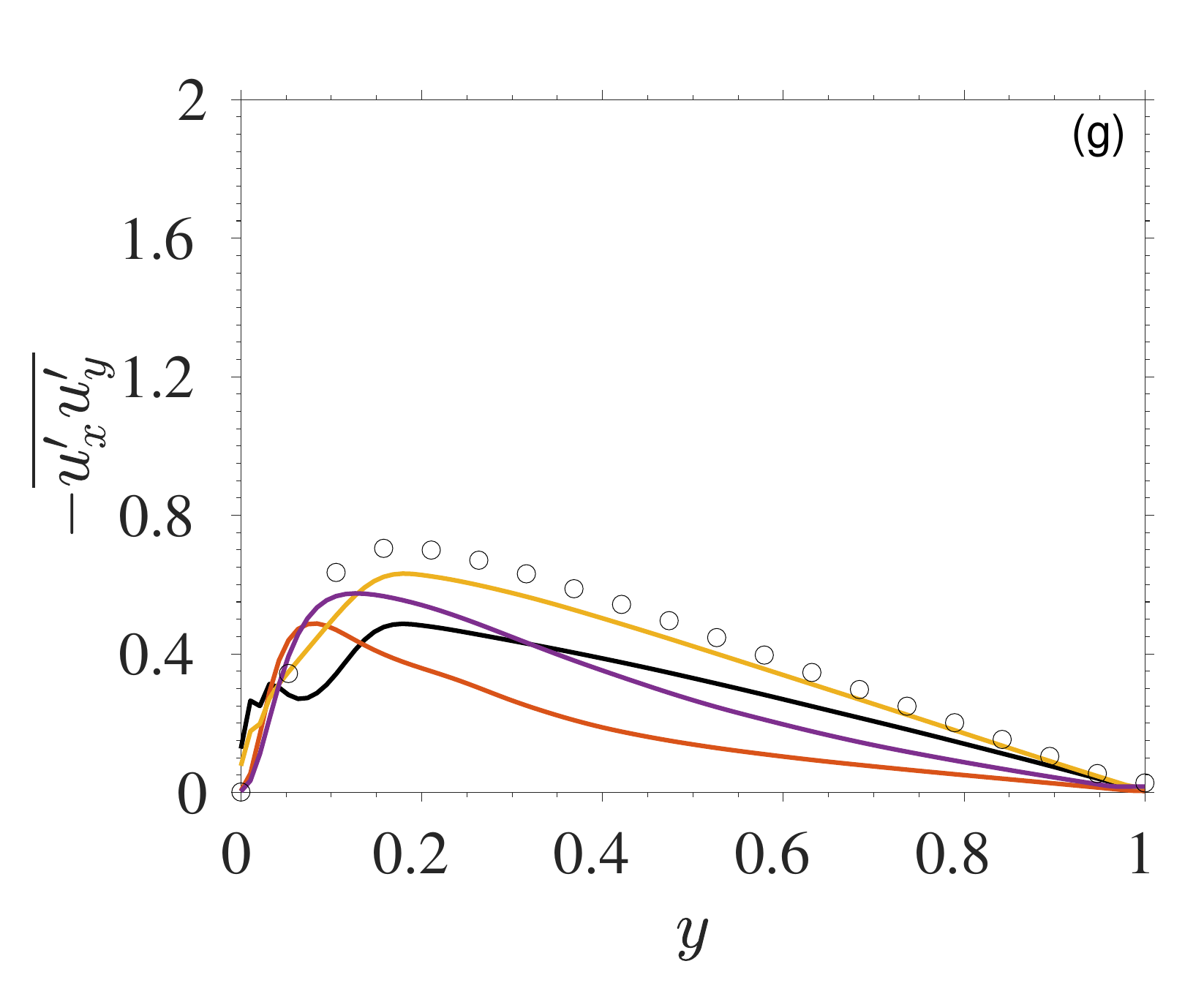}
    \end{subfigure}%
    \hspace{-5mm}
    \begin{subfigure}[b]{0.5\textwidth}
        \centering
        \includegraphics[height=5cm,keepaspectratio]{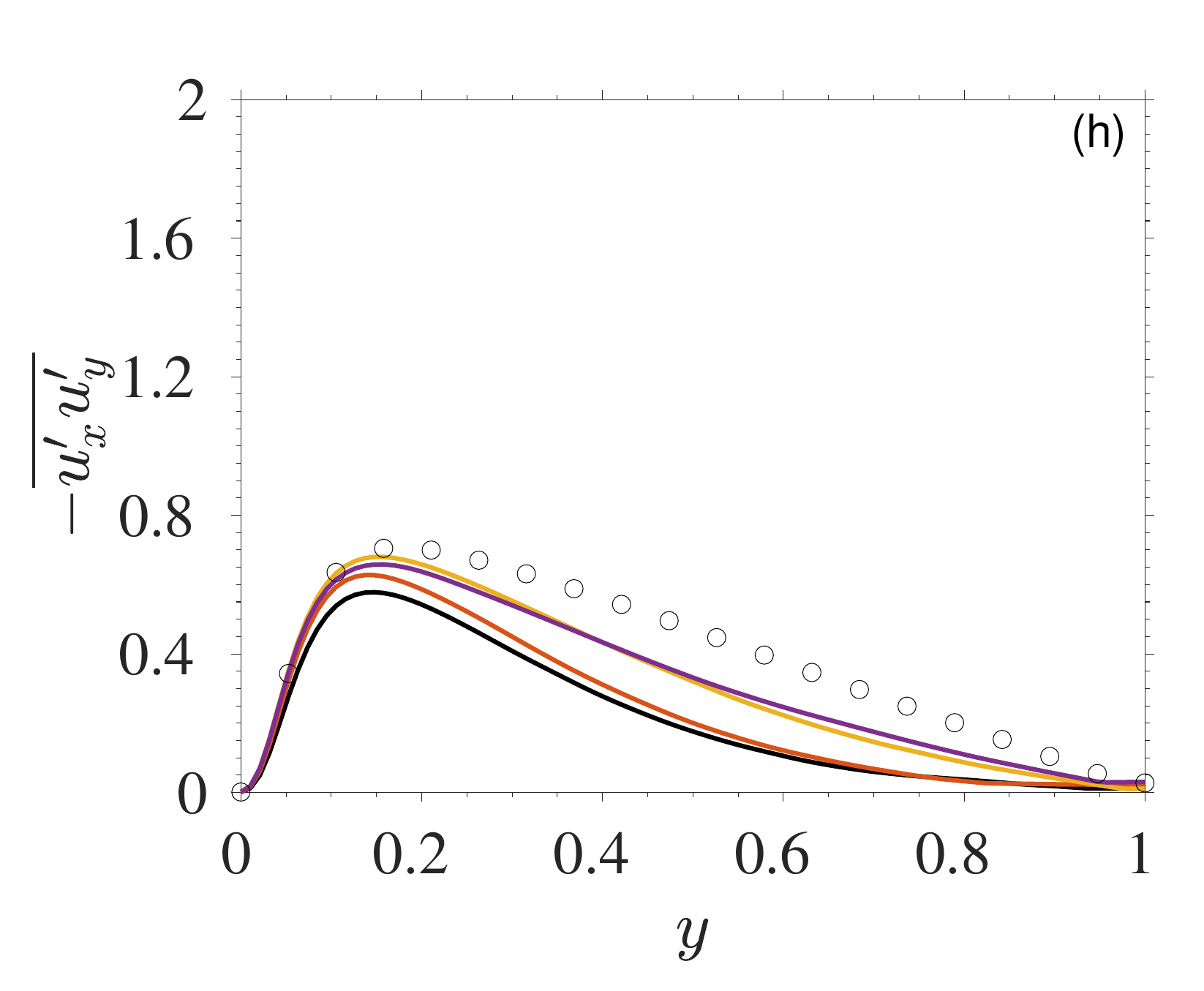}
    \end{subfigure}
    
    \caption{Velocity fluctuations \( \overline{u'_{x}u'_{x}} \) (a), \( \overline{u'_{y}u'_{y}} \) (c), \( \overline{u'_{z}u'_{z}} \) (e), and \( -\overline{u'_{x}u'_{y}} \) (g) of the liquid phase for the upward case using phase averaging compared with the single-phase flow. Similarly, \( \overline{u'_{x}u'_{x}} \) (b), \( \overline{u'_{y}u'_{y}} \) (d), \( \overline{u'_{z}u'_{z}} \) (f), and \( -\overline{u'_{x}u'_{y}} \) (h) correspond to the downward case.} 
    \label{fig:Velocity_fluctuations}
\end{figure*}

\begin{figure*}		
    \centering
    \begin{subfigure}[b]{0.5\textwidth} 
        \centering
        \includegraphics[height=6.8cm,keepaspectratio]{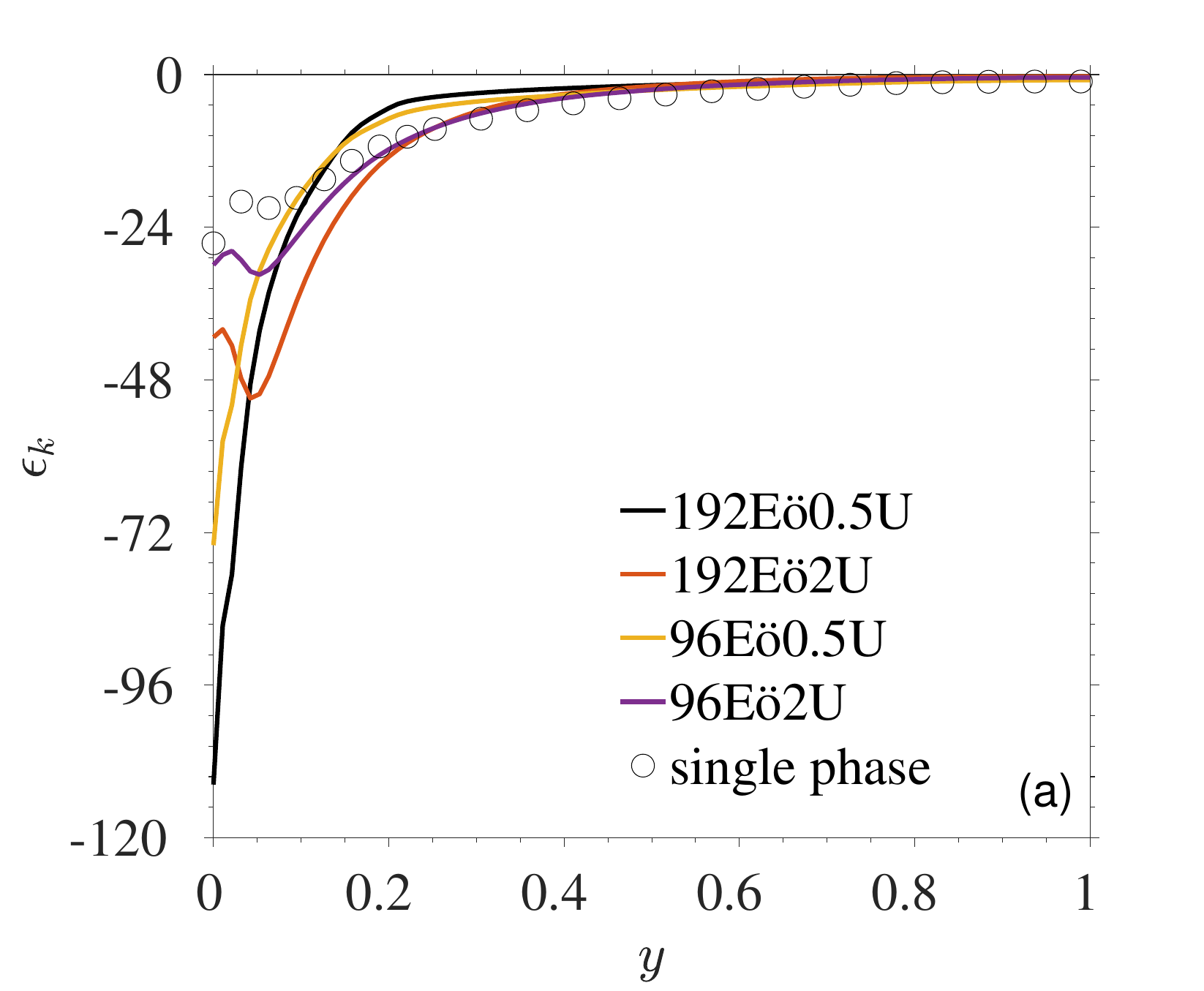}
    \end{subfigure}%
    \hspace{-5mm} 
    \begin{subfigure}[b]{0.5\textwidth} 
        \centering
        \includegraphics[height=6.8cm,keepaspectratio]{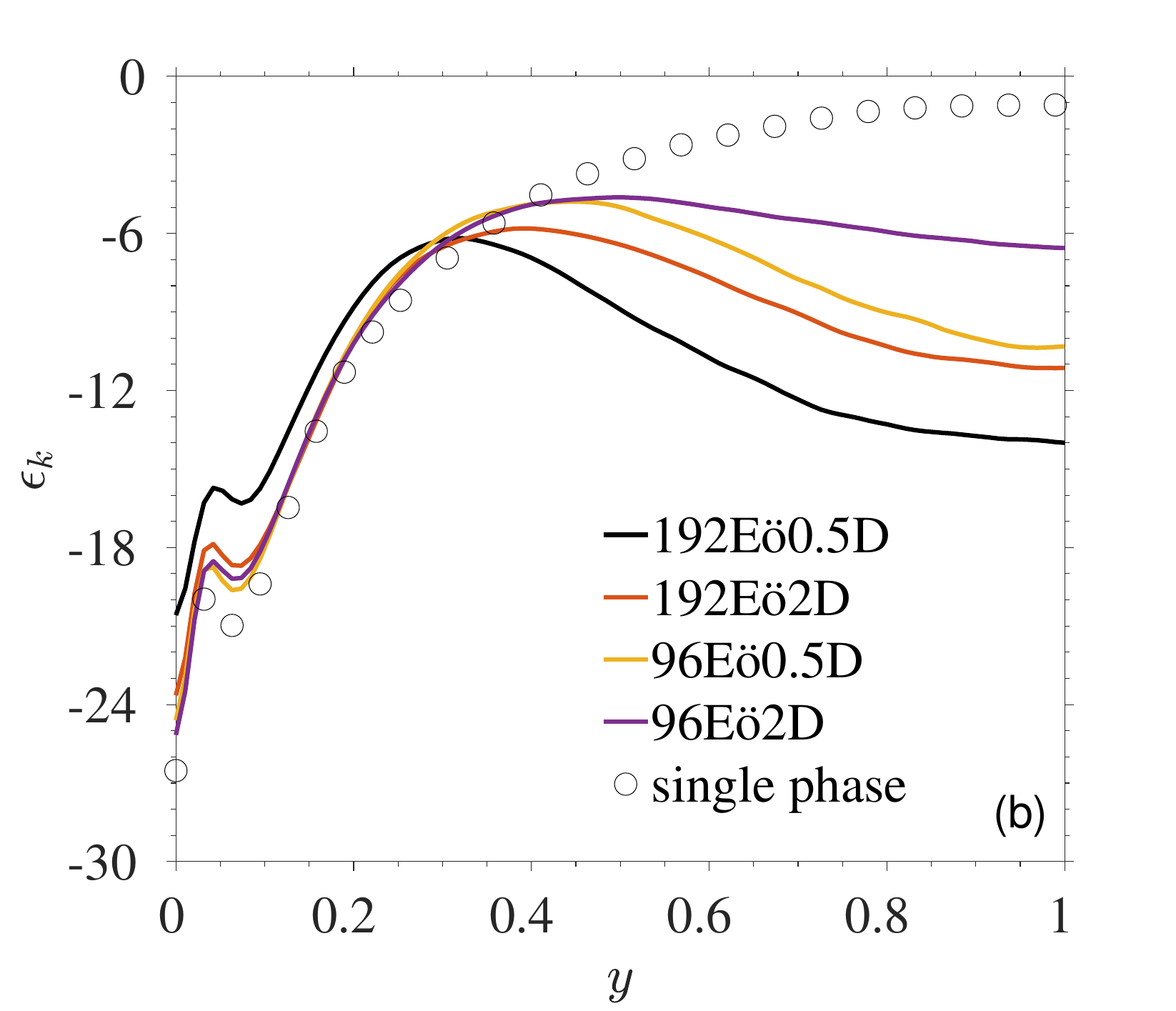}
    \end{subfigure}
    \caption{(a) Dissipation in the upward case; (b) Dissipation in the downward case.}
    \label{fig:Dissipation}
\end{figure*}

The turbulent intensity components $\overline{u'_{x}u'_{x}}$, $\overline{u'_{y}u'_{y}}$, $\overline{u'_{z}u'_{z}}$ and Reynolds shear stress $-\overline{u'_{x}u'_{y}}$ of the liquid phase, calculated using phase averaging, are depicted for both upward and downward cases in FIG.~\ref{fig:Velocity_fluctuations}, compared with the single-phase DNS channel flow. The inner peaks of streamwise turbulence intensity, denoted by $\overline{u'_{x}u'_{x}}$ in FIG.~\ref{fig:Velocity_fluctuations}(a), within the buffer layer at \textit{y} = 0.1, are diminished in bubbly flows, particularly exhibiting a more substantial decrease in the upward cases. In this case, the scenario with 192 bubbles is closer to the wall compared to the case with 96 bubbles, due to the higher bubble density. More bubbles serve as a medium for energy dissipation, helping to absorb and disperse the energy generated during fluid motion. This observation is consistent with the dissipation distribution shown in FIG.~\ref{fig:Dissipation}(a). Additionally, the case with an Eötvös number of 0.5 exhibits a greater attenuation in peak, consistent with the observations by \citet{Liu2023}. This is due to the inverse relationship between surface tension and the Eötvös number. A smaller Eötvös number results in a shape closer to a sphere, leading to more isotropic characteristics. In the range \(0.2 < y < 1\), the case with an Eötvös number of 0.5 more closely resembles single phase flow, due to the absence of bubbles. Meanwhile, in the range \(0.2 < y < 0.8\), the presence of bubbles in the case with an Eötvös number of 2 acts as a buffer, leading to a decline in \(\overline{u'_{x}u'_{x}}\).  In the downward examples, as depicted in FIG.~\ref{fig:Velocity_fluctuations}(b), \(\overline{u'_{x}u'_{x}}\)shows an increase at the channel center when compared to the single-phase flow. The enhanced energy in the channel center indicates that bubble clustering in that region induces extra turbulence, also referred to as pseudo-turbulence \cite{chu2020turbulence,pandey2020liquid}.

In the upward cases, the rise of bubbles near the wall induces a stirring motion of the liquid, specifically leading to an increase in the fluctuations of \(\overline{u'_{y}u'_{y}}\) and \(\overline{u'_{z}u'_{z}}\) within the near-wall region. In the cases studied, the void fraction for 192 bubbles closely aligns with the observations reported by \citet{lu2013dynamics}, showing similar phenomena: the longitudinal vorticity is essentially zero in the middle of the channel but peaks near the walls. For 96 bubbles, within \(0.2 < y < 1\), the \(\overline{u'_{x}u'_{x}}\) is larger than in the 192-bubble cases, but the peak near the wall is lower, which is likely related to differences in the void fraction. In downward cases, an identical increase in kinetic energy at the channel centre is also evident in the wall-normal fluctuation \(\overline{u'_{y}u'_{y}}\) and spanwise fluctuation \(\overline{u'_{z}u'_{z}}\), exhibiting a similar trend to the behaviour observed in \(\overline{u'_{x}u'_{x}}\). 

Regarding the Reynolds stress \(\overline{-u'_{x}u'_{y}}\), the intensity is diminished throughout the entire region, which contrasts with the central part of the channel where an increasing trend is observed in the three turbulence intensities. In single-phase turbulent flow, the Reynolds shear stress, \(\overline{-u'_{x}u'_{y}}\) demonstrates a linear profile across a significant portion of the channel \(0.2 < y < 1\), reflecting the equilibrium between total shear stress and the pressure gradient, while the viscous contribution to the total shear stress is negligible in this region \cite{Liu2023}. For downward bubbly flows, the linear decrease of \(\overline{-u'_{x}u'_{y}}\) is more pronounced near the wall and diminishes more rapidly. It is notable that cases with 192 bubbles exhibit this behavior closer to the wall compared to those with 96 bubbles. Moreover, \(\overline{-u'_{x}u'_{y}}\) approaches almost zero in the channel's central region (y > 0.8), where the bubble volume fraction peaks and remains relatively constant. Therefore, \(\overline{-u'_{x}u'_{y}}\) is closely associated with the local bubble volume fraction. In upward cases, the case with 96 bubbles is more similar to a single-phase flow, whereas the case with 192 bubbles exhibits a slight linear variation. As the E\"otv\"os number decreases, the Reynolds stress approaches zero except in a layer near the walls. Conversely, with a higher E\"otv\"os number, the Reynolds stress maintains a linear profile throughout the channel, similar to that observed in single-phase turbulent flow. These phenomena are consistent with the observations made by \citet{Dabiri2013}.




As mentioned in \citet{Klein2022}, an increase in void fraction can increase the anisotropy, and specifically, the streamwise direction may exhibit a pronounced dominance. Therefore, it is crucial to investigate the extent of the impact of anisotropy in different cases, which will also assist in appropriately parameterizing models for RANS.

The normalized Reynolds stress anisotropy tensor, \(b_{ij}\), is defined by Eq.~(\ref{eq:bij}), which can be utilized to describe the turbulence characteristics near the wall.


\begin{equation}
b_{ij} = \frac{\overline{u_i u_j}}{\overline{u_k u_k}} - \frac{\delta_{ij}}{3}
\label{eq:bij}
\end{equation}

Using the invariants of a symmetric second-order tensor, the Reynolds stress anisotropy tensor can be characterized by three principal invariants, denoted \text{I}, \text{II}, and \text{III}.

\begin{equation}I = 0, \text{II} = -b_{ij}b_{ji}/2, \text{III} = b_{ij}b_{jk}b_{ki}/3. \quad
\end{equation}

Since the anisotropy tensor possesses a zero trace \((b_{ii} = 0)\), the characterization of the Reynolds stress anisotropy tensor is fully described by its second and third invariants. The condition of anisotropy can be described using only two variables, represented by \(\xi\) and \(\eta\), which are defined by Eq.~(\ref{eq:eta}) and Eq.~(\ref{eq:xi}), respectively. The anisotropic state of the Reynolds stresses at any point along the wall-normal direction can be represented by plotting on the \(\xi\)-\(\eta\) plane \cite{CHOI_LUMLEY_2001}.

\begin{equation}
\eta^2 = -\frac{\text{II}}{3} = \frac{b_{ii}^2}{6} = \frac{b_{ij}b_{ji}}{6},
\label{eq:eta}
\end{equation}

\begin{equation}
\xi^{3} = \frac{\text{III}}{2} = \frac{b_{ii}^{3}}{6} = \frac{b_{ij}b_{jk}b_{ki}}{6}.
\label{eq:xi}
\end{equation}

\begin{figure*}		
	\centering
	\includegraphics[width=10cm,keepaspectratio]{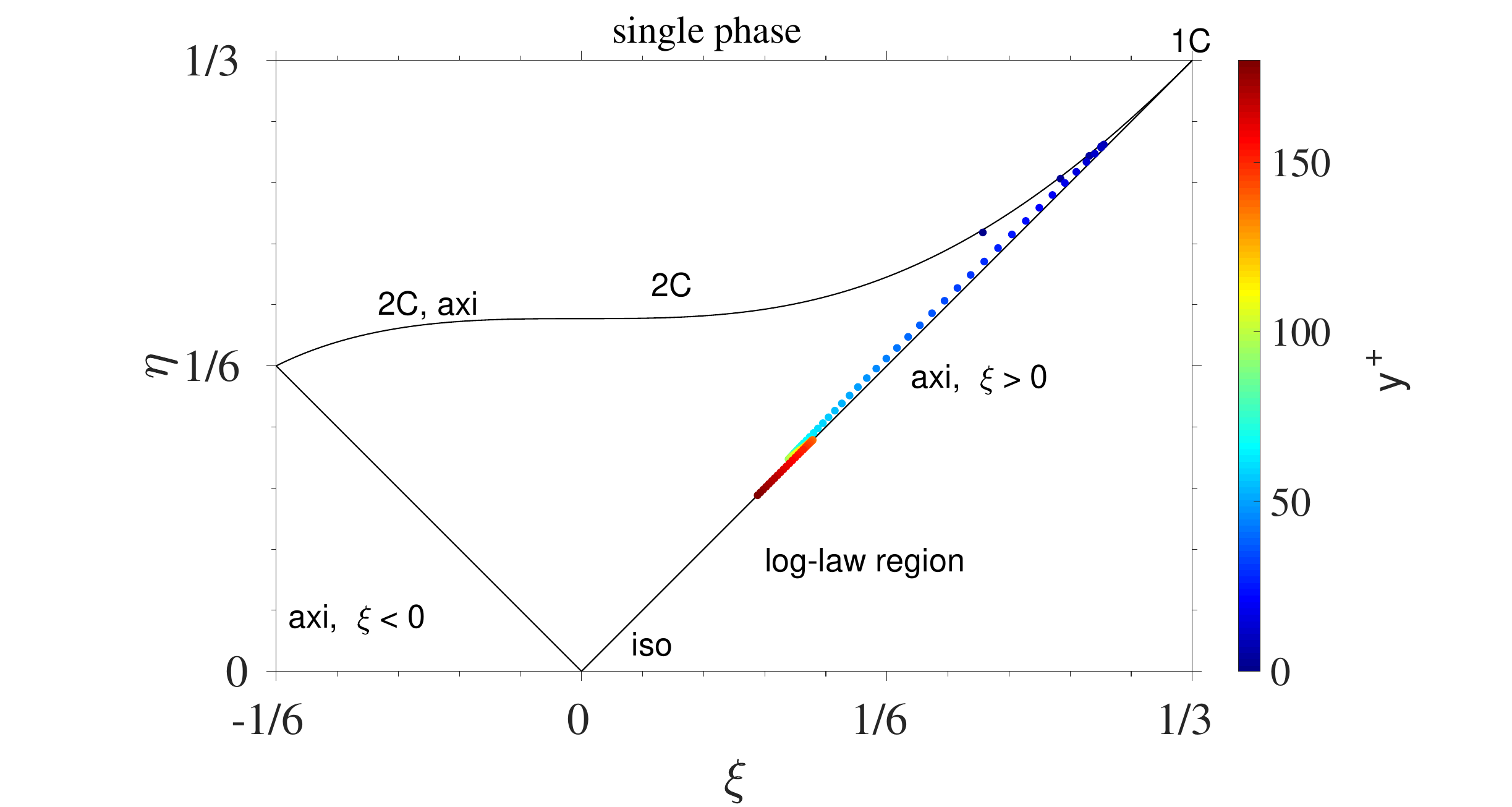}
	\caption{Sketch of the Lumley triangle on the plane of the invariants \(\xi\) and \(\eta\).1C, one-component; 2C, two-component.}
	\label{fig.lumley}
\end{figure*}

FIG. \ref{fig.lumley} represents the \(\xi\)-\(\eta\) plane, also known as the single-phase Lumley triangle, presents the values of \((\xi,\eta)\) derived from turbulent channel flow and a turbulent mixing layer. The Lumley triangle has three borders, two of which are straight lines connecting the origin to the points \((-1/6, 1/6)\) and \((1/3, 1/3)\). The other border is a curved line defined by \(\eta = \left(\frac{1}{27} + 2\xi^{3}\right)^{\frac{1}{2}}\). In this figure, `iso', `2C', and `axi' correspond to the isotropic state, two-component state, and axisymmetric state, respectively. The colorbar on the right corresponds to \(y^+\), where \(y^+ = \frac{u_{\tau} y}{\nu}\) and \(u_{\tau}\) is the friction velocity. In the single-phase turbulent channel flow near the wall, specifically where \(y^+ < 5\), the turbulence exhibits a two-component state. The anisotropy peaks at \(y^+ \approx 7\). Further along the channel, the Reynolds stress maintains a nearly axisymmetric state with a positive \(\xi\). At the core of the turbulent mixing layer, the Reynolds stresses remain nearly axisymmetric with a positive \(\xi\) yet exhibit slightly less anisotropy compared to the log-law region of the channel flow.

\begin{figure*}		
    \centering
    \begin{subfigure}[b]{0.5\textwidth}
        \centering
        \includegraphics[height=5cm,keepaspectratio]{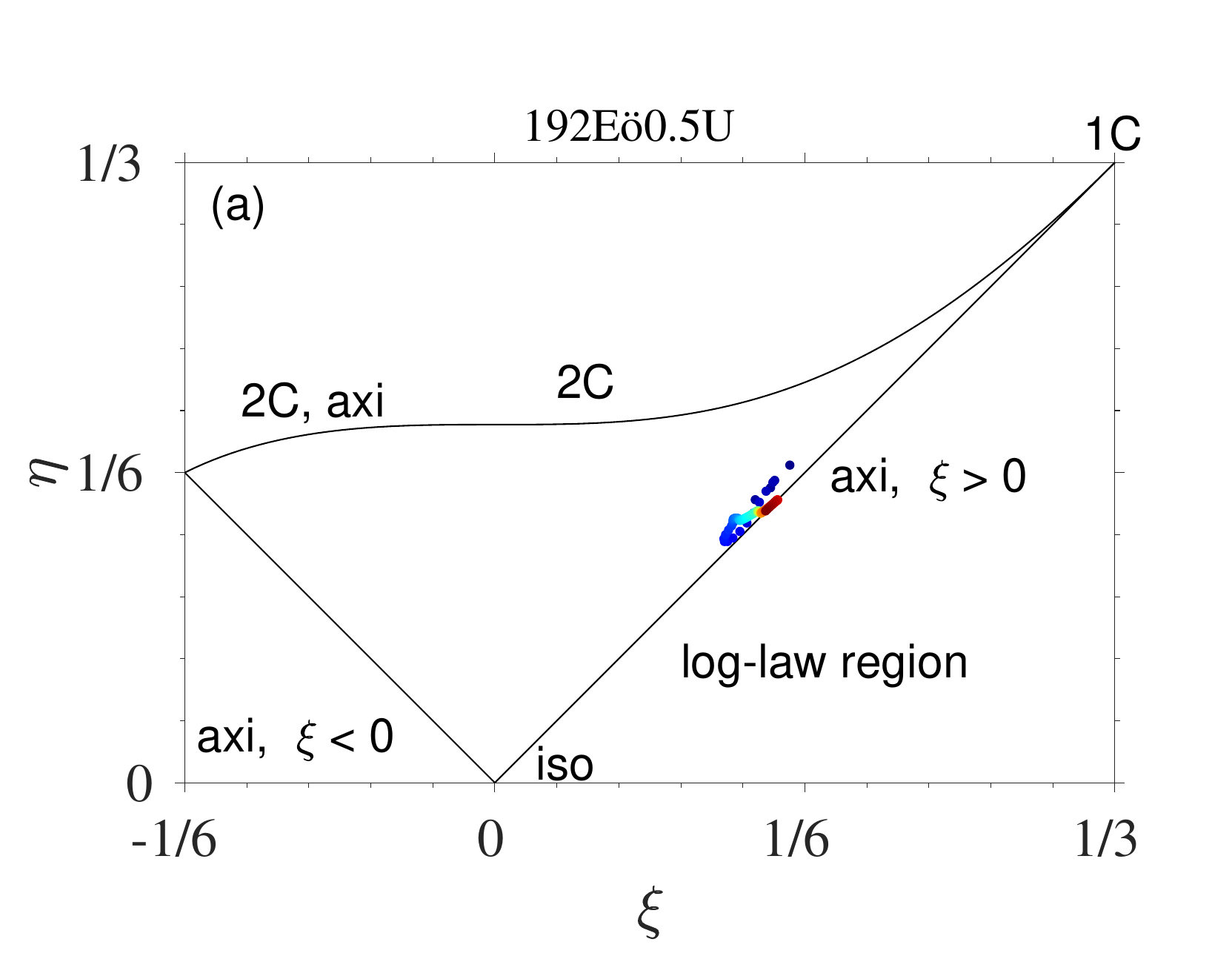}
    \end{subfigure}%
    \hspace{-5mm}
    \begin{subfigure}[b]{0.5\textwidth}
        \centering
        \includegraphics[height=5cm,keepaspectratio]{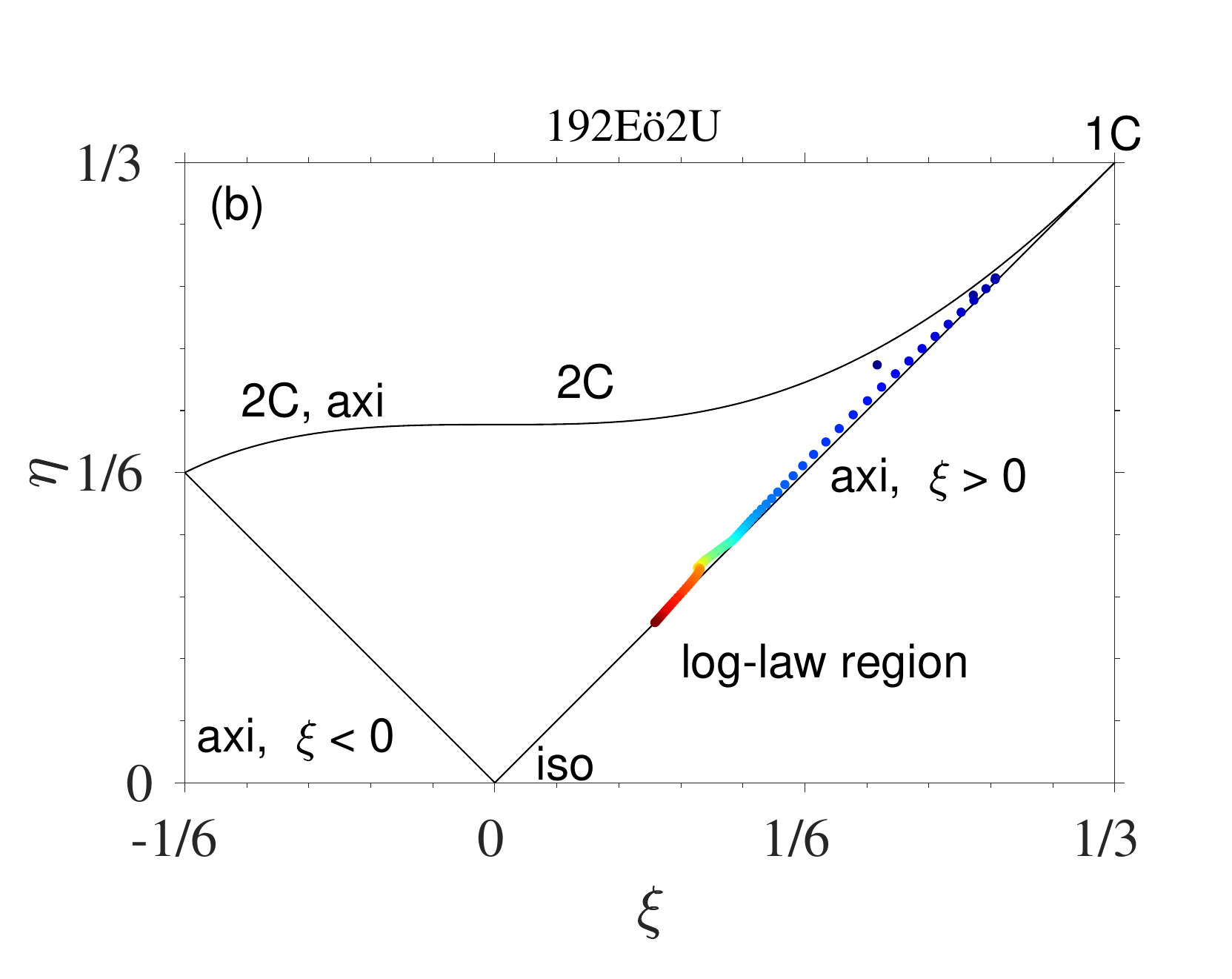}
    \end{subfigure}
    
    \begin{subfigure}[b]{0.5\textwidth}
        \centering
        \includegraphics[height=5cm,keepaspectratio]{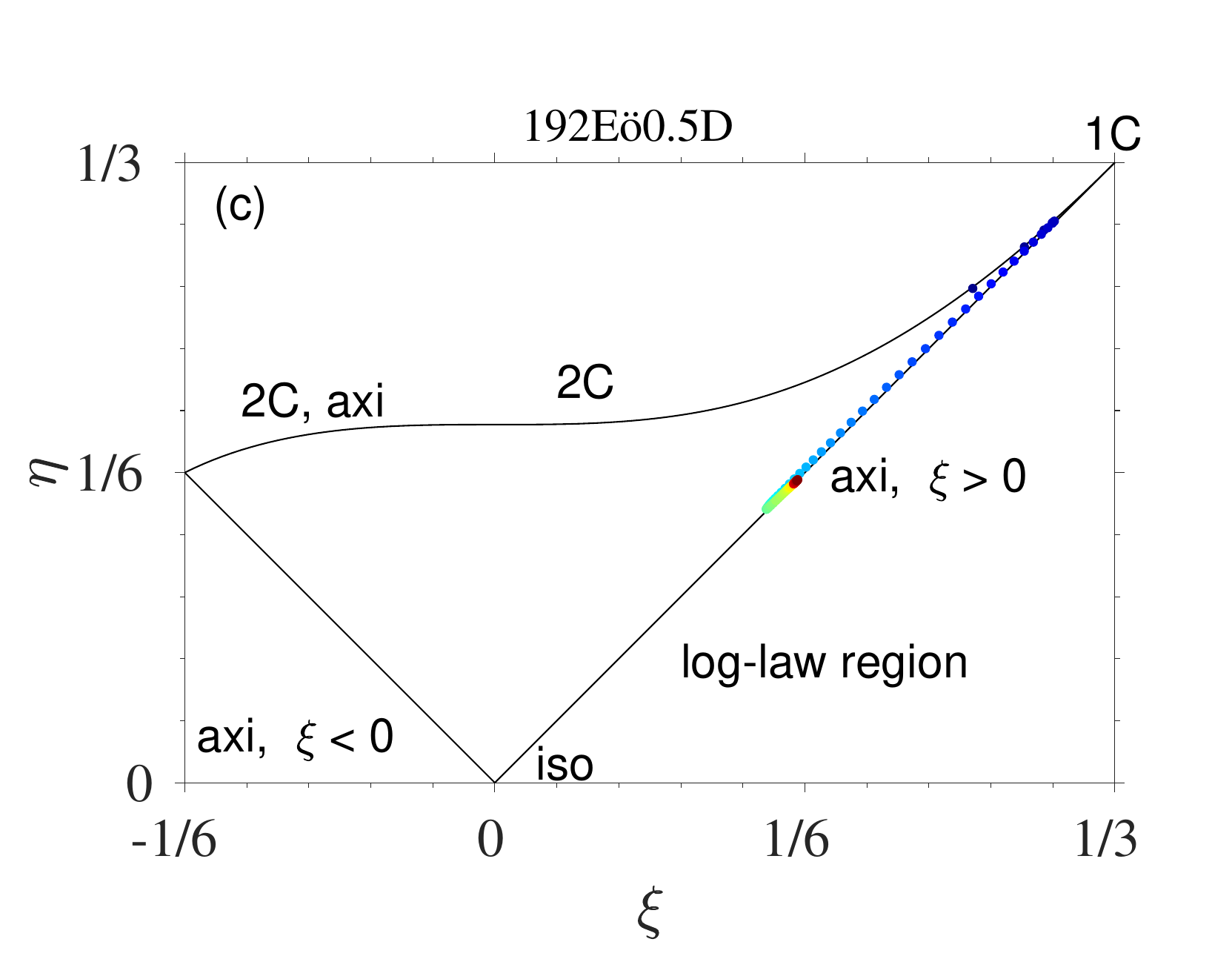}
    \end{subfigure}%
    \hspace{-5mm}
    \begin{subfigure}[b]{0.5\textwidth}
        \centering
        \includegraphics[height=5cm,keepaspectratio]{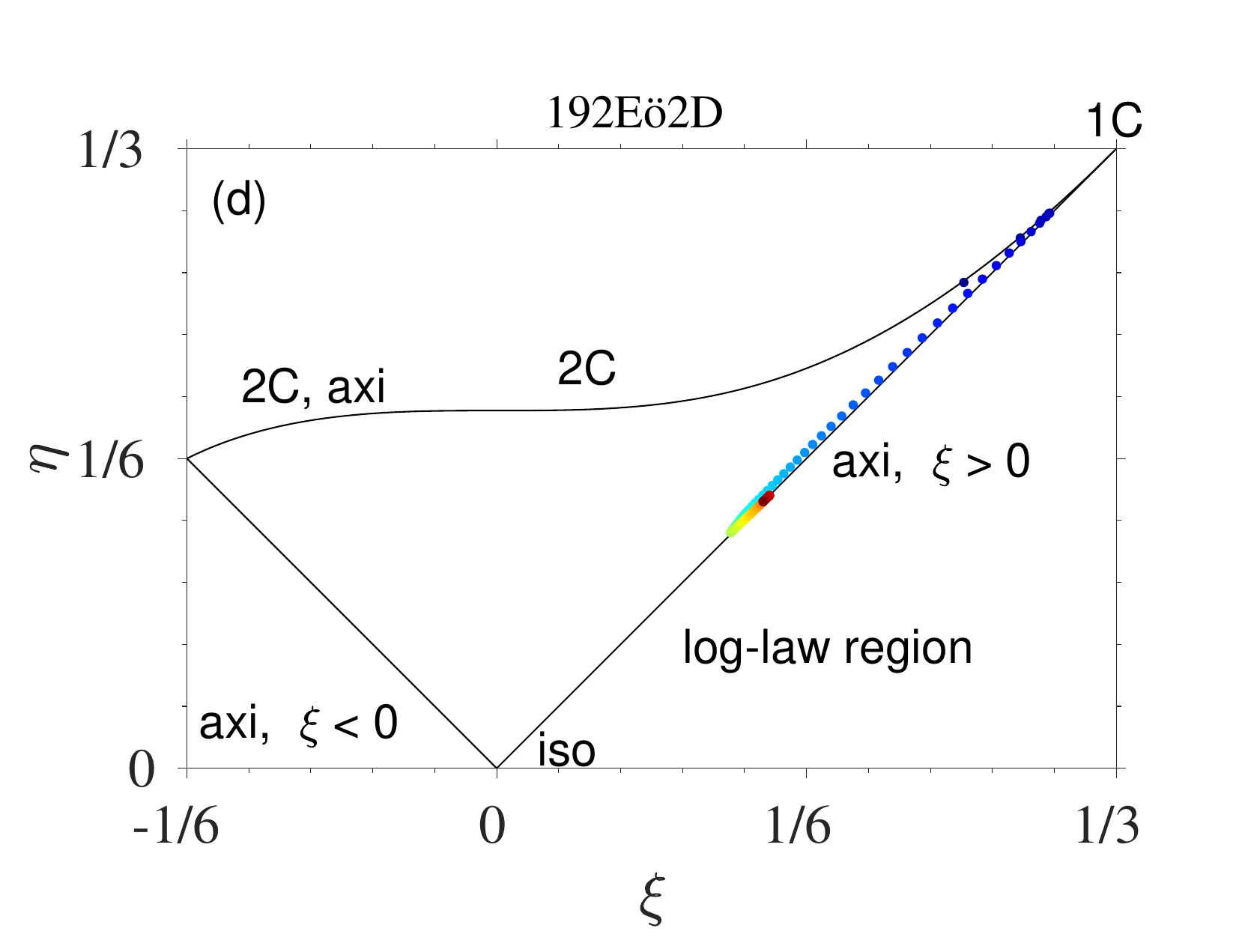}
    \end{subfigure}
    
    \begin{subfigure}[b]{0.5\textwidth}
        \centering
        \includegraphics[height=5cm,keepaspectratio]{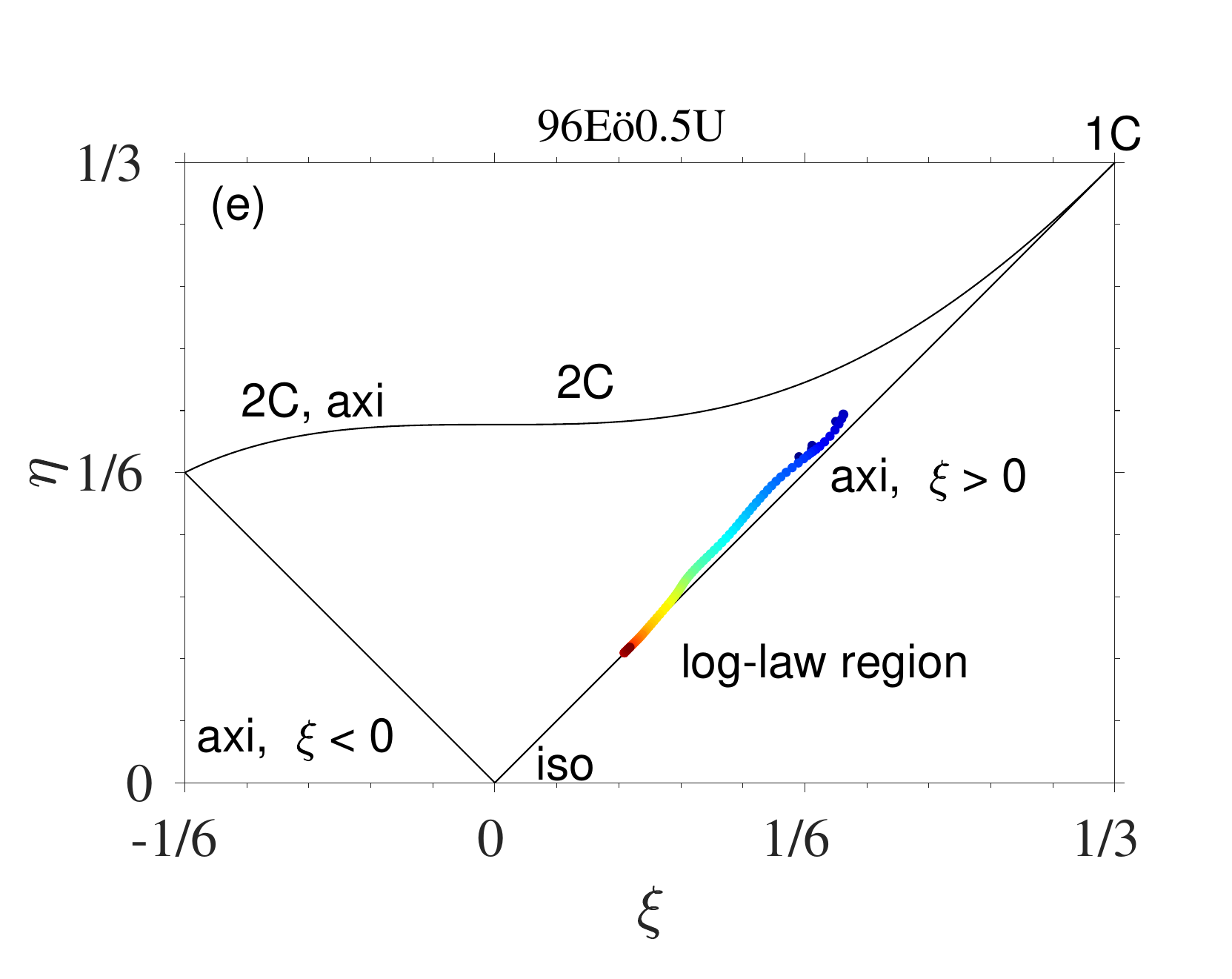}
    \end{subfigure}%
    \hspace{-5mm}
    \begin{subfigure}[b]{0.5\textwidth}
        \centering
        \includegraphics[height=5cm,keepaspectratio]{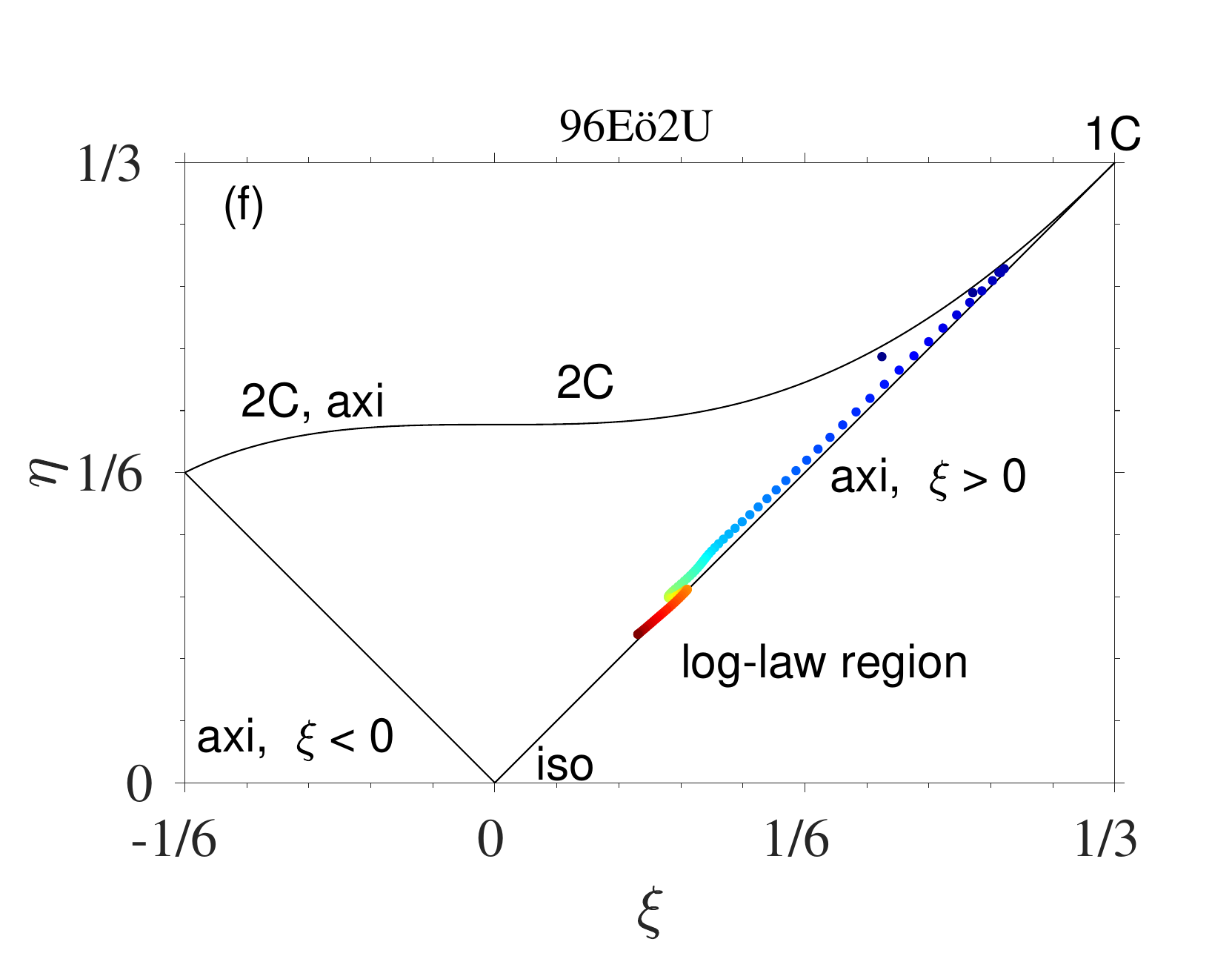}
    \end{subfigure}
    
    \begin{subfigure}[b]{0.5\textwidth}
        \centering
        \includegraphics[height=5cm,keepaspectratio]{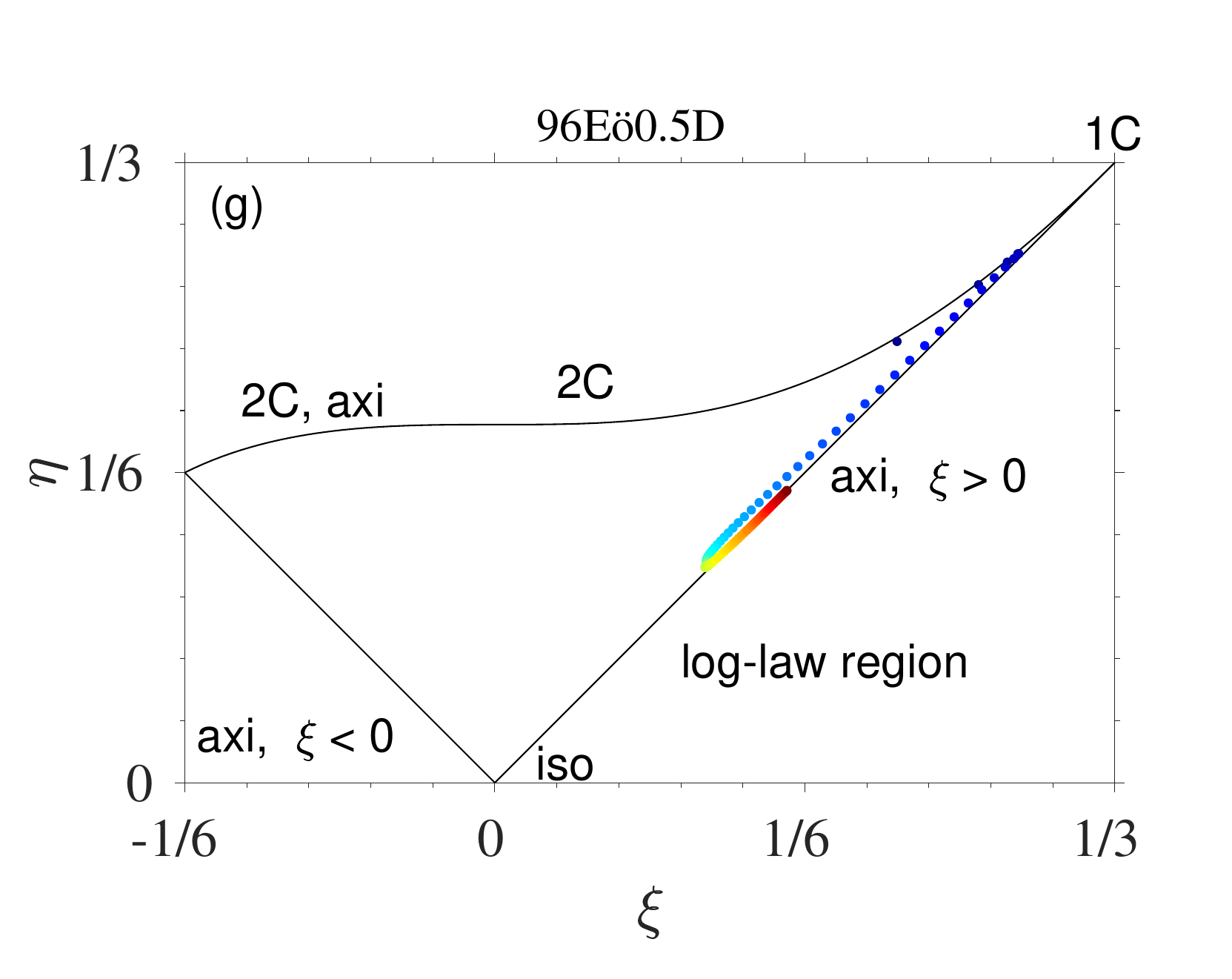}
    \end{subfigure}%
    \hspace{-5mm}
    \begin{subfigure}[b]{0.5\textwidth}
        \centering
        \includegraphics[height=5cm,keepaspectratio]{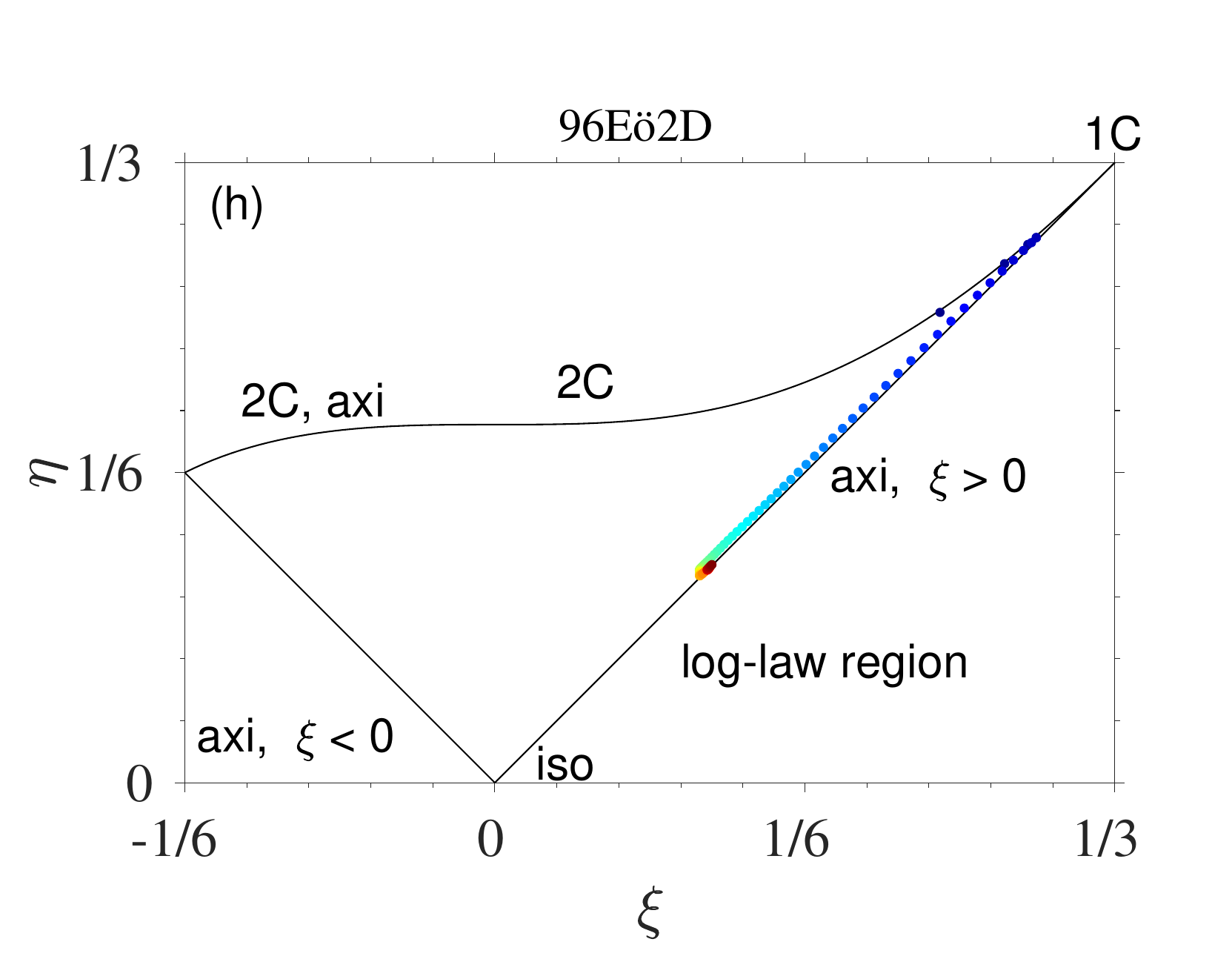}
    \end{subfigure}
    
    \caption{Lumley triangle for 8 cases} 
    \label{fig:Lumley_Triangle_8}
\end{figure*}


FIG. \ref{fig:Lumley_Triangle_8} presents the analysis of the anisotropy of the Reynolds stresses across all cases within the Lumley triangle. The general trends observed are largely consistent with single-phase DNS data. Notably, very close to the wall, within the viscous sublayer, the turbulence is predominantly two-component. Anisotropy peaks at a dimensionless wall distance, \(y^+ \approx 7\), nearing the 1C state, and it becomes progressively more isotropic towards the channel center\cite{Pope2000}. In the upward case example, compared to the instance where the E\"otv\"os number equals 0.5, the example with the E\"otv\"os number equal to 2 exhibits anisotropy that is closer to that of a single phase. FIG. \ref{fig:Lumley_Triangle_8}(a) and (e) showcase the differences in bubble count under upward flow with an Eötvös number of 0.5. FIG. \ref{fig:Lumley_Triangle_8}(a) and (b) display variations under upward flow with 192 bubbles at different Eötvös numbers, while FIG. \ref{fig:Lumley_Triangle_8}(e) and (f) show the differences for 96 bubbles at varying Eötvös numbers. When the Eötvös number is 0.5, bubbles are notably closer to the wall. The stirring action of the bubbles, especially in regions dense with bubbles, disrupts the orderly structure of the flow, effectively breaking up large-scale vortices and redistributing their energy into smaller-scale vortices that are more isotropic. Additionally, the upward movement of the bubbles induces extra motion in directions perpendicular to the flow, aiding in the dispersion of energy across multiple directions. This is evident from FIG. \ref{fig:Velocity_fluctuations}(c) and (e), where in the case of 192 bubbles in upward flow, \(\overline{u'_{y}u'_{y}}\) and \(\overline{u'_{z}u'_{z}}\) are higher near the wall compared to other cases. Furthermore, bubble-induced turbulence enhances the isotropy of the turbulence. The distribution and motion of bubbles in the fluid increase the exchange of kinetic energy between wall-normal and spanwise directions, helping to balance velocity fluctuations across different orientations. It is noteable that in downward cases, near the wall-center region, unlike single-phase flow, the liquid tends to deviate further from isotropic behavior. This observation is associated with additional pseudo-turbulence caused by bubble aggregation, as shown in FIG. \ref{fig:Velocity_fluctuations}(b), (d), and (f).




To further investigate the impact of the Eötvös number, void fraction, and flow direction on the turbulent kinetic energy (TKE), the budget terms from its transport equation are evaluated \cite{ma2017direct,Kataoka1989}. The transport equation for the turbulence kinetic energy of the liquid phase is expressed as



\begin{equation}\rho_L\frac{D\overline{\varphi_L}k}{Dt}=P_k+D_k+\epsilon_k-\underbrace{\overline{p_L^{\prime}u_{L,i}^{\prime}n_iI}+\overline{\tau_{L,ij}^{\prime}u_{L,i}^{\prime}n_jI}}_{S_k},
\label{eq:TKE}
\end{equation}

with \begin{equation}P_k=-\rho_L\overline{\varphi_L} \overline{\overline{ u_i^{\prime}u_j^{\prime}}}\frac{\partial\overline{\overline{u_i}}}{\partial u_j},
\label{eq:production}
\end{equation}

\begin{equation}\begin{aligned}D_k&=-\frac\partial{\partial x_i}\left(\overline{\varphi_L}\overline{\overline{p^{\prime}u_i^{\prime}}}\right)-\rho_L\frac\partial{\partial x_j}\left(\overline{\varphi_L}\overline{\overline{u_i^{\prime}u_i^{\prime}u_j^{\prime}}}\right)+\frac\partial{\partial x_j}\left(\overline{\varphi_L}\overline{\overline{u_i^{\prime}\tau_{ij}^{\prime}}}\right)
\label{eq:diffusion}
,\end{aligned}\end{equation}

\begin{equation}\epsilon_k=-\overline{\varphi_L}\overline{\overline{\tau_{ij}'\frac{\partial u_i'}{\partial x_j}}},
\label{eq:dissipation}
\end{equation}

\begin{equation}k=\frac12\overline{\overline{u_i^{\prime}u_i^{\prime}}}\end{equation}

\begin{equation}\overline{\tau_{ij}}=\overline{\nu{\left(\frac{\partial u_i}{\partial u_j}+\frac{\partial u_j}{\partial u_i}\right)}},
\label{eq:tau}
\end{equation}


where \(\overline{\varphi_L}\) is the indicator of the liquid phase, as shown in Eq.~\ref{eq:19}. The simple statistical averaging is indicated by a single overbar, while the double overbar denotes the phase-weighted averaging, as demonstrated in Eq.~\ref{eq:18}. Both averaging procedures involve spatial averaging in the wall-normal direction and time averaging. The variables \( P_k \), \( D_{k} \), \( \epsilon_k \), and \( S_k \) represent shear production, turbulent diffusion, dissipation, and the interfacial transfer of turbulent energy between bubbles and the liquid, respectively. Additionally, \( \rho \), \( u \),\( k \) and \( p \) are the density, velocity, TKE of liquid phase and pressure of the liquid phase, respectively. The terms \( p'_L \), \( u'_{L,i} \), and \( \tau'_{L,ij} \) denote the fluctuations of pressure, the \(i\)th velocity component, and the viscous stress tensor in the liquid phase. Furthermore, \( n_i \) represents the normal vector at the phase boundary directed towards the gas phase, and \( I \) is the interfacial area concentration.

\begin{equation}
 I n_i= -\frac{\partial \varphi}{\partial x_i}
\label{eq:interfacial_concentration}
\end{equation}

\begin{figure*}		
    \centering
    \begin{subfigure}[b]{0.5\textwidth} 
        \centering
        \includegraphics[height=6.8cm,keepaspectratio]{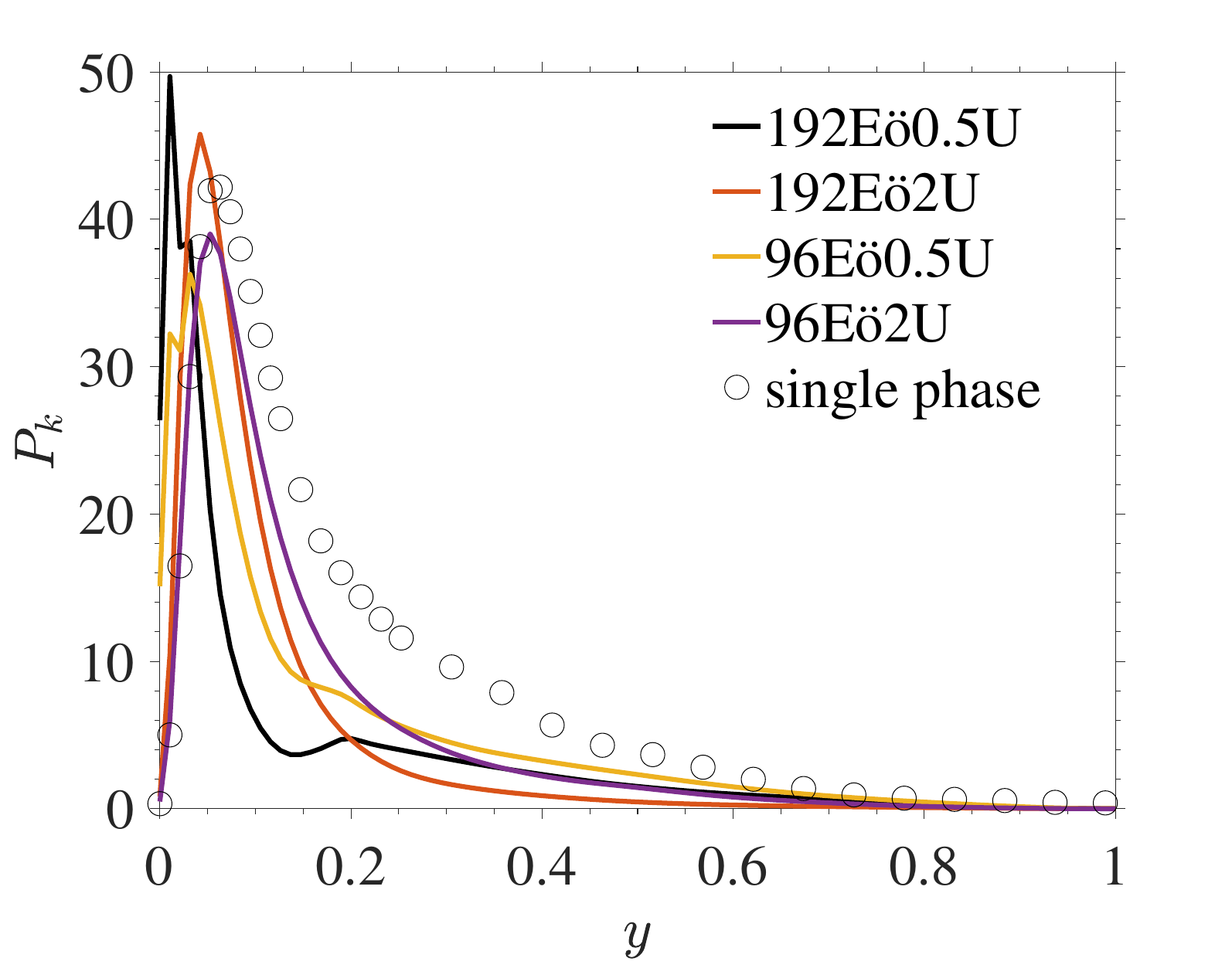}
    \end{subfigure}%
    \hspace{-5mm} 
    \begin{subfigure}[b]{0.5\textwidth} 
        \centering
        \includegraphics[height=6.8cm,keepaspectratio]{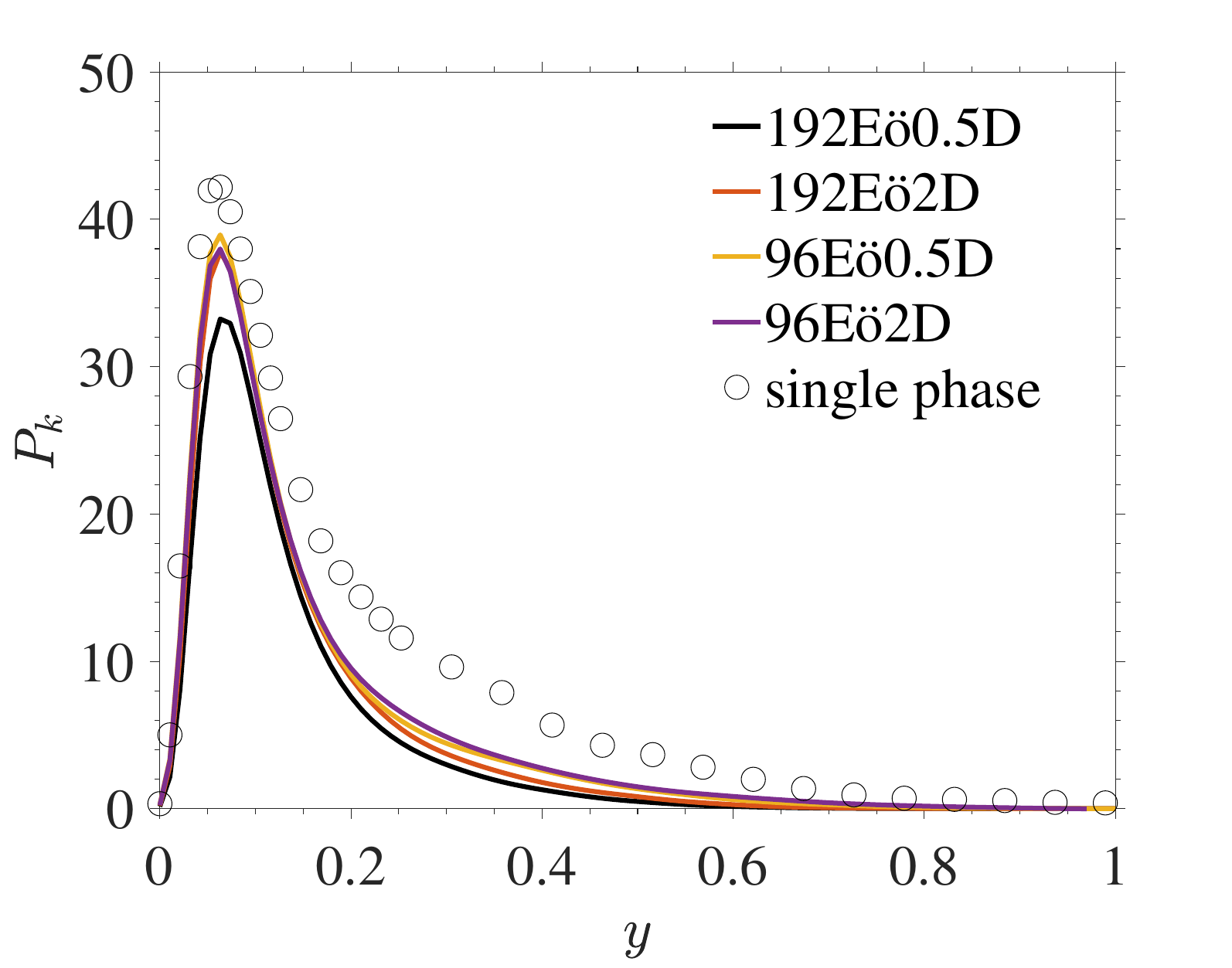}
    \end{subfigure}
    \caption{(a) Production in the upward case; (b) Production in the downward case.}

    \label{fig:Production}
\end{figure*}

Production is demonstrated in FIG.~\ref{fig:Production} for eight cases. Figure~\ref{fig:Production}(b), which corresponds to the downward cases, shows a decay across the entire domain compared to single-phase flow. This trend is consistent with that observed in Figure~\ref{fig:Velocity_fluctuations}(h). Figure~\ref{fig:Production}(a) corresponds to the upward cases, where the case with an E\"otv\"os number equal to 0.5 shows Production closer to the wall. This is associated with greater disturbances generated in the spanwise and wall-normal directions, consistent with the trends observed in Figures~\ref{fig:Velocity_fluctuations}(c) and (e). Dissipation, as illustrated by Eq.~(\ref{eq:dissipation}), is demonstrated in FIG. \ref{fig:Dissipation} for eight cases. FIG. \ref{fig:Dissipation}(a) represents the upward cases. For the case with an E\"otv\"os number of 2, the curve initially rises near the wall, then decreases, reaching a minimum near \(y=0.1\), and subsequently monotonically increases, approaching zero after \(y=0.5\). For the case with an E\"otv\"os number of 0.5, there is an overall trend of monotonic increase, and compared to the case with an E\"otv\"os number of 2, it is closer to the wall. Additionally, the amplitude of Dissipation is greater than that of single-phase flow for \( y < 0.2 \). FIG. \ref{fig:Dissipation}(b) corresponds to the downward cases. For \( y > 0.4 \), the amplitude of Dissipation exceeds that of the single-phase flow, which can be attributed to bubble clustering.





In Eq.~(\ref{eq:TKE}), the single-phase term can be closed using the corresponding terms from the shear stress transport (SST) model\cite{Menter1994}. Additionally, a source term $S_k^{\mathrm{RANS}}$ is introduced to represent the production of BIT

\begin{equation}\frac{D(\alpha_L\rho_Lk)}{Dt}=P_k^\mathrm{RANS}+D_k^\mathrm{RANS}\underbrace{-\alpha_LC_\mu\rho_L\omega k}_{\varepsilon_k^\mathrm{RANS}}+\underbrace{C_I\boldsymbol{F}_D(\overline{u_{G}}-\overline{u_{L}})}_{S_k^\mathrm{RANS}},
\label{eq:SST}
\end{equation}

Here, \( \alpha_L = \overline{\varphi}\) is the void fraction of liquid, and \( \omega \) is the turbulent eddy frequency. \( C_I \) is a coefficient that can be determined through modeling, usually \( C_I \leq 1 \).
Specifically, it is assumed that the drag force, acting as the sole contributor to turbulence generation, results in all energy lost by the bubble being converted into turbulent kinetic energy in the bubble's wake. The expression for \( S_k^{\mathrm{RANS}} \) is given as:

\begin{equation}S_k^{\mathrm{RANS}}=C_I\boldsymbol{F}_D(\overline{u_{G}}-\overline{u_{L}}),
\label{eq:Skrans}
\end{equation}
where $\boldsymbol{F}_D$ is:

\begin{equation}\boldsymbol{F}_D=\frac3{4d_p}C_D\rho_L\alpha^G|\overline{u_{G}}-\overline{u_{L}}|(\overline{u_{G}}-\overline{u_{L}}).\end{equation}

with \( d_p \), the average bubble diameter. The drag coefficient \( C_D \) is expressed as a function of the bubble Reynolds number $Re_p=|\overline{u_{G}}-\overline{u_{L}}|d_p/\nu$ and the Eötvös number \(Eo \), where $\nu$ is the kinematic viscosity\cite{COLOMBO2021,Ishii1979}.

\begin{equation}C_D=max\big(C_{D,sphere},min\big(C_{D,ellipse},C_{D,cap}\big)\big)
\label{eq:CD}
\end{equation}
where

\begin{equation}\left.\left\{\begin{aligned}C_{D,sphere}&=\frac{24}{Re}\big(1+0.1Re^{0.75}\big)\\\\C_{D,ellipse}&=\frac{2}{3}\sqrt{Eo}\\\\C_{D,cap}&=\frac{8}{3}\end{aligned}\right.\right.\end{equation}

\begin{figure*}		
    \centering
    \begin{subfigure}[b]{0.45\textwidth} 
        \centering
        \includegraphics[height=6.5cm,keepaspectratio]{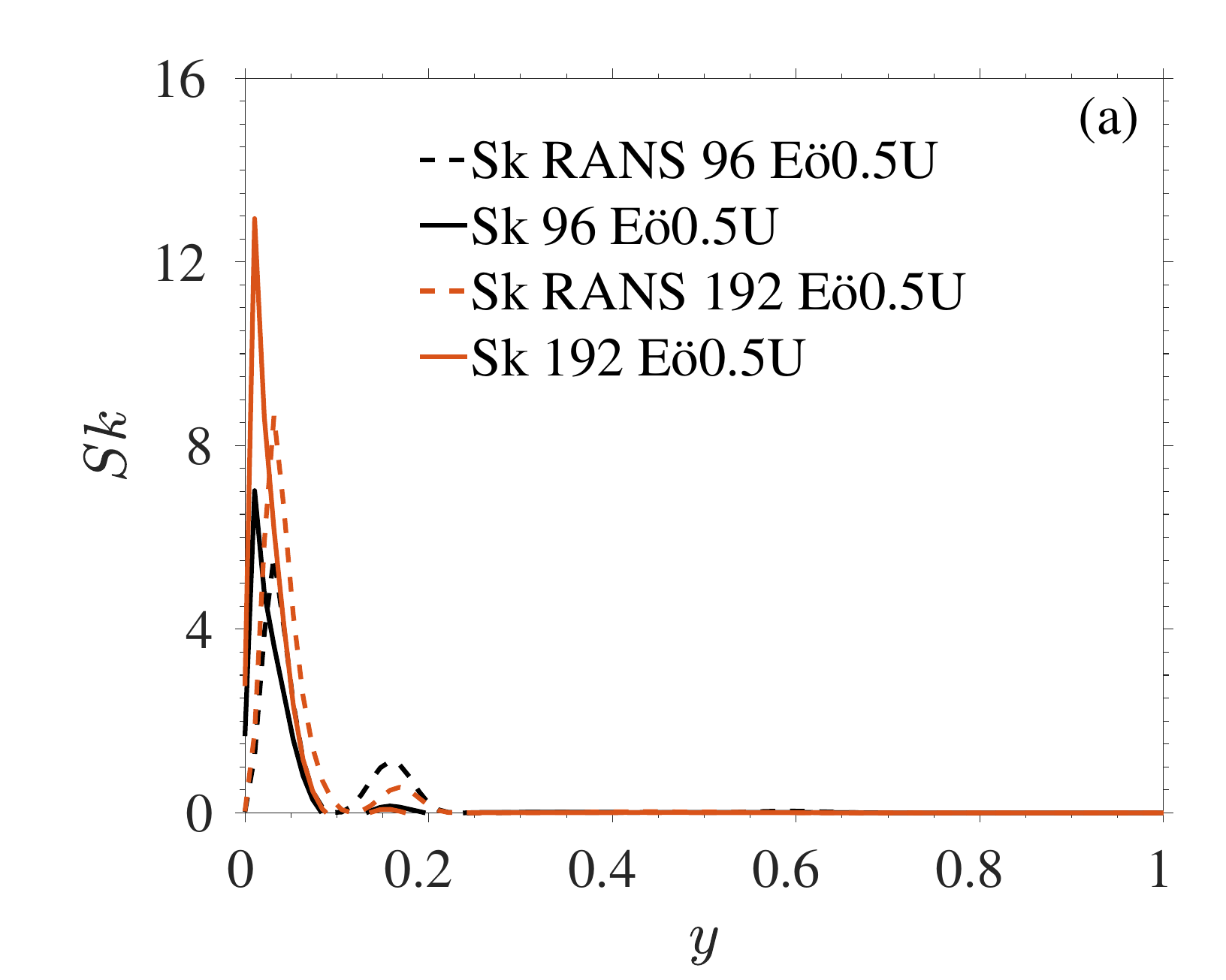}
        \label{fig:upward_eo0.5}
    \end{subfigure}%
    \hspace{5mm} 
    \begin{subfigure}[b]{0.45\textwidth} 
        \centering
        \includegraphics[height=6.5cm,keepaspectratio]{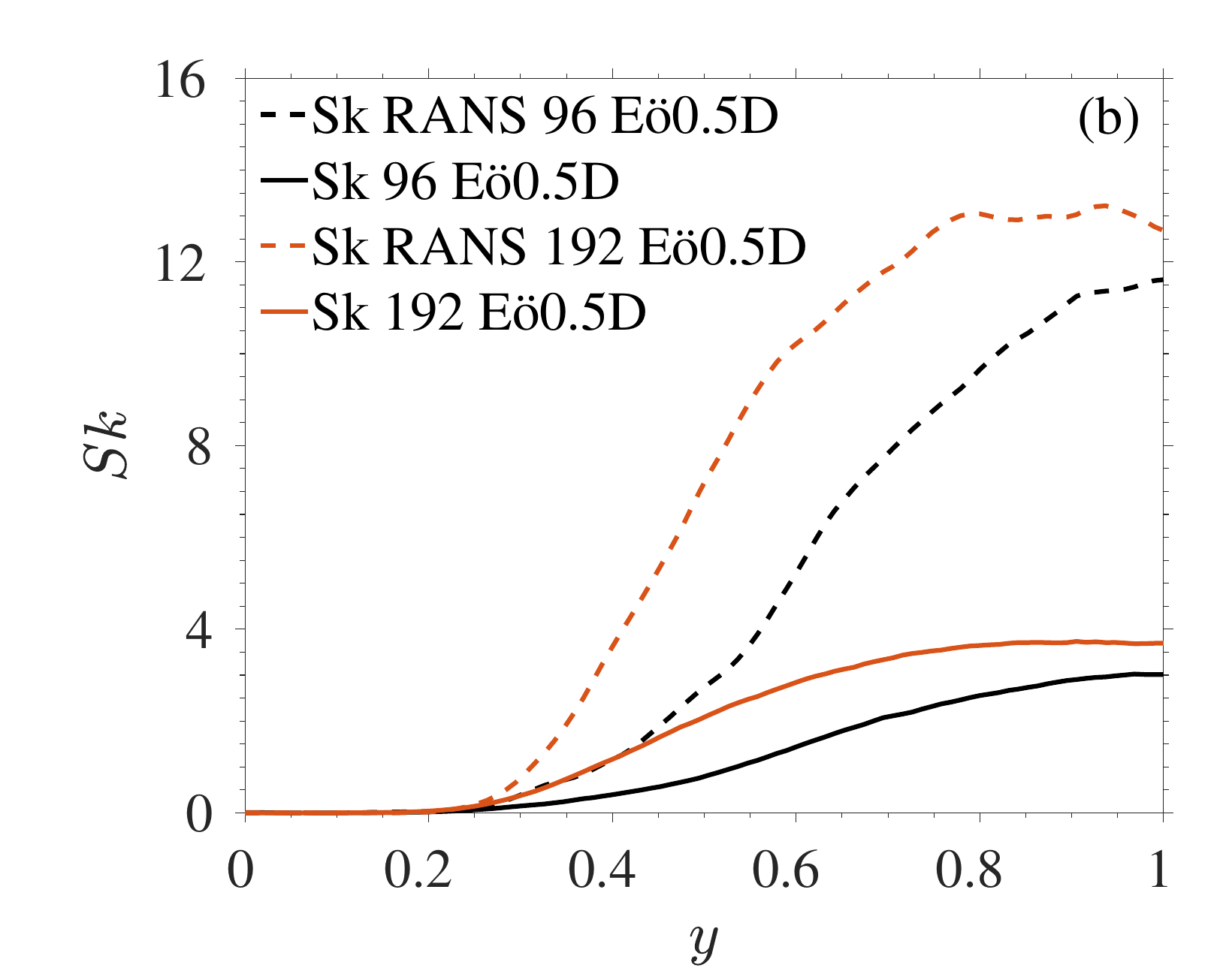}
        \label{fig:downward_eo0.5}
    \end{subfigure}
    
    \begin{subfigure}[b]{0.45\textwidth}
        \centering
        \includegraphics[height=6.5cm,keepaspectratio]{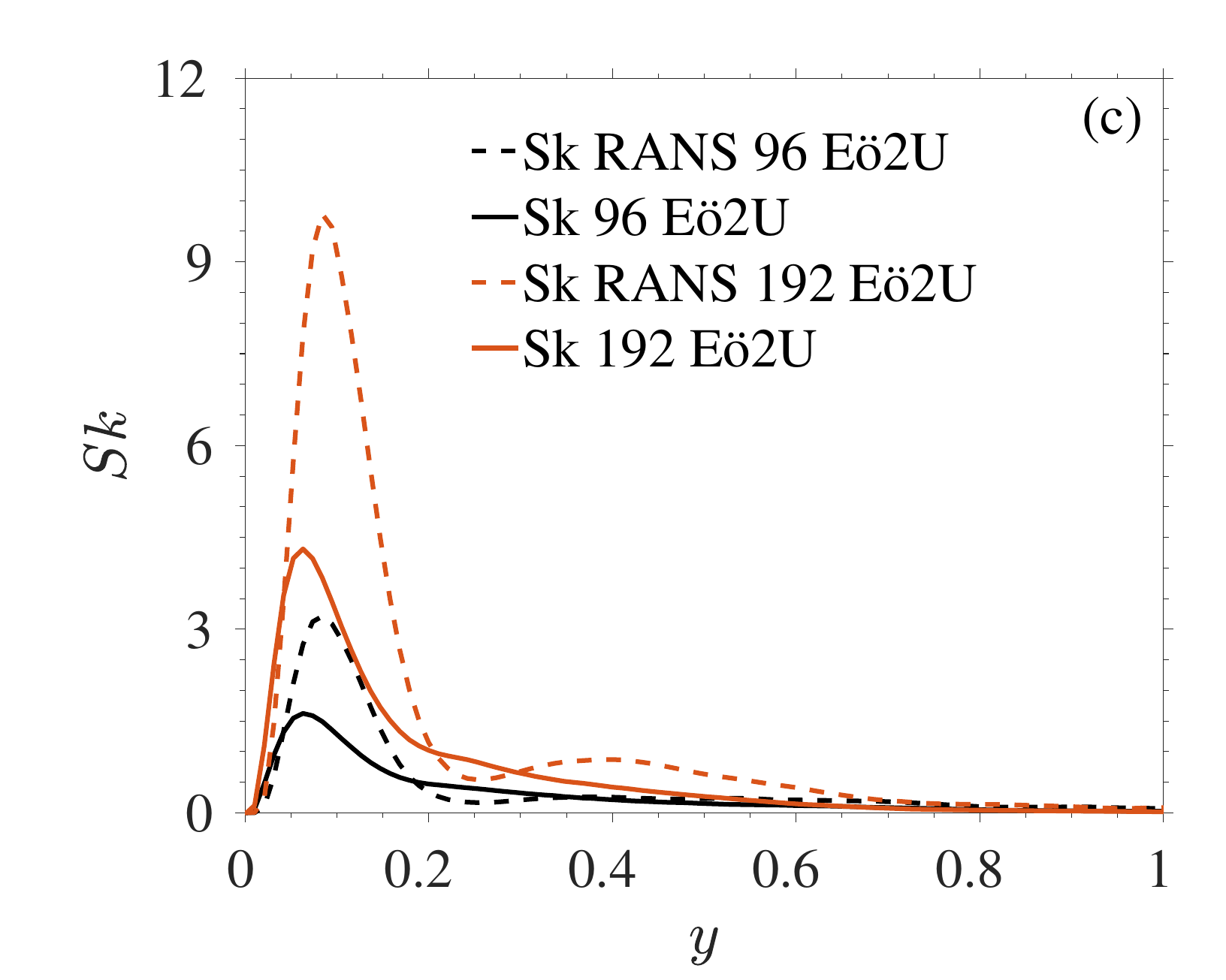}
        \label{fig:upward_eo2}
    \end{subfigure}%
    \hspace{5mm} 
    \begin{subfigure}[b]{0.45\textwidth}
        \centering
        \includegraphics[height=6.5cm,keepaspectratio]{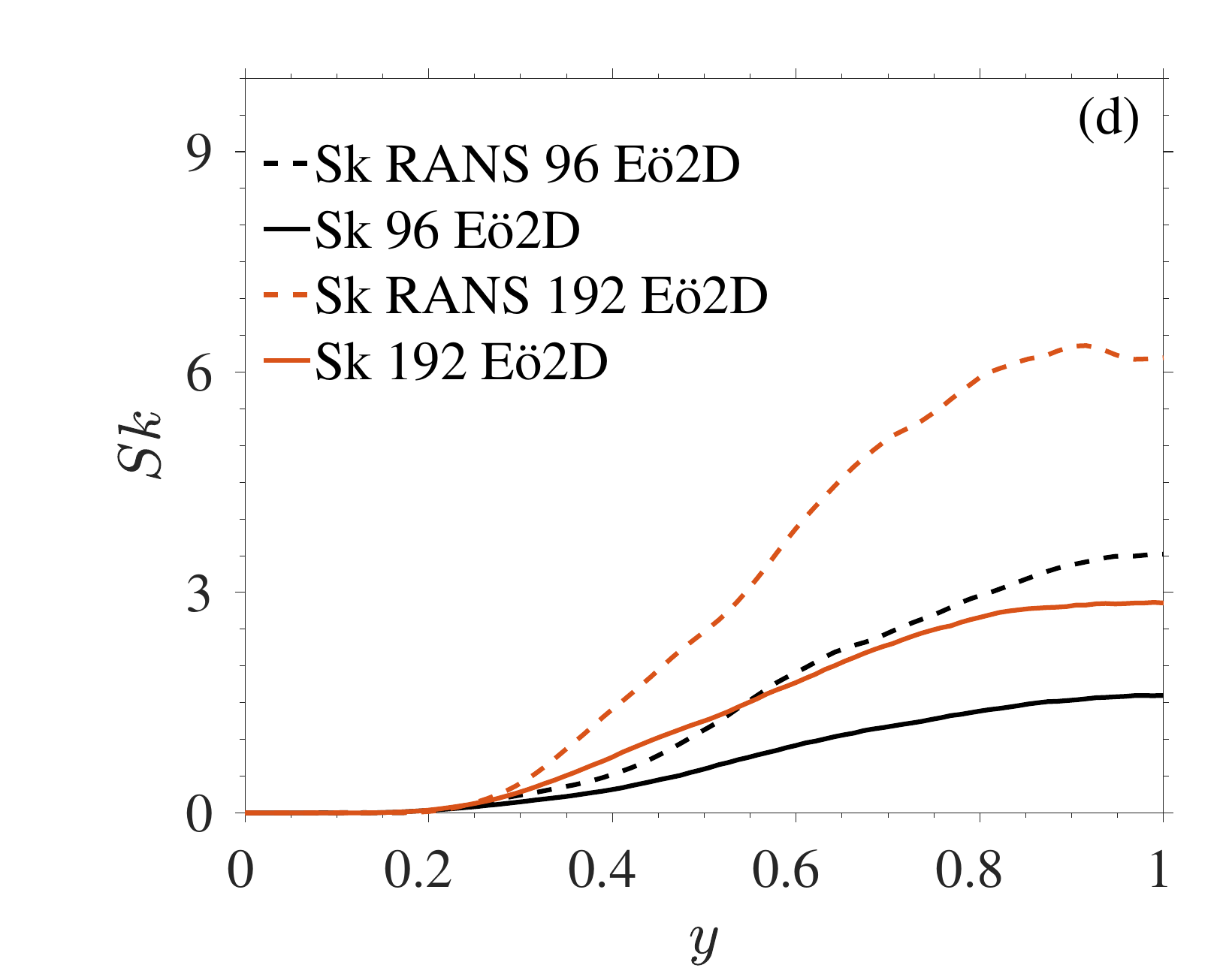}
        \label{fig:downward_eo2}
    \end{subfigure}
    
    \caption{Comparative analysis of bubble dynamics in different flow orientations and Eötvös numbers, showcasing the comparison between the interfacial term $S_k$ and $S_k^{\text{RANS}}$  based on \textit{a priori} tests: (a) upward flow with Eötvös number of 0.5, (b) downward flow with Eötvös number of 0.5, (c) upward flow with Eötvös number of 2, and (d) downward flow with Eötvös number of 2. }

    \label{fig:Sk}
\end{figure*}

FIG.\ref{fig:Sk} includes four subfigures, each illustrating the distribution of the interfacial term $S_k$ under different conditions and its comparison with $S_k^{\text{RANS}}$, based on \textit{a priori} tests. From FIG.\ref{fig:Sk}(a) and (c), it can be observed that compared to $S_k^{\text{RANS}}$, the values of $S_k$ are closer to the wall. This is because $S_k$ represents the interfacial energy transfer between bubbles, and thus the wall-normal averaging of $S_k$ reflects the position of the bubble interface. On the other hand, $S_k^{\text{RANS}}$, as seen from Eq.~(\ref{eq:Skrans}), is obtained through the difference in average velocities between the liquid and the gas, therefore it is closer to the center of the bubbles. Consequently, $S_k^{\text{RANS}}$ does not align well with the $S_k$ in the upward case. From Eq.~(\ref{eq:TKE}), it can be seen that $S_k$ is composed of two terms, one of which is related to the fluctuation of pressure and velocity, while the other is associated with the fluctuation of shear stress and velocity. Considering that shear stress originates from velocity gradients, $S_k$ exhibits higher values near the wall-side interface. Furthermore, by comparing FIG.~\ref{fig:Sk}(a) and (c), it is evident that in the case where the Eötvös number equals 0.5, which is closer to the wall, there is an increase in the fluctuation of both velocity and shear stress due to the proximity of the bubbles to the wall. $S_k$ also neutralizes a portion of the dissipation.


Thus, discussing the changes in the stress components becomes more crucial. Consequently, we introduce the exact balance equation for the Reynolds stresses, which is given as follows \cite{Kataoka1992}:

\begin{equation}\frac{\mathrm{D}}{\mathrm{D}t}(\overline{\varphi}\overline{\overline{u_i^{\prime}u_j^{\prime}}})=\mathcal{P}_{ij}+\mathcal{D}_{ij}+\phi_{ij}+\varepsilon_{ij}+\mathcal{S}_{R,ij},
\label{eq:BE_reynolds_stress}
\end{equation}

To distinguish from Equation~\ref{eq:TKE}, an upright font is used in Equation~\ref{eq:BE_reynolds_stress}. with the r.h.s. terms written as follows:

production:
\begin{equation}
\mathcal{P}_{ij}=-\overline{\varphi}\overline{\overline{u_i^{\prime}u_k^{\prime}}}\frac{\partial\overline{\overline{u_j}}}{\partial x_k}-\overline{\varphi}\overline{\overline{u_j^{\prime}u_k^{\prime}}}\frac{\partial\overline{\overline{u_i}}}{\partial x_k},\end{equation}

diffusion:
\begin{equation}\mathcal{D}_{ij}=-\frac\partial{\partial x_k}\left(\frac1\rho_L\overline{\varphi}\overline{\overline{p^{\prime}(\delta_{jk}u_i^{\prime}+\delta_{ik}u_j^{\prime})}}+\overline{\varphi}\overline{\overline{u_i^{\prime}u_j^{\prime}u_k^{\prime}}}-\frac1\rho_L\overline{\varphi}(\overline{\overline{u_j^{\prime}\tau_{ik}^{\prime}}}+\overline{\overline{u_i^{\prime}\tau_{jk}^{\prime}}})\right)\end{equation}

pressure-strain:
\begin{equation}\phi_{ij}=\frac1\rho_L\overline{\varphi}\overline{\overline{p^{\prime}\left(\frac{\partial u_i^{\prime}}{\partial x_j}+\frac{\partial u_j^{\prime}}{\partial x_i}\right)}},
\label{eq:pressurestrain}
\end{equation}

dissipation:
\begin{equation}\varepsilon_{ij}=-\frac1\rho_L\overline{\varphi}\overline{\overline{\tau_{ik}^{\prime}\left(\frac{\partial u_{j}^{\prime}}{\partial x_{k}}\right)}}-\frac1\rho_L\overline{\varphi}\overline{\overline{\tau_{jk}^{\prime}\left(\frac{\partial u_{i}^{\prime}}{\partial x_{k}}\right)}.}\end{equation}

the interfacial energy transfer term:
\begin{equation}S_{R,ij}=-\frac1\rho_L(\overline{p_L^{\prime}u_{L,j}^{\prime}n_iI}+\overline{p_L^{\prime}u_{L,i}^{\prime}n_jI})+\frac1\rho_L(\overline{\tau_{L,ik}^{\prime}u_{L,j}^{\prime}n_kI}+\overline{\tau_{L,jk}^{\prime}u_{L,i}^{\prime}n_kI})\end{equation}

\begin{figure*}		
    \centering
    \begin{subfigure}[b]{0.5\textwidth}
        \centering
        \includegraphics[height=5cm,keepaspectratio]{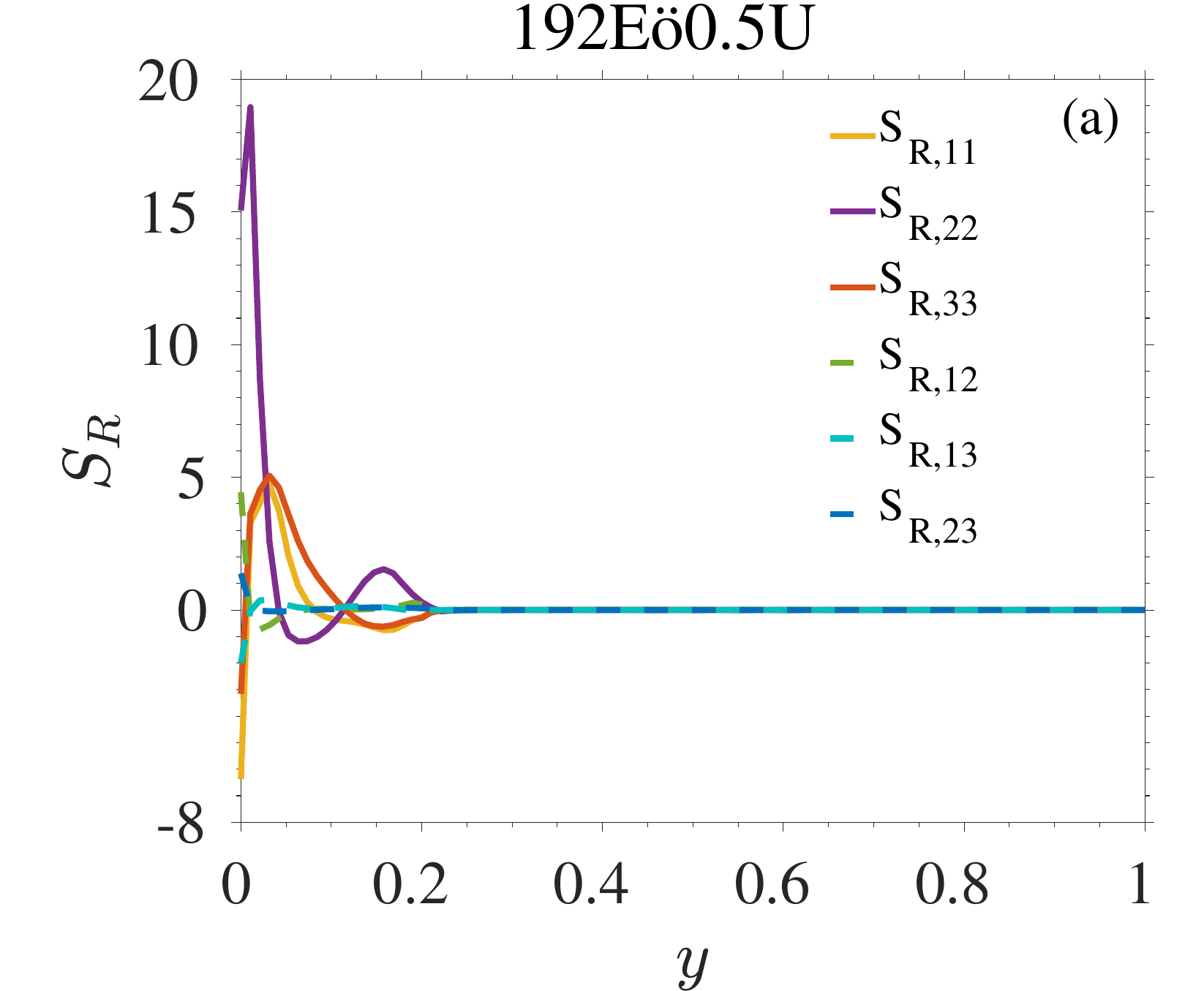}
    \end{subfigure}%
    \hspace{-5mm}
    \begin{subfigure}[b]{0.5\textwidth}
        \centering
        \includegraphics[height=5cm,keepaspectratio]{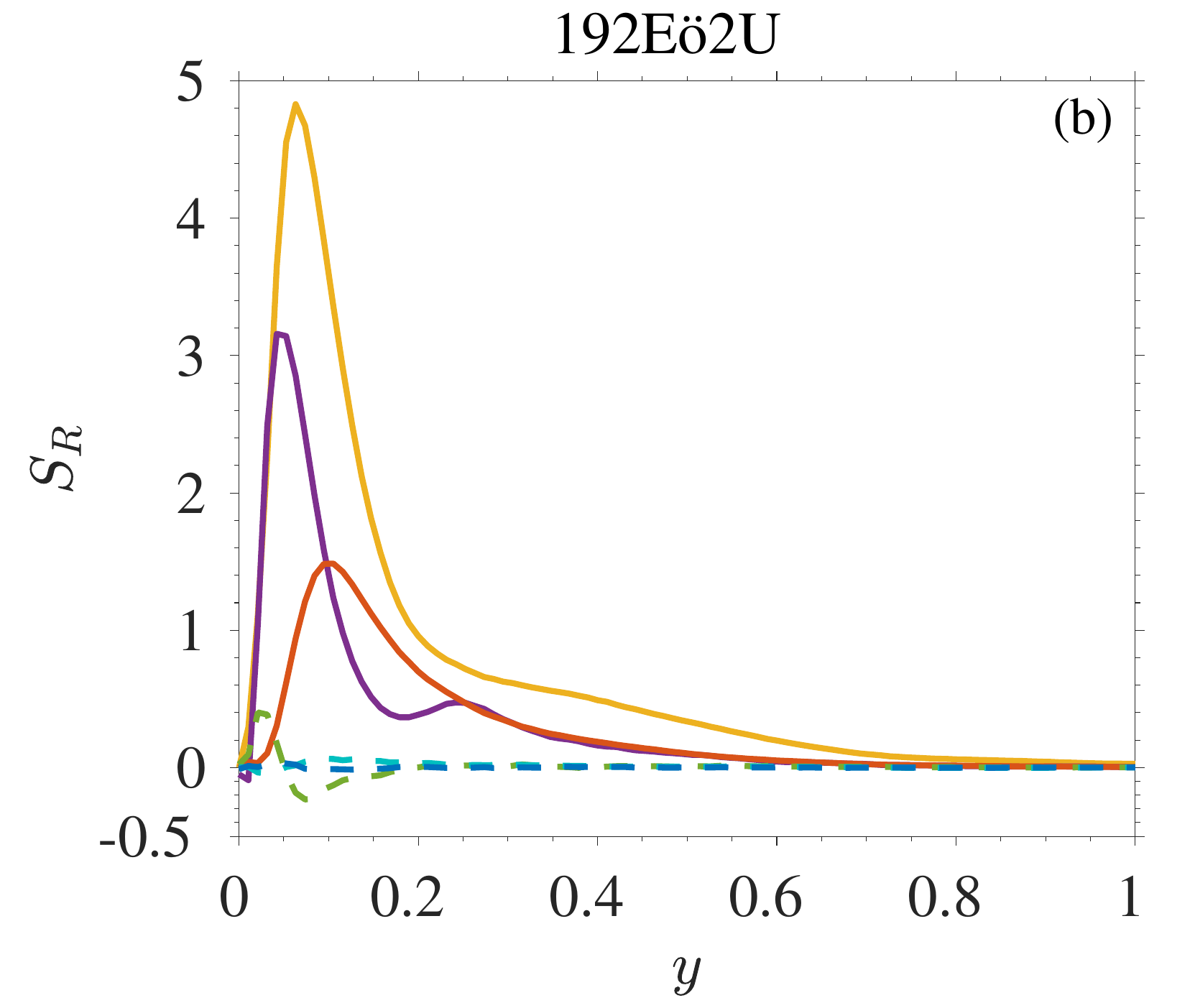}
    \end{subfigure}
    
    \begin{subfigure}[b]{0.5\textwidth}
        \centering
        \includegraphics[height=5cm,keepaspectratio]{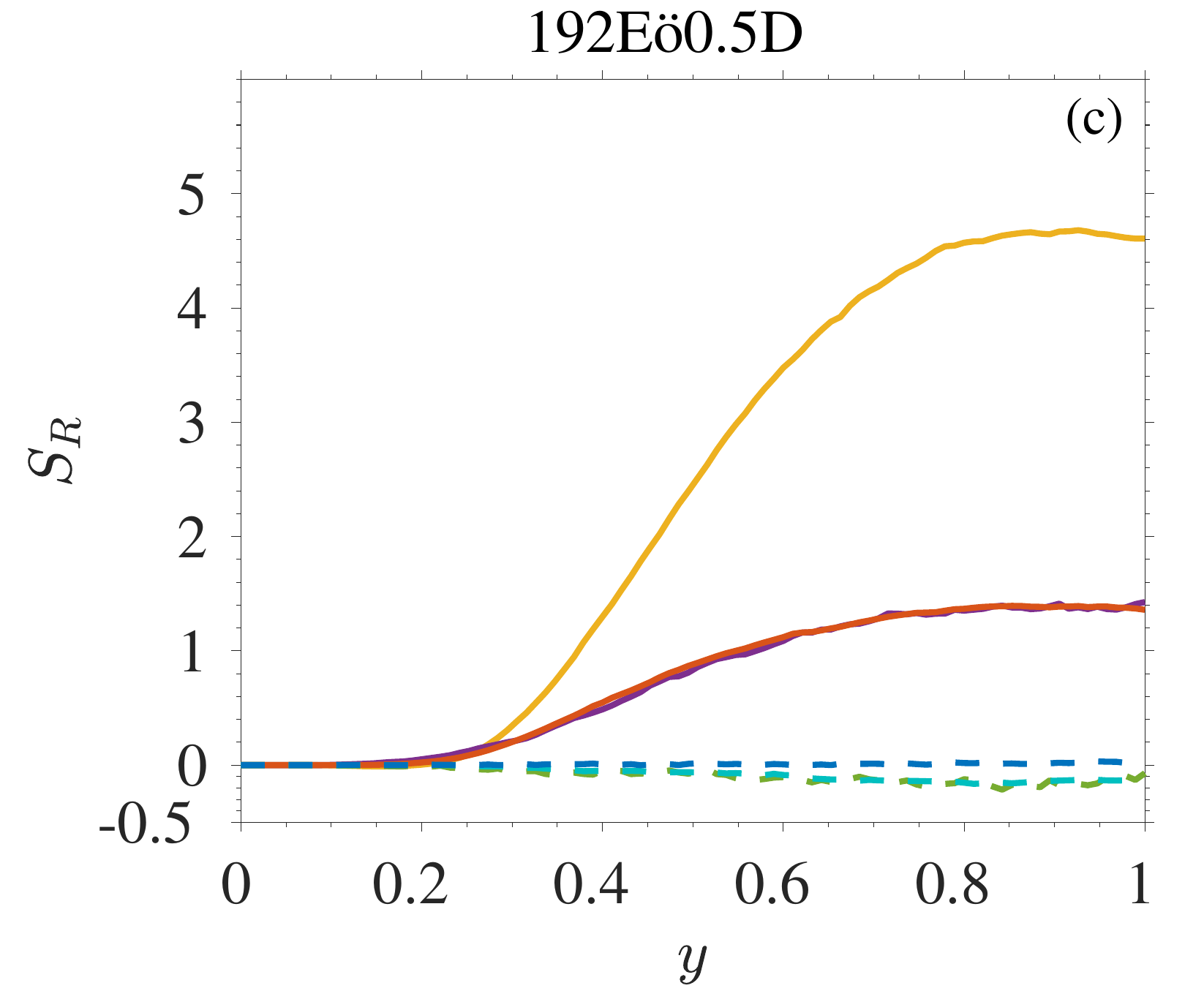}
    \end{subfigure}%
    \hspace{-5mm}
    \begin{subfigure}[b]{0.5\textwidth}
        \centering
        \includegraphics[height=5cm,keepaspectratio]{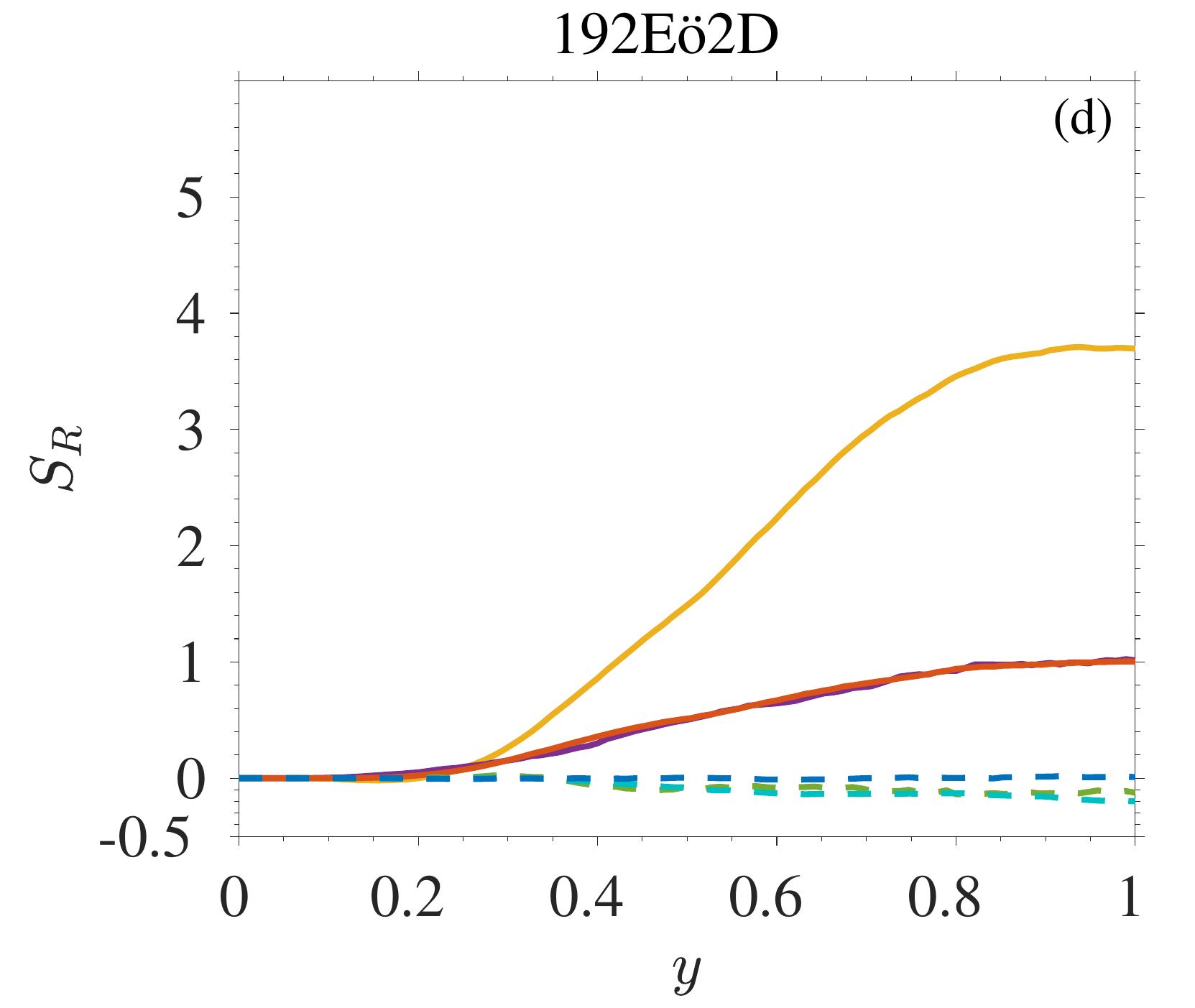}
    \end{subfigure}
    
    \begin{subfigure}[b]{0.5\textwidth}
        \centering
        \includegraphics[height=5cm,keepaspectratio]{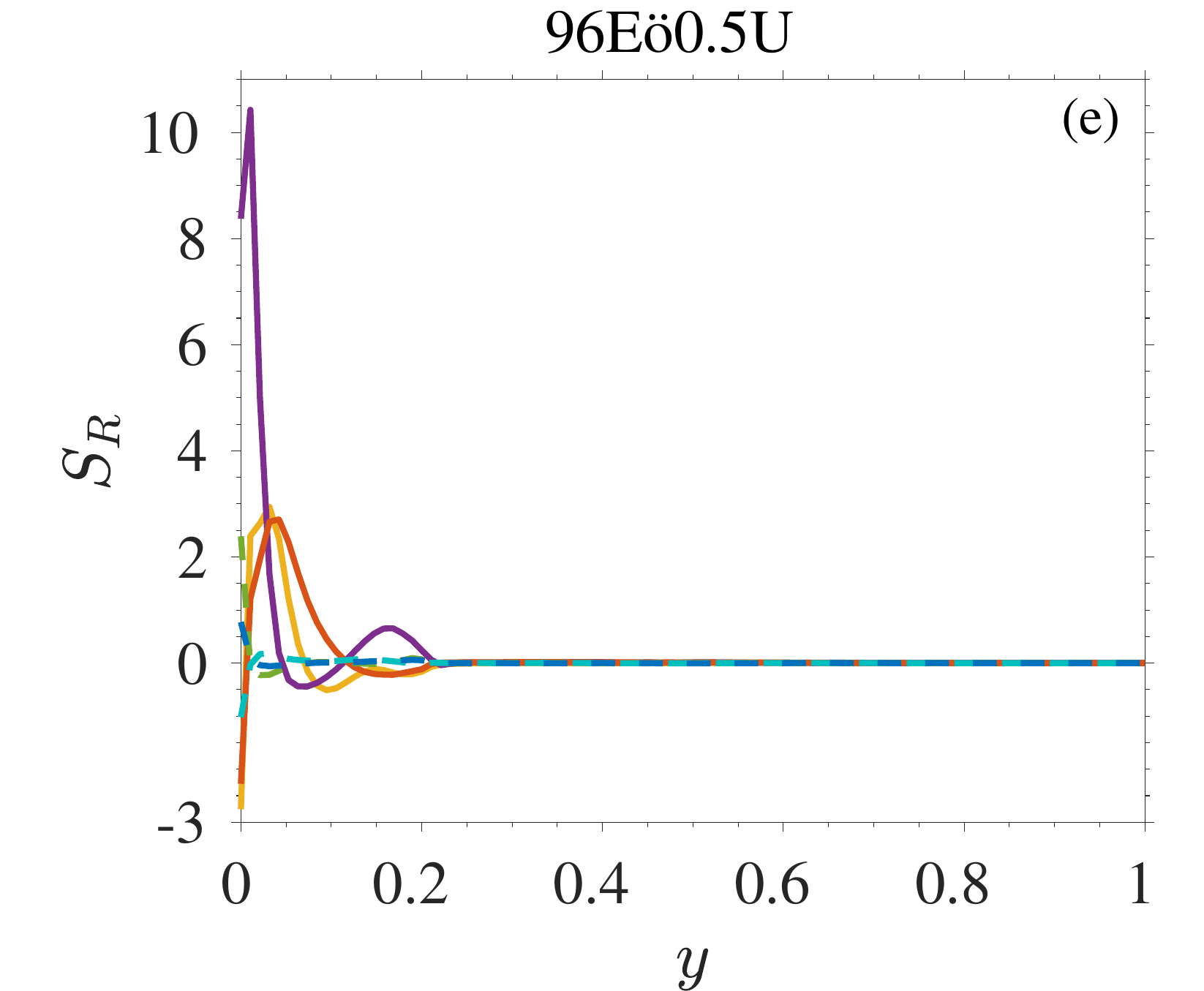}
    \end{subfigure}%
    \hspace{-5mm}
    \begin{subfigure}[b]{0.5\textwidth}
        \centering
        \includegraphics[height=5cm,keepaspectratio]{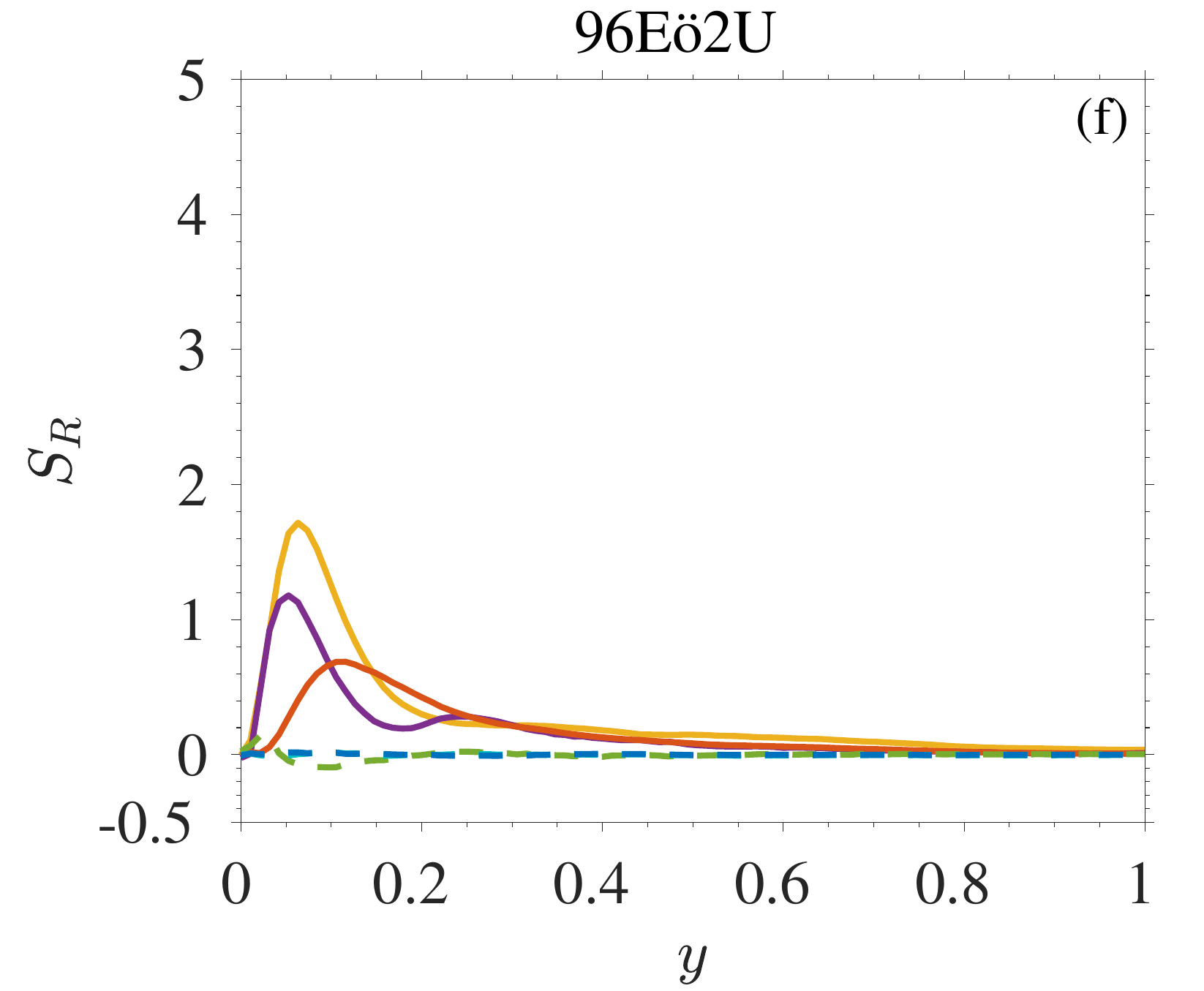}
    \end{subfigure}
    
    \begin{subfigure}[b]{0.5\textwidth}
        \centering
        \includegraphics[height=5cm,keepaspectratio]{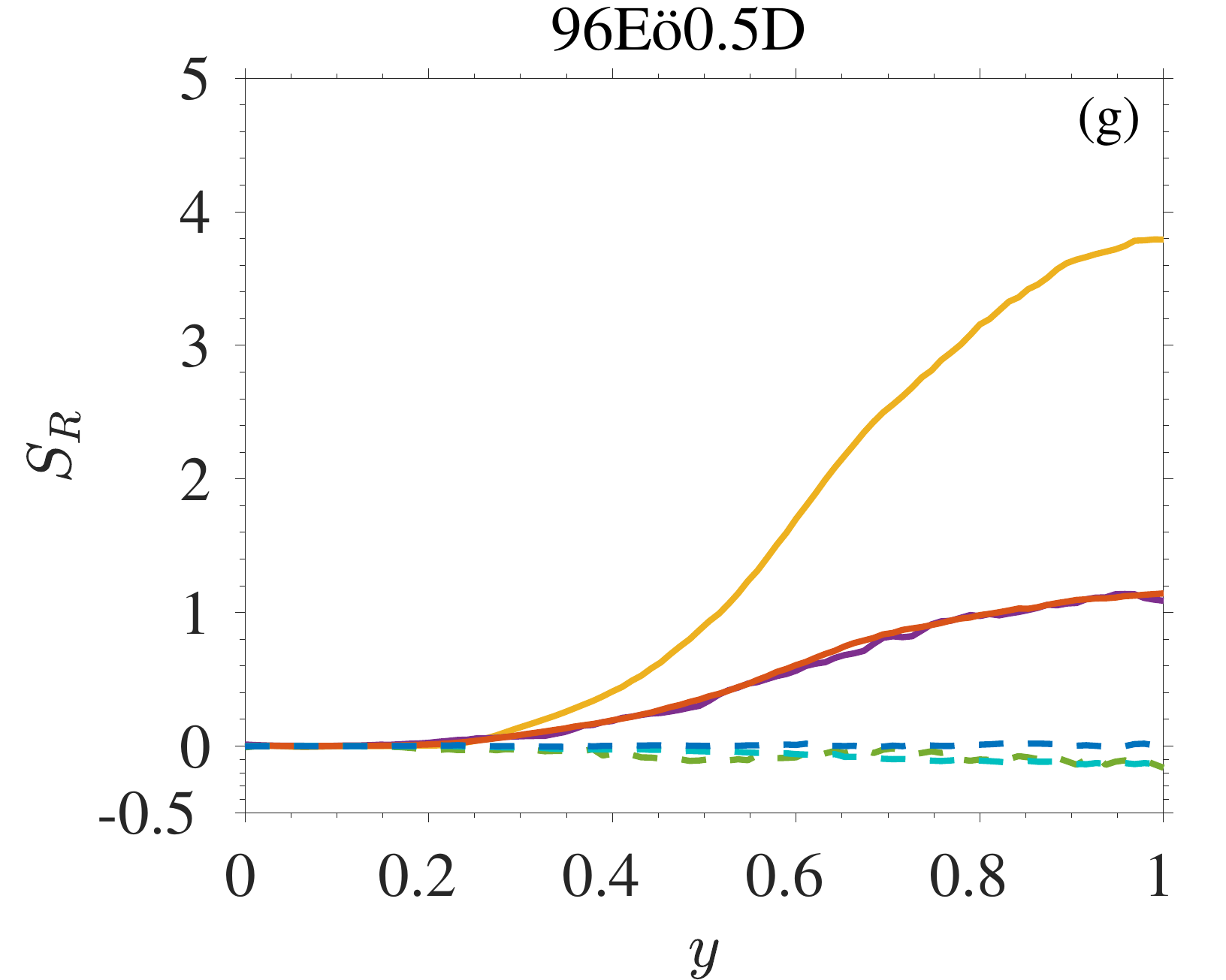}
    \end{subfigure}%
    \hspace{-5mm}
    \begin{subfigure}[b]{0.5\textwidth}
        \centering
        \includegraphics[height=5cm,keepaspectratio]{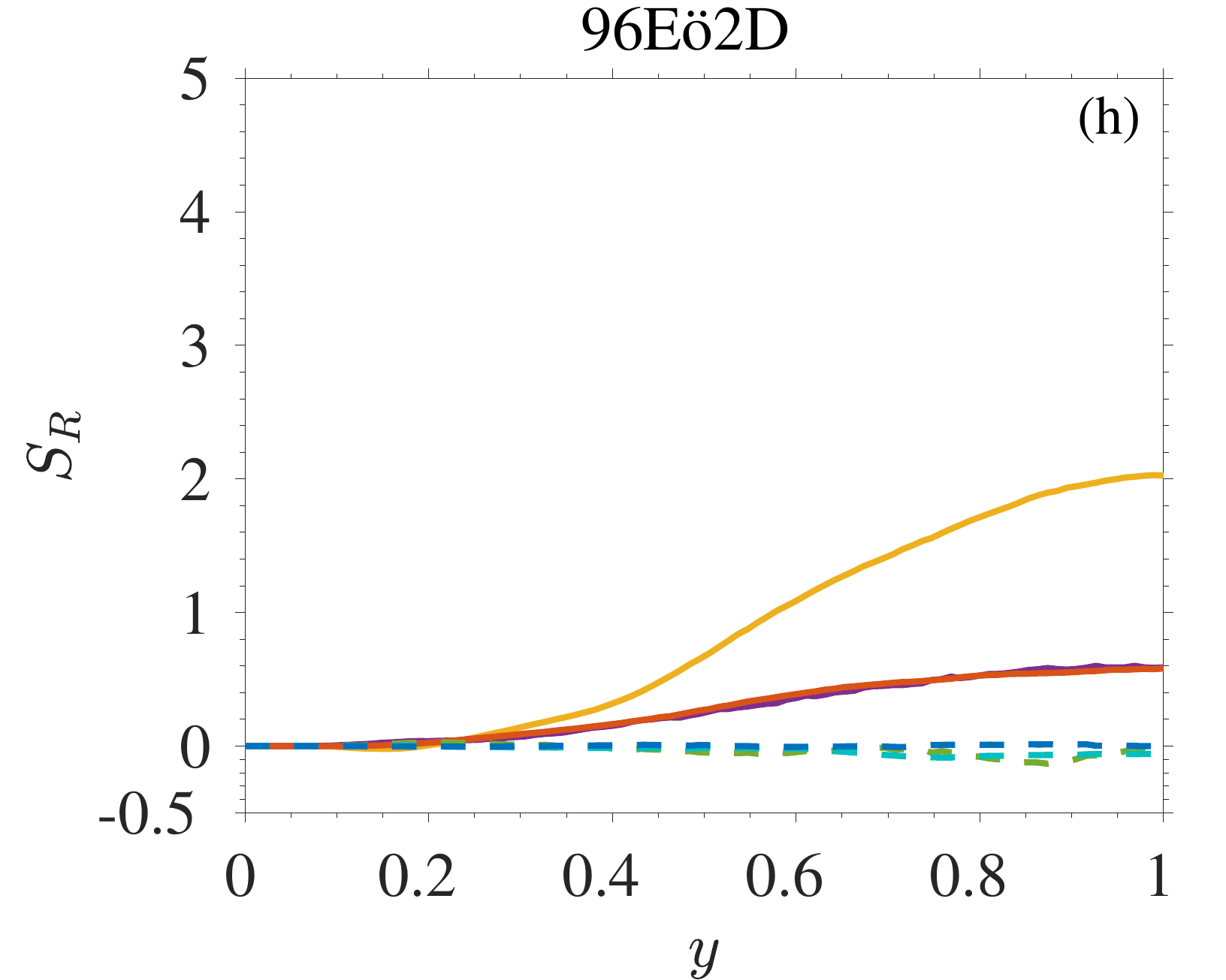}
    \end{subfigure}
    
    \caption{The interfacial energy transfer $S_{R,ij}$ for eight cases} 
    \label{fig:SRij}
\end{figure*}

FIG.~\ref{fig:SRij} presents a set of eight subfigures illustrating interfacial energy transfer. In the four downward cases, $S_{R,ij}$ is concentrated around the center of the channel. Among these, $S_{R,11}$ is the dominant component, while $S_{R,22}$ and $S_{R,33}$ have similar values and the other three quantities are close to zero. In cases where the Eötvös number is 0.5, $S_{R,ij}$ is greater than in cases with an Eötvös number of 2. This is associated with the pseudo-turbulence generated by bubble clustering in the channel center, as evident from FIG.~\ref{fig:Velocity_fluctuations}(b), (d), and (f). In upward cases, the term related to shear stress makes a main contribution to $S_{R,ij}$. The motion of bubbles near the wall generates substantial velocity gradients, particularly in the wall-normal direction. This increases the shear stress in that direction. Thus, in upward cases, $S_{R,22}$ exceeds $S_{R,33}$ in magnitude and is also closer to the wall. When the Eötvös number is 0.5, due to the bubbles being closer to the wall, $S_{R,22}$ even surpasses $S_{R,11}$, becoming the dominant component in $S_{R,ij}$.

\begin{figure*}		
    \centering
    \begin{subfigure}[b]{0.5\textwidth}
        \centering
        \includegraphics[height=5cm,keepaspectratio]{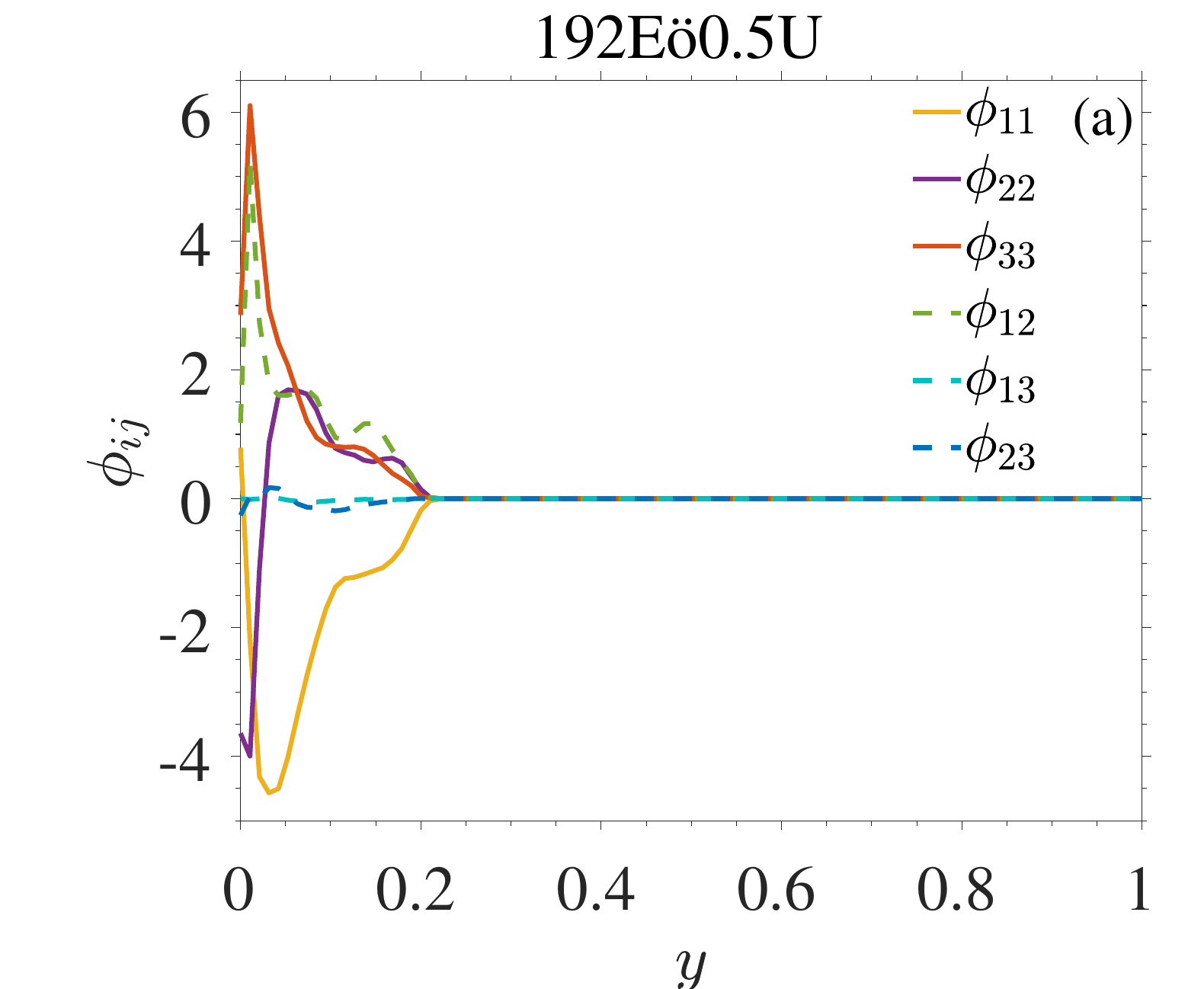}
    \end{subfigure}%
    \hspace{-5mm}
    \begin{subfigure}[b]{0.5\textwidth}
        \centering
        \includegraphics[height=5cm,keepaspectratio]{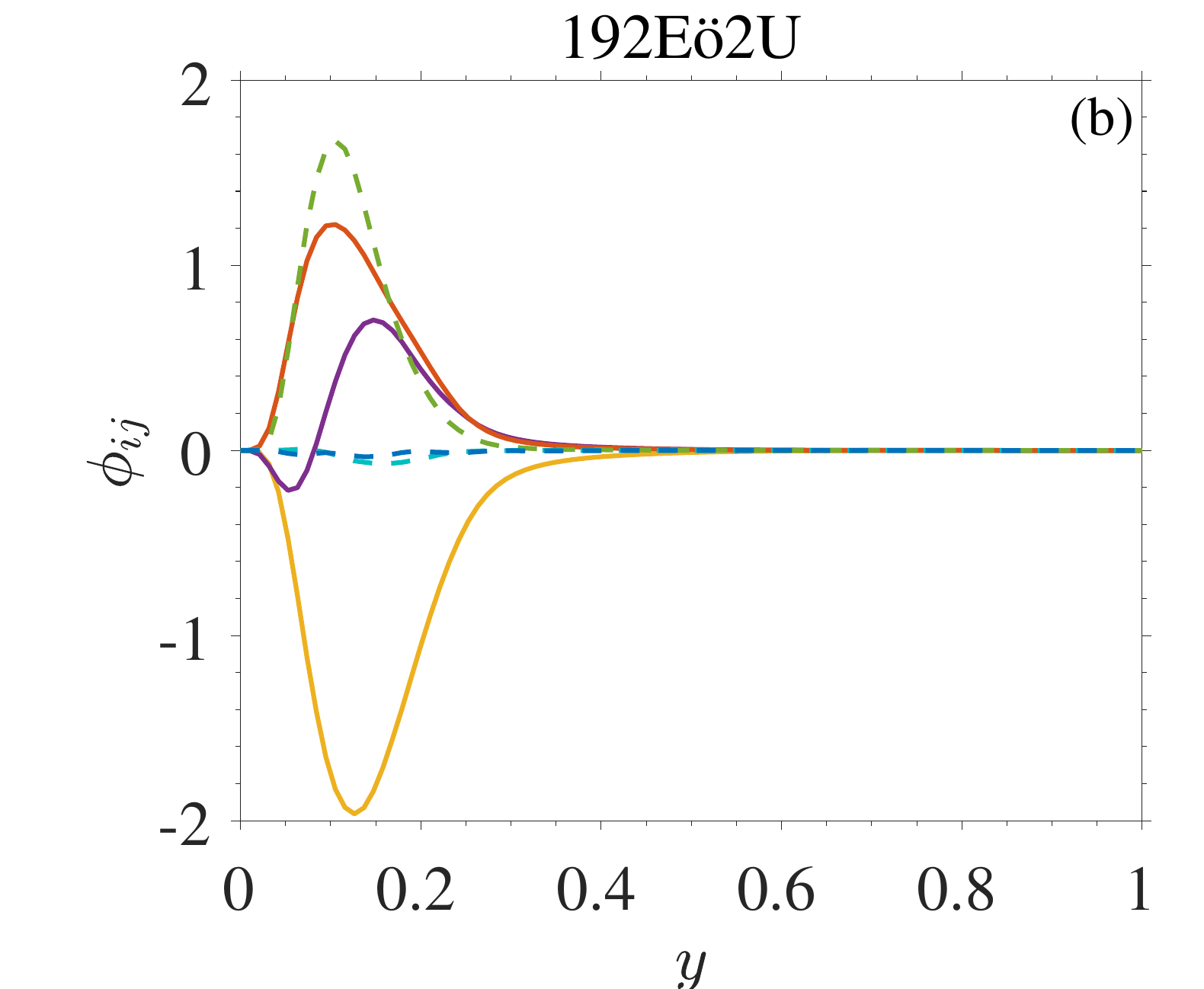}
    \end{subfigure}
    
    \begin{subfigure}[b]{0.5\textwidth}
        \centering
        \includegraphics[height=5cm,keepaspectratio]{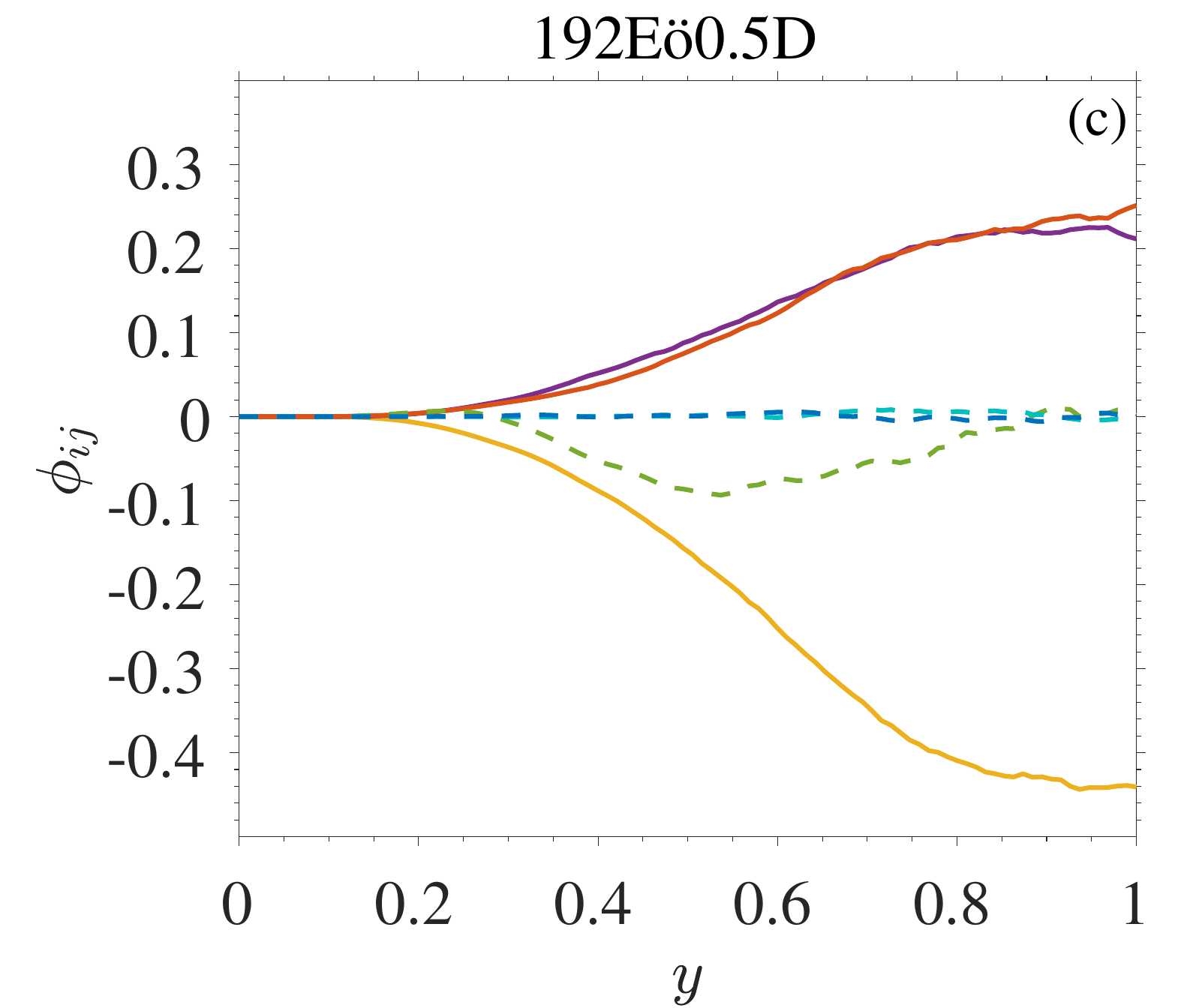}
    \end{subfigure}%
    \hspace{-5mm}
    \begin{subfigure}[b]{0.5\textwidth}
        \centering
        \includegraphics[height=5cm,keepaspectratio]{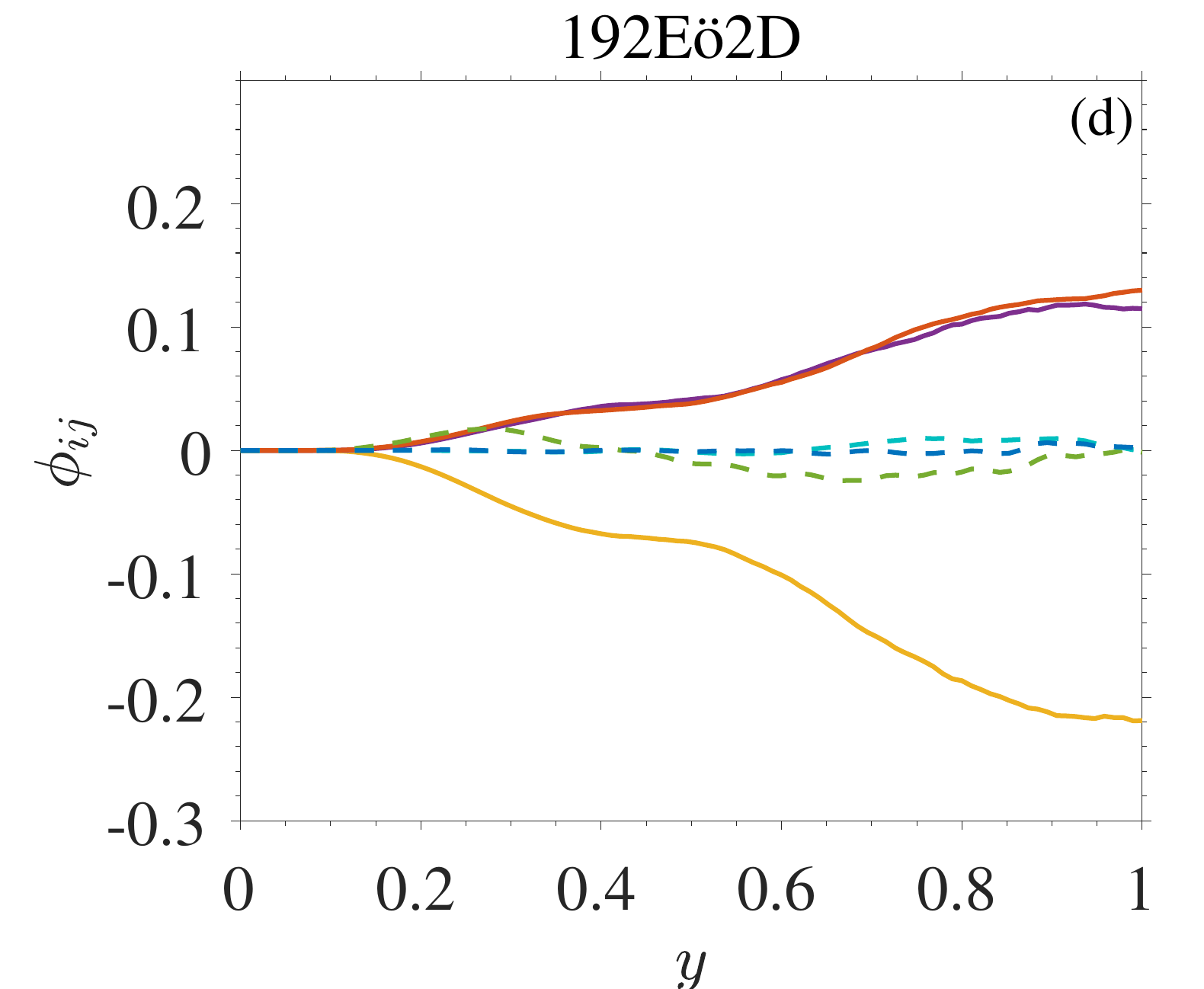}
    \end{subfigure}
    
    \begin{subfigure}[b]{0.5\textwidth}
        \centering
        \includegraphics[height=5cm,keepaspectratio]{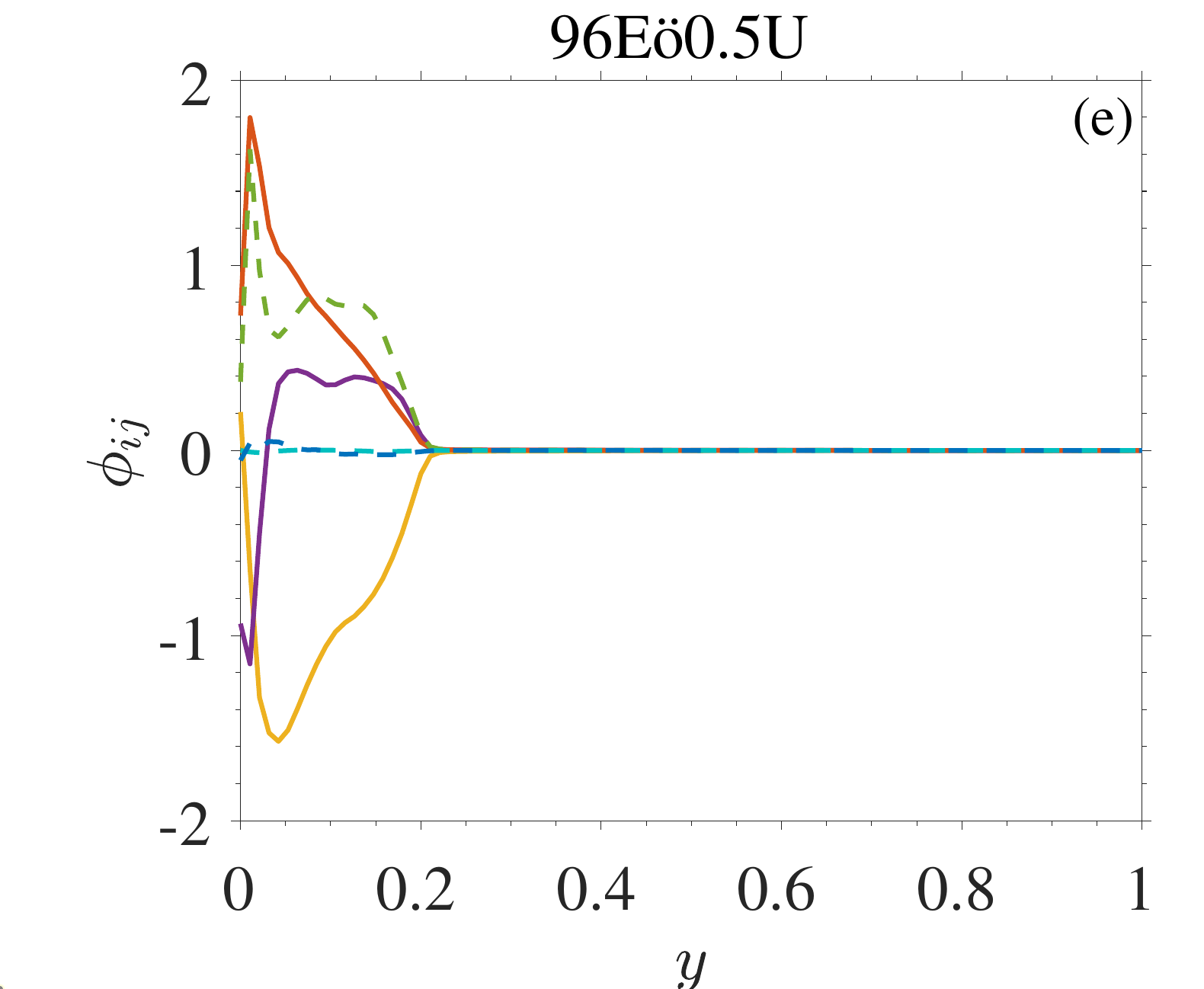}
    \end{subfigure}%
    \hspace{-5mm}
    \begin{subfigure}[b]{0.5\textwidth}
        \centering
        \includegraphics[height=5cm,keepaspectratio]{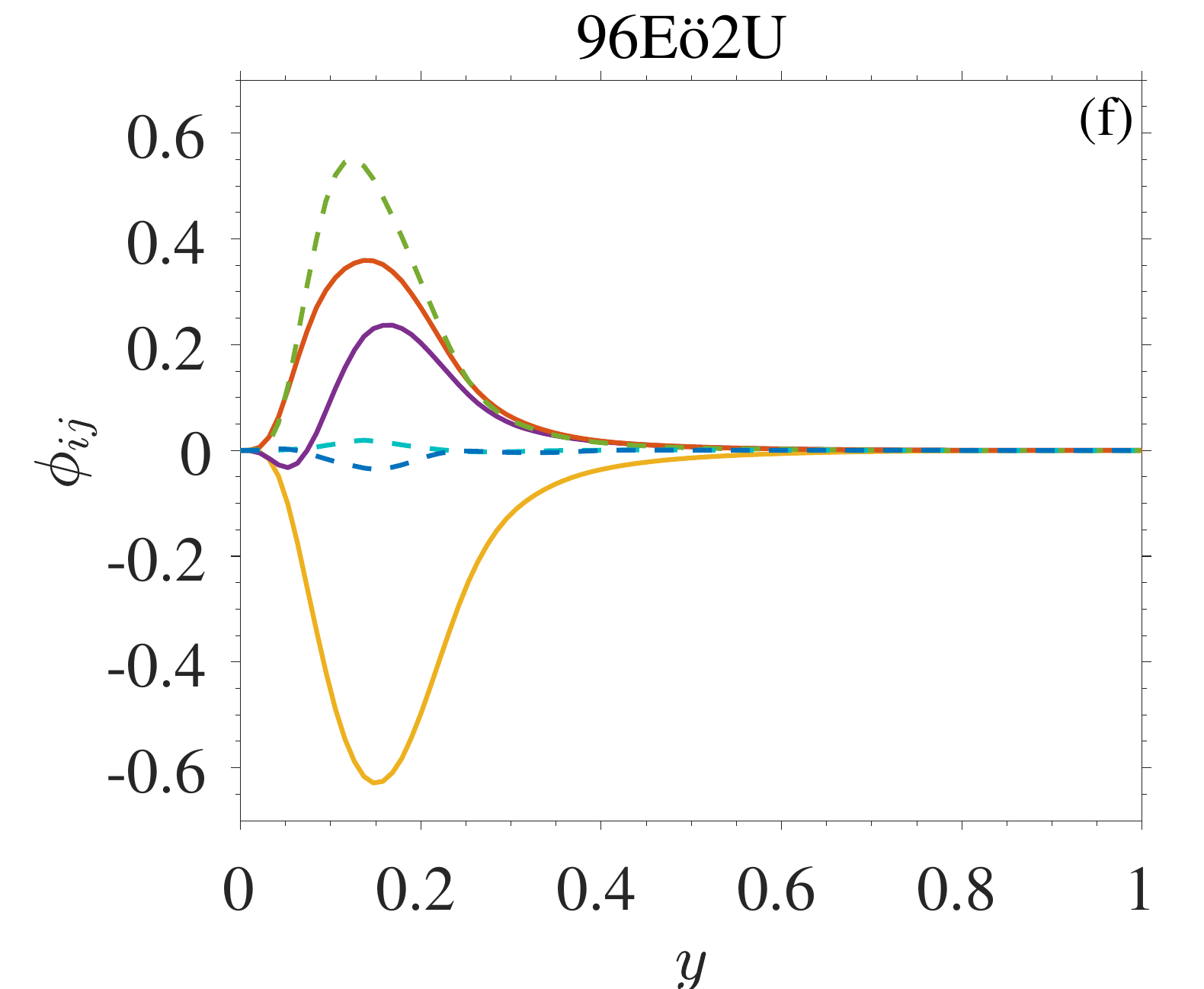}
    \end{subfigure}
    
    \begin{subfigure}[b]{0.5\textwidth}
        \centering
        \includegraphics[height=5cm,keepaspectratio]{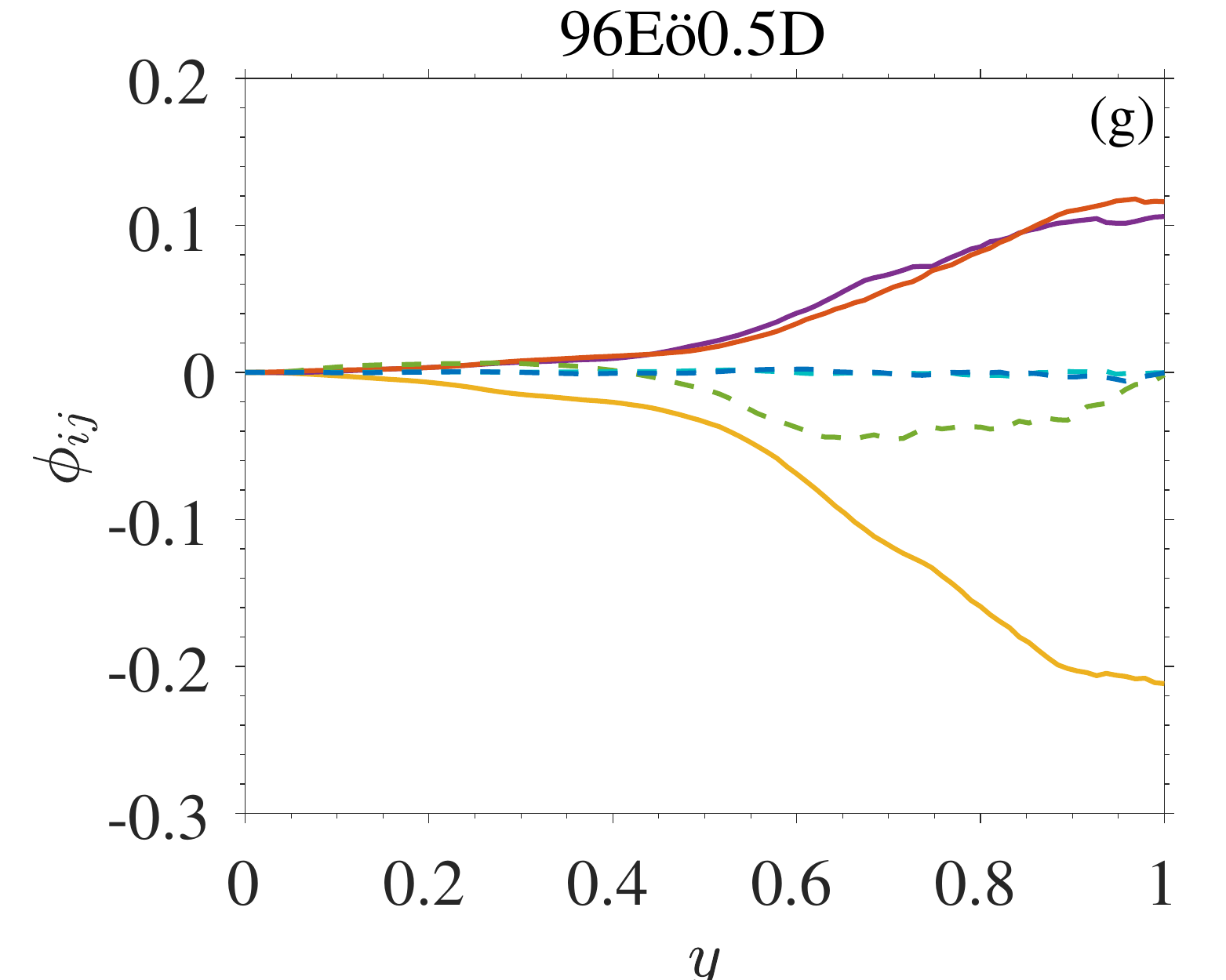}
    \end{subfigure}%
    \hspace{-5mm}
    \begin{subfigure}[b]{0.5\textwidth}
        \centering
        \includegraphics[height=5cm,keepaspectratio]{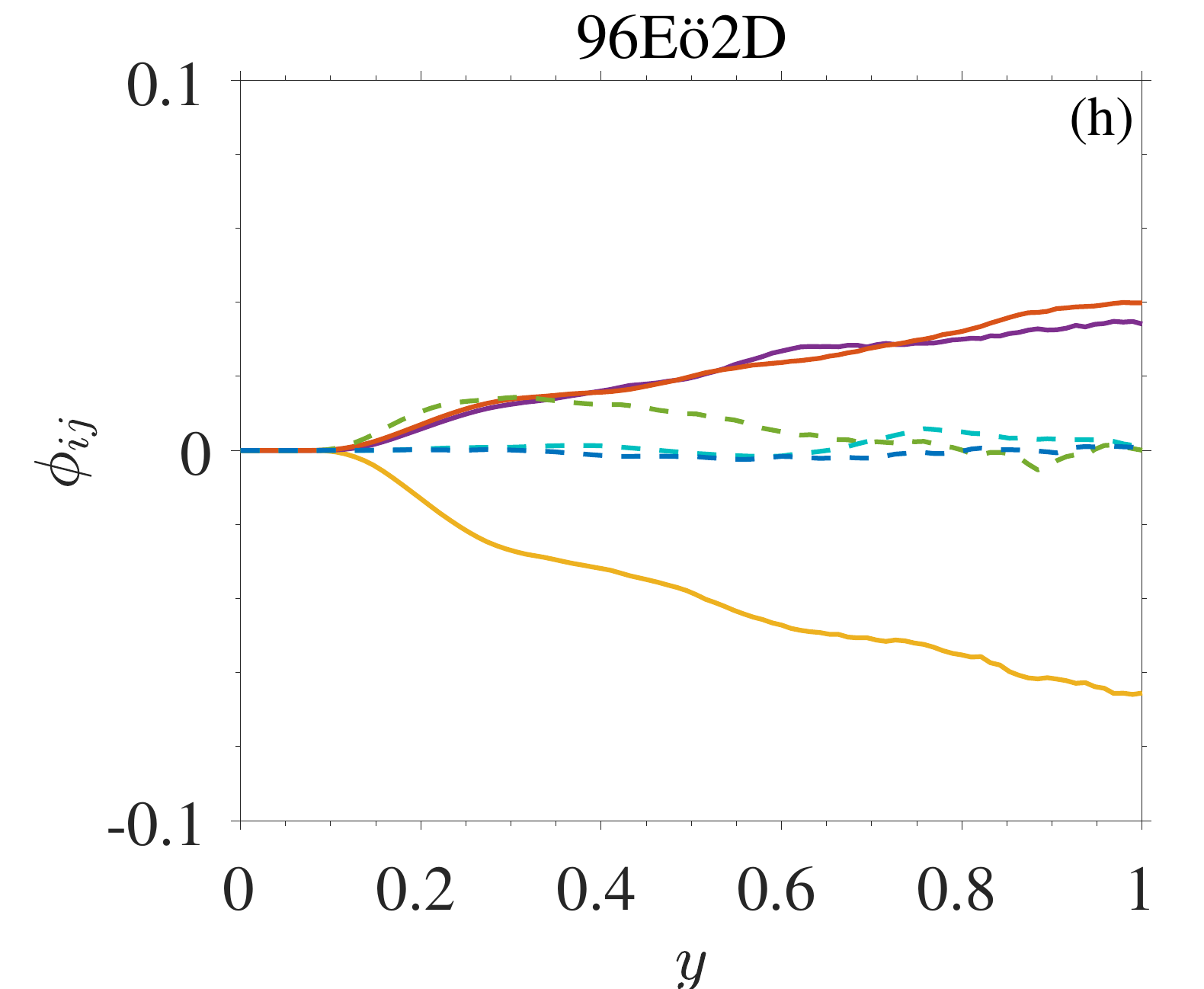}
    \end{subfigure}
    
    \caption{Pressure strain for 8 cases} 
    \label{fig:Pressure_strain}
\end{figure*}

Apart from the interfacial term, the pressure-strain term $\phi_{ij}$, which includes the gradients of velocity fluctuations, stands as the sole correlation containing directional information and plays a crucial role in capturing the anisotropy of Reynolds stress. FIG.~\ref{fig:Pressure_strain} displays the pressure strain for eight cases, corresponding to an \textit{a priori} evaluation of the pressure-strain terms from Eq.~(\ref{eq:pressurestrain}). In the figures, $\phi_{11}$, $\phi_{22}$, and $\phi_{33}$ are depicted using solid lines, while the other three quantities are shown with dashed lines. Here, $\phi_{11}$ is negative, whereas $\phi_{22}$ and $\phi_{33}$ are positive. In downward scenarios, $\phi_{22}$ and $\phi_{33}$ display comparable values, with $\phi_{11}$ having a magnitude nearly twice that of $\phi_{22}$. The terms $\phi_{13}$ and $\phi_{23}$ are close to zero. The cases with an Eötvös number of 0.5 exhibit a larger amplitude compared to those with an Eötvös number of 2, which is attributed to the additional pseudo-turbulence caused by bubble clustering.

In the upward cases for Eötvös number 2, $\phi_{11}$, $\phi_{22}$, $\phi_{33}$, and $\phi_{12}$ peak around $y = 0.15$, with $\phi_{12}$ being more dominant than $\phi_{22}$ and $\phi_{33}$. The term $\phi_{12}$ describes the momentum exchange between the streamwise and wall-normal directions, indicating significant changes in the velocity gradient in the wall-normal direction. This can be seen in FIG.~\ref{fig:Velocity_fluctuations}(g). Additionally, both $\phi_{22}$ and $\phi_{22}$ exhibit high values near the wall, with substantial gradients, as illustrated in FIG.~\ref{fig:Velocity_fluctuations}(c) and (e).



\section{Conclusions}

Dispersed bubbly turbulent flow in a channel is investigated through interface-resolved direct numerical simulation. An efficient CLSVOF solver is utilized to simulate bubble channel flow, where the initial diameter of each bubble remains constant and each bubble is represented by 20 grids per diameter. Depending on the number of bubbles (96 and 192), variations in flow direction and Eötvös number, eight cases are investigated via DNS to study the impact of these variables. The primary research findings are as follows.

In upward flow, bubbles accumulate near the wall. The smaller the Eötvös number, the closer the bubbles are to the wall, and the greater the attenuation of the liquid phase velocity. More bubbles near the wall serve as a medium for energy dissipation, helping to absorb and disperse the energy produced during fluid motion. Moreover, a smaller Eötvös number results in bubbles that are closer to spherical, producing turbulence that is more isotropic. When the Eötvös number is 0.5, \(\overline{u'_{y}u'_{y}}\) and \(\overline{u'_{z}u'_{z}}\) increase near the wall, as the wall itself is a source of disturbance, and the proximity of bubbles to the wall increases nearby turbulence and vorticity.

With an Eötvös number of 0.5, bubbles are closer to the wall. The stirring action of the bubbles, especially in areas dense with bubbles, disrupts the orderly structure of the flow, effectively breaking up large-scale vortices and redistributing their energy into smaller-scale vortices that are more isotropic. Additionally, the upward motion of the bubbles induces extra motion in directions perpendicular to the flow, aiding in the dispersion of energy across multiple directions.

Analyzing the interfacial energy transfer \(S_k\) from the k-equation, \(S_k\) consists of two components, one related to fluctuations of pressure and velocity, and the other associated with shear stress and velocity fluctuations. Given that shear stress originates from velocity gradients, \(S_k\) exhibits higher values near the wall-side interface. Consequently, the current model source term \(S_k^{\text{RANS}}\) in upward cases does not align well with \(S_k\).

Analyzing the exact balance equation for the Reynolds stresses and its interfacial energy transfer term \(S_{R,ij}\) and pressure-strain \(\phi_{ij}\), the motion of bubbles near the wall generates significant velocity gradients, especially in the wall-normal direction. This results in \(S_{R,22}\) becoming a major contributor to \(S_{R,ij}\). \(\phi_{ij}\) points to the same conclusion.

In downward flow scenarios, bubbles are observed to cluster in the middle of the channel. The streamwise velocity of the bubbles is significantly lower than that of the liquid in the center of the channel, a consequence of the buoyancy effect due to the density difference. Bubbles in the center of the channel induce additional turbulence kinetic energy and also attenuate the energy in the buffer layer. This phenomenon is reflected in velocity fluctuations, Reynolds shear stress, the Lumley triangle, \(S_{R,ij}\), and \(\phi_{ij}\).

\section{Acknowledgments}

The study is funded by Deutsche Forschungsgemeinschaft (DFG, German Research Foundation) under Germany’s Excellence
Strategy-EXC2075-390740016. We also thank the Deutsche Forschungsgemeinschaft (DFG, German Research Foundation) for
supporting this work by funding DFG-SFB 1313, Project No.327154368. X. Zhang and Y. Liu acknowledges the support from the Chinese Scholarship Council (CSC). G. Yang kindly acknowledges the support from Natural Science Foundation of China (NSFC: 52276013).
The authors gratefully appreciate the access to the high performance computing facility Hawk at HLRS, Stuttgart of Germany.

\bibliography{references}

\appendix

\end{document}